\documentclass[smallextended]{svjour3}     

\usepackage[utf8]{inputenc}
\usepackage[T1]{fontenc}

\usepackage{amsmath,amssymb,amsfonts}
\usepackage{array}
\usepackage{graphicx}
\usepackage{subfig}
\usepackage{balance}
\usepackage{esvect}
\usepackage{booktabs}
\usepackage{tabularx}
\usepackage{anyfontsize}
\usepackage[all]{nowidow}
\usepackage[normalem]{ulem}
\usepackage{multirow, makecell}
\usepackage{enumitem}
\usepackage[hidelinks]{hyperref}
\usepackage{wrapfig}

\usepackage{natbib}
\usepackage{xcolor}
\colorlet{rev}{black}
\colorlet{rev2}{black}
\colorlet{rev3}{black}

\newcommand{\var}[1]{\mathit{#1}}
\newcommand{\fun}[1]{\mathit{#1}}

\newcolumntype{Y}{>{\centering\arraybackslash}X}
\newcolumntype{R}{>{\raggedleft\arraybackslash}X}

\usepackage{framed}
\usepackage{tikz}
\definecolor{light-gray}{gray}{0.9}
\usepackage[framemethod=TikZ]{mdframed}
\mdfsetup{skipabove=5pt,skipbelow=3pt}
\usepackage{lipsum}
\mdfdefinestyle{RQFrame}{%
	linecolor=black,
	outerlinewidth=0.15pt,
	roundcorner=3pt,
	innertopmargin=2pt,
	innerbottommargin=2pt,
	innerrightmargin=4pt,
	innerleftmargin=4pt,
	backgroundcolor=light-gray}

\usepackage{colortbl}
\usepackage{listings}

\definecolor{javared}{rgb}{0.6,0,0} 
\definecolor{javagreen}{rgb}{0.25,0.5,0.35} 
\definecolor{javapurple}{rgb}{0.5,0,0.35} 
\definecolor{javadocblue}{rgb}{0.25,0.35,0.75} 
\definecolor{javagrey}{rgb}{0.46,0.45,0.48} 

\lstdefinestyle{Alg}{
  basicstyle=\ttfamily\footnotesize,
  breaklines=true,
  tabsize=2,
  mathescape,
  numbers=left,
  xleftmargin=2.5em,
  xrightmargin=0.5em,
  frame=tb,
  framexleftmargin=2em,
  emph={Algorithm,Input,Output,for,each,do,if,else,Function,while,let,be,repeat,until,return,times,and,or,break,in,then,},
  emphstyle={\textbf},
  escapechar=?,
  morecomment=[l][\color{javagreen}]{//},
  columns=flexible,
}

\journalname{Empirical Software Engineering}
\begin{document}

\setcounter{page}{1}



\title{Optimal Priority Assignment for Real-Time Systems:\\A Coevolution-Based Approach}
\titlerunning{Optimal Priority Assignment for Real-Time Systems}

\author{Jaekwon Lee \and Seung Yeob Shin \and \hbox{Shiva Nejati} \and Lionel C. Briand}

\institute{
            Jaekwon Lee \at
            {SnT, University of Luxembourg, Luxembourg} \\
            {University of Ottawa, Canada} \\
            \email{jaekwon.lee@uni.lu} \\
        \and
            Seung Yeob Shin \at
            {SnT, University of Luxembourg, Luxembourg} \\
            \email{seungyeob.shin@uni.lu} \\
        \and
            Shiva Nejati \at 
            {University of Ottawa, Canada} \\
            {SnT, University of Luxembourg, Luxembourg} \\
            \email{snejati@uottawa.ca} \\
        \and
            Lionel C. Briand \at
            {SnT, University of Luxembourg, Luxembourg} \\
            {University of Ottawa, Canada} \\
            \email{lionel.briand@uni.lu} \\
}

%

\date{Received: date / Accepted: date}

\maketitle

\newcommand{\APPROACH}{OPAM}
\newcommand{\PARTNER}{LuxSpace}

\begin{abstract}
In real-time systems, priorities assigned to real-time tasks determine the order of task executions, by relying on an underlying task scheduling policy. Assigning optimal priority values to tasks is critical to allow the tasks to complete their executions while maximizing safety margins from their specified deadlines. This enables real-time systems to tolerate unexpected overheads in task executions and still meet their deadlines. In practice, priority assignments result from an interactive process between the development and testing teams. In this article, we propose an automated method that aims to identify the best possible priority assignments in real-time systems, accounting for multiple objectives regarding safety margins and engineering constraints. Our approach is based on a multi-objective, competitive coevolutionary algorithm mimicking the interactive priority assignment process between the development and testing teams. We evaluate our approach by applying it to six industrial systems from different domains and several synthetic systems. The results indicate that our approach significantly outperforms both our baselines, i.e., random search and sequential search, and solutions defined by practitioners.Our approach scales to complex industrial systems as an offline analysis method that attempts to find near-optimal solutions within acceptable time, i.e., less than 16 hours.
\end{abstract}

\section{Introduction}
\label{sec:intro}

Mission-critical systems are found in many different application domains, such as aerospace, automotive, and healthcare domains. The success of such systems depends on both functional and temporal correctness. For functional correctness, systems are required to provide appropriate outputs in response to the corresponding stimuli. Regarding temporal correctness, systems are supposed to generate outputs within specified time constraints, often referred to as deadlines. The systems that have to comply with such deadlines are known as real-time systems~\citep{Liu2000}. Real-time systems typically run multiple tasks in parallel and rely on a real-time scheduling policy to decide which tasks should have access to processing cores, i.e., CPUs, at any given time. 

While developing a real-time system, one of the most common problems that engineers face is the assignment of priorities to real-time tasks in order for the system to meet its deadlines. Based on priorities of real-time tasks, the system's task scheduler determines a particular order for allocating real-time tasks to processing cores. Hence, a priority assignment that is poorly designed by engineers makes the system scheduler execute tasks in an order that is far from optimal. In addition, the system will likely violate its performance and time constraints, i.e., deadlines, if a poor priority assignment is used. 

In real-time systems, the problem of optimally assigning priorities to tasks is important not only to avoid deadline misses but also to maximize \emph{safety margins} from task deadlines and is subject to \emph{engineering constraints}. Tasks may exceed their expected execution times due to unexpected interrupts. For example, it is infeasible to test an aerospace system exhaustively on the ground such that potential environmental uncertainties, e.g., those related to space radiations, are accounted for. Hence, engineers assign optimal priorities to tasks such that the remaining times from tasks' completion times to their deadlines, i.e., safety margins, are maximized to cope with potential uncertainties. Furthermore, engineers typically have to account for additional engineering constraints, e.g., they assign higher priorities to critical tasks that must always meet their deadlines compared to the tasks that are less critical or non-critical.

A brute force approach to find an optimal priority assignment would have to examine all $n!$ distinct priority assignments, where $n$ denotes the number of tasks. Furthermore, for a given priority assignment, schedulability analysis is, in general, known as a hard problem~\citep{Audsley2001}, which determines whether or not tasks will always complete their executions within their specified deadlines. Thus, optimizing priority assignments is also a hard problem because the space of all possible system states to explore in order to find optimal priority assignments is very large. Most of the prior works on optimizing priority assignments provide analytical methods~\citep{Fineberg1967,Leung1982,Audsley1991,Davis2007,Chu2008,Davis2009,Davis2012}, which rely on well-defined system models and are very restrictive. For example, they assume that tasks are independent, i.e., tasks do not share resources~\citep{Davis2016,Zhao2017}. Industrial systems, however, are typically not compatible with such (simple) system models. In addition, none of the existing work addresses the problem of optimizing priority assignments by simultaneously accounting for multiple objectives, such as safety margins and engineering constraints, as discussed above.

Search-based software engineering (SBSE) has been successfully applied in many application domains, including software testing~\citep{Wegener1997,Wegener1998,Lin2009,Arcuri2010,Shin2018}, program repair~\citep{Weimer2009,Tan2016,Abdessalem2020}, and self-adaptation~\citep{Andrade2013,Chen2018,Shin2020},  where the search spaces are very large. Despite the success of SBSE, engineering problems in real-time systems have received much less attention in the SBSE community. In the context of real-time systems, there exists limited work on finding stress test scenarios~\citep{Briand2005} and predicting worst-case execution times~\citep{Lee2020}, which complements our work.

In practice, priority assignments result from an interactive process between the development and testing teams. While developing a real-time system, developers assign priorities to real-time tasks in the system and then testers stress the system to check whether or not the system meets its specified deadlines. If testers find a problematic condition under which any of the tasks violates its deadline, developers have to modify the priority assignment  to address the problem. The back-and-forth between the development and testing teams continues until a priority assignment that does not lead to any deadline miss is found or the one that yields the least critical deadline misses is identified. The process is, however, not automated.

In this article, we use metaheuristic search algorithms to automate the process of assigning priorities to real-time tasks. To mimic the interactive back-and-forth between the development and testing teams, we use competitive coevolutionary algorithms~\citep{Luke2013}. Coevolutionary algorithms are a specialized class of evolutionary search algorithms. They simultaneously coevolve two populations (also called species) of (candidate) solutions for a given problem. They can be cooperative or competitive.
Such competitive coevolution is similar to what happens in nature between predators and preys. For example, faster preys escape predators more easily, and hence they have a higher probability of generating offspring. This impacts the predators, because they need to evolve as well to become faster if they want to feed and survive~\citep{Meneghini2016}. Hence, the two species, i.e., predators and preys, have coevolved competitively. We note that no species has the competing traits of predators and preys simultaneously as such species could not evolve to survive.
In our context, priority assignments defined by developers can be seen as preys and stress test scenarios as predators. The priority assignments need to evolve so that stress testing is not able to push the system into breaking its real-time constraints. Dually, stress test scenarios should evolve to be able to break the system when there is a chance to do so.

\noindent\textbf{Contributions.}
We propose an \underline{O}ptimal \underline{P}riority \underline{A}ssignment \underline{M}ethod for real-time systems (OPAM). Specifically, we apply multi-objective, two-population competitive coevolution~\citep{Popovici2012} to address the problem of finding near-optimal priority assignments, aiming at maximizing the magnitude of safety margins from deadlines and constraint satisfaction. In OPAM, two species relate to priority assignment and stress testing coevolve synchronously, and compete against each other to find the best possible solutions. We evaluated OPAM by applying it to six complex, industrial systems from different domains, including the aerospace, automotive, and avionics domains, and several synthetic systems. Our results show that: (1)~OPAM finds significantly better priority assignments compared to our baselines, i.e., random search and sequential search, (2)~the execution time of OPAM scales linearly with the number of tasks in a system and the time required to simulate task executions, and (3)~OPAM priority assignments significantly outperform those manually defined by engineers based on domain expertise.

We note that OPAM is the first attempt to apply coevolutionary algorithms to address the problem of priority assignment. Further, it enables engineers to explore trade-offs among different priority assignments with respect to two objectives: maximizing safety margins and satisfying engineering constraints. Our full evaluation package is available online~\citep{Artifacts}.

\noindent\textbf{Organization.}
The remainder of this article is structured as follows: Section~\ref{sec:motivation} motivates our work. Section~\ref{sec:problem} defines our specific problem of priority assignment in practical terms. Section~\ref{sec:relatedwork} discusses related work. Sections~\ref{sec:overview} and \ref{sec:approach} describe OPAM. Section~\ref{sec:eval} evaluates OPAM. Section~\ref{sec:conclusion} concludes this article.

\section{Motivating case study}
\label{sec:motivation}

We motivate our work using an industrial case study from the satellite domain. Our case study concerns a mission-critical real-time satellite, named ESAIL~\citep{ESAIL}, which has been developed by LuxSpace -- a leading system integrator for microsatellites and aerospace system. ESAIL tracks vessels' movements over the entire globe as the satellite orbits the earth. The vessel-tracking service provided by ESAIL requires real-time processing of messages received from vessels in order to ensure that their voyages are safe with the assistance of accurate, prompt route provisions. Also, as ESAIL orbits the planet, it must be oriented in the proper position on time in order to provide services correctly. Hence, ESAIL's key operations, implemented as real-time tasks, need to be completed within acceptable times, i.e., deadlines.

Engineers at LuxSpace analyze the schedulability of ESAIL across different development stages. At an early design stage, the engineers use a priority assignment method that extends the rate monotonic scheduling policy~\citep{Fineberg1967}, which is a theoretical priory assignment algorithm used in real-time systems. At a later development stage, if the engineers found that any real-time task of ESAIL cannot complete its execution within its deadline, the engineers, in our study context, reassign priorities to tasks in order to address the problem of deadline violations.

The rate monotonic policy assigns priorities to tasks that arrive to be executed periodically and must be completed within a certain amount of time, i.e., periodic tasks with hard deadlines. According to the policy, periodic tasks that arrive frequently have higher priorities than those of other tasks that arrive rarely. In ESAIL, for example, if the vessel-tracking task arrives every 100ms and the satellite-position control task arrives every 150ms, the former has a higher priority than the latter. However, the rate monotonic policy does not account for tasks that arrive irregularly and should be completed within a reasonable amount of time, i.e., aperiodic tasks with soft deadlines. ESAIL contains aperiodic tasks with soft deadlines as well, such as a task for updating software. Hence, the engineers extend the rate monotonic policy to assign priorities to all tasks of ESAIL. The extensions are as follows: First, the engineers assign priorities to periodic tasks based on the rate monotonic policy. Second, the engineers assign lower priorities to aperiodic tasks than those of periodic tasks. As aperiodic tasks with soft deadlines are typically considered less critical than periodic tasks with hard deadlines, the engineers aim to ensure that periodic tasks complete their executions within their deadlines by assigning lower priorities to aperiodic tasks while periodic tasks have higher priority. Engineers use a heuristic to assign priorities to aperiodic tasks. They treat aperiodic tasks as (pseudo-)periodic tasks by setting aperiodic tasks' (expected) minimum arrival rates as their fixed arrival periods, making the tasks frequently arrive. The engineers then apply the rate monotonic policy for the aperiodic tasks with the synthetic periods while ensuring that aperiodic tasks have lower priorities than those of periodic tasks. 

A priority assignment made at an early design stage keeps changing while developing ESAIL due to various reasons, such as changes in requirements and implementation constraints. At a development stage, instead of relying on the extended rate monotonic policy, the engineers assign priorities based on their domain expertise, manually inspecting schedulability analysis results. Hence, a priority assignment at later development stages often does not follow the extended rate monotonic policy. For example, as aperiodic tasks are also expected to be completed within a reasonable amount of time, some aperiodic tasks may have higher priorities than some periodic tasks as long as they are schedulable.

\begin{sloppypar}
Engineers at LuxSpace, however, are still faced with the following issues: (1)~Their priority assignment method, which extends the rate monotonic scheduling policy, assigns priorities to tasks in order to ensure only that tasks are to be schedulable. However, engineers have a pressing need to understand the quality of priority assignments in detail as they impact ESAIL operations differently. For example, once ESAIL is launched into orbit, the satellite operates in the space environment, which is inherently impossible to be fully tested on the ground. Unexpected space radiations may trigger unusual system interrupts, which hasn't been observed on the ground, resulting in overruns of ESAIL tasks' executions. In such cases, a priority assignment assessed on the ground may not be able to tolerate such unexpected uncertainties. Hence, engineers need a priority assignment that enables ESAIL tasks to tolerate unpredictable uncertainties as much as possible and to be schedulable. (2)~Engineers at LuxSpace assign priorities to tasks without any systematic assistance. Instead, they rely on their expertise and the current practices described above to manually assign priorities to ensure that tasks are to be schedulable. To this end, we are collaborating with LuxSpace to develop a solution for addressing these issues in assigning task priority.
\end{sloppypar}   
\section{Problem description}
\label{sec:problem}

\begin{figure}[t]
    \centering\includegraphics[width=.9\columnwidth]{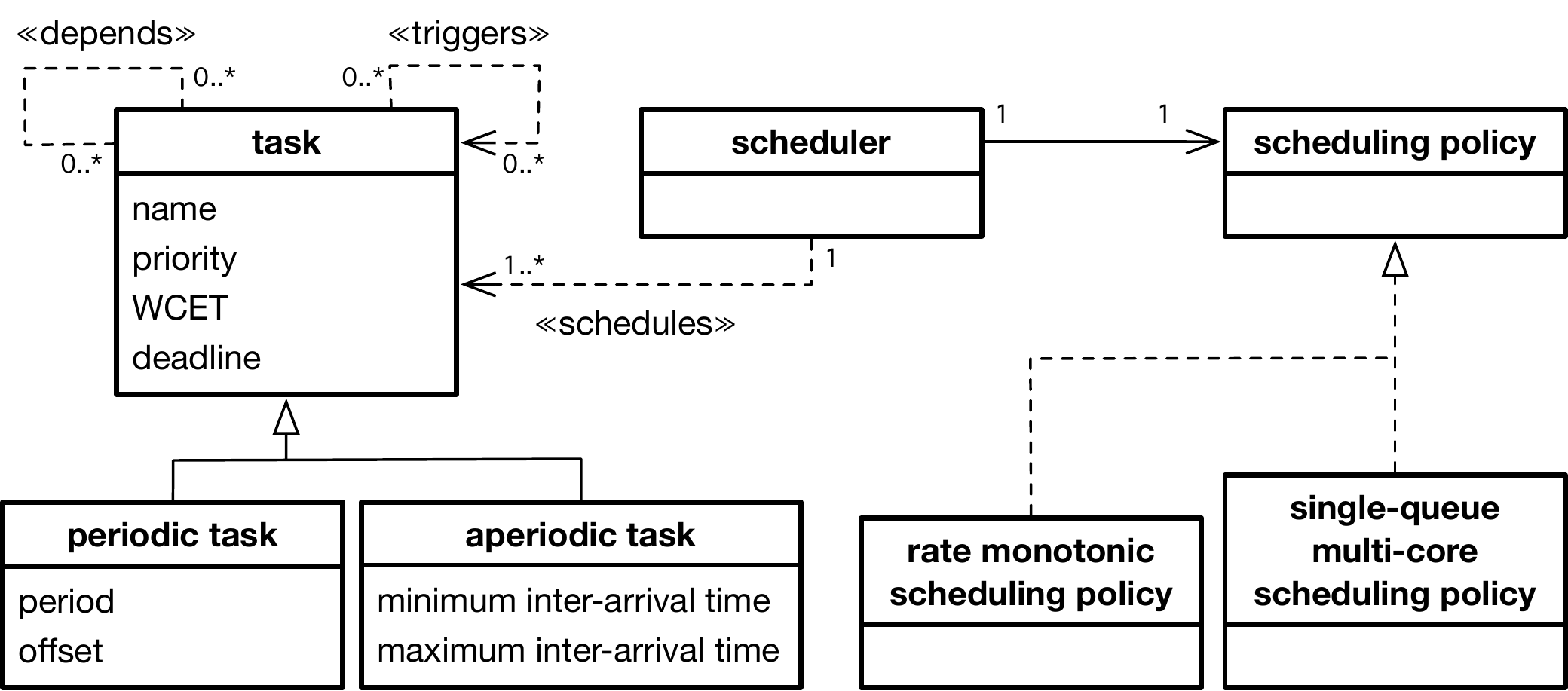}
    \caption{A conceptual model representing the key abstractions to analyze optimal priority assignments.}
    \label{fig:conceptual model}
\end{figure}

This section defines the task, scheduler, and schedulability concepts, which extend the concepts defined in our previous work~\citep{Lee2020} by augmenting our previous definitions with the notions of safety margins, constraints in assigning priorities, and relationships between real-time tasks. We then describe the problem of optimizing priority assignments such that we maximize the magnitude of safety margins and the degree of constraint satisfaction. Figure~\ref{fig:conceptual model} shows an overview of the conceptual model that represents the key abstractions required to analyze optimal priority assignments for real-time systems. The entities in the conceptual model are described below.

\textbf{Task.} We denote by $j$ a real-time task that should complete its execution within a specified deadline after it is activated (or arrived). Every real-time task $j$ has the following properties: priority denoted by $\fun{pr}(j)$, deadline denoted by $\fun{dl}(j)$, and worst-case execution time (WCET) denoted by $\fun{wcet}(j)$. Task priority $\fun{pr}$ determines if an execution of a task is preempted by another task. Typically, a task $j$ preempts the execution of a task $j^\prime$ if the priority of $j$ is higher than the priority of $j^\prime$, i.e., $\fun{pr}(j) > \fun{pr}(j^\prime)$.
\textcolor{rev3}{The $\fun{pr}(j)$ priority is a fixed value assigned to task $j$. Such fixed priorities are determined offline; hence, they are not changed online for any reason. Note that a real-time task scheduler that relies on fixed priorities is applied in all the study subjects in this article (see Section~\ref{subsec:industrial subjects}) and is commonly used in industrial systems~\citep{Briand2005, Guan2009, Lin2009, Anssi2011, Zeng2014, Alesio2015, Durr2019, Lee2020Panda}.}

The $\fun{dl}(j)$ function determines the deadline of a task $j$ relative to its arrival time. A task deadline can be either \emph{hard} or \emph{soft}. A hard deadline of a task $j$ constrains that $j$ \emph{must} complete its execution within a deadline $\fun{dl}(j)$ after $j$ is activated. 
While violations of hard deadlines are not acceptable, depending on the operating context of a system, violating soft deadlines may be to some extent tolerated. Note that we use a metaheuristic search relying on fitness functions quantifying the degrees of deadline misses, safety margins, and constraint satisfaction. Such functions do not depend on the nature of the deadlines. Our approach outputs a set of priority assignments that are Pareto optimal with respect to safety margins and constraint satisfaction. Engineers then perform domain-specific trade-off analysis among Pareto solutions. Hence, in this article, we handle hard and soft deadline tasks in the same manner.

Real-time tasks are either \emph{periodic} or \emph{aperiodic}. Periodic tasks, which are typically triggered by timed events, are invoked at regular intervals specified by their \emph{period}. We denote by $\fun{pd}(j)$ the period of a periodic task $j$, i.e., a fixed time interval between subsequent activations (or arrivals) of $j$. Any task that is not periodic is called aperiodic. Aperiodic tasks have irregular arrival times and are activated by external stimuli which occur irregularly.
In real-time analysis, based on domain knowledge, we typically specify a minimum inter-arrival time denoted by $\fun{pmin}(j)$ and a maximum inter-arrival time denoted by $\fun{pmax}(j)$ indicating the minimum and maximum time intervals between two consecutive arrivals of an aperiodic task $j$. In real-time analysis, sporadic tasks are often separately defined as having irregular arrival intervals and hard deadlines~\citep{Liu2000}. In our conceptual definitions, however, we do not introduce new notations for sporadic tasks because the deadline and period concepts defined above sufficiently characterize sporadic tasks. Note that for periodic tasks $j$, we have $\fun{pmin}(j) = \fun{pmax}(j) = \fun{pd}(j)$. Otherwise, for aperiodic tasks $j$, we have $\fun{pmax}(j) > \fun{pmin}(j)$. 

\textbf{Task relationships.} The execution of a task $j$ depends not only on its own parameters described above, e.g., priority $\fun{pr}(j)$ and period $\fun{pd}(j)$, but also on its relationships with other tasks. Relationships between tasks are typically determined by task interactions related to accessing shared resources and triggering arrivals of other tasks~\citep{Alesio2012}. Specifically, if two tasks $j$ and $j^\prime$ access a shared resource $r$ in a mutually exclusive way, $j$ may be blocked from executing for the period during which $j^\prime$ accesses $r$. We denote by $\fun{dp}(j,j^\prime)$ the resource-dependency relation between tasks $j$ and $j^\prime$ that holds if $j$ and $j^\prime$ have mutually exclusive access to a shared resource $r$ such that they cannot be executed in parallel or preempt each other, but one can execute only after the other has completed accessing $r$.

The other type of relationship between tasks is related to a task $j$ triggering the arrival of another task $j^\prime$. This is a common interaction between tasks~\citep{Locke1990,Anssi2011,Alesio2015}. For example, $j$ may hand over some of its workload to $j^\prime$ due to performance or reliability reasons. We denote by $\fun{tr}(j,j^\prime)$ the triggering relation between tasks $j$ and $j^\prime$ that holds if $j$ triggers the arrival of $j^\prime$. We note that both relationships are defined at the level of tasks, following prior works~\citep{Locke1990,Anssi2011,Alesio2015} describing the five industrial case study systems used in our experiments (see Section~\ref{subsec:industrial subjects}). 

\textbf{Scheduler.} Let $J$ be a set of tasks to be scheduled by a real-time scheduler. A scheduler then dynamically schedules executions of tasks in $J$ according to the tasks' arrivals and the scheduler's scheduling policy over the scheduling period $\mathbb{T} = [0,\mathbf{T}]$. We denote by $\fun{at}_k(j)$ the $k$th arrival time of a task $j \in J$. The first arrival of a periodic task $j$ does not always occur immediately at the system start time ($0$). Such offset time from the system start time to the first arrival time $\fun{at}_1(j)$ of $j$ is denoted by $\fun{offset}(j)$. For a periodic task $j$, the $k$th arrival of $j$ within $\mathbb{T}$ is $\fun{at}_k(j) \leq \mathbf{T}$ and is computed by $\fun{at}_k(j) = \fun{offset}(j) + (k-1) \cdot \fun{pd}(j)$. For an aperiodic task $j^\prime$, $\fun{at}_k(j^\prime)$ is determined based on the $k{-}1$th arrival time of $j^\prime$ and its minimum and maximum arrival times. Specifically, for $k > 1$, $\fun{at}_k(j^\prime) \in [\fun{at}_{k-1}(j^\prime)+\fun{pmin}(j^\prime), \fun{at}_{k-1}(j^\prime)+\fun{pmax}(j^\prime)]$ and, for $k = 1$, $\fun{at}_1(j^\prime) \in [\fun{pmin}(j^\prime), \fun{pmax}(j^\prime)]$, where $\fun{at}_k(j^\prime) < \mathbf{T}$.

A scheduler reacts to a task arrival at $\fun{at}_k(j)$ by scheduling the execution of $j$. Depending on a scheduling policy (e.g., rate monotonic scheduling policy for single-core systems~\citep{Fineberg1967} and single-queue multi-core scheduling policy~\citep{Arpaci2018}), an arrived task $j$ may not start its execution at the same time as it arrives when higher priority tasks are executing on all processing cores. Also, task executions may be interrupted due to preemption. We denote by $\fun{et}_k(j)$ the completion time for the $k$th arrival of a task $j$. According to the worst-case execution time of a task $j$, we have: $\fun{et}_k(j) \ge \fun{at}_k(j) + \fun{wcet}(j)$. 

During system operation, a scheduler generates a \emph{schedule scenario} which describes a sequence of task arrivals and their completion time values. We define a schedule scenario as a set $S$ of tuples $(j, \fun{at}_k(j), \fun{et}_k(j))$ indicating that a task $j$ has arrived at $\fun{at}_k(j)$ and completed its execution at $\fun{et}_k(j)$. Due to a degree of randomness in task execution times and aperiodic task arrivals, a scheduler may generate a different schedule scenario for different runs of a system. 

\begin{figure}[t]
\begin{center}
    \subfloat[Schedule scenario $S$]{
        \parbox{1\columnwidth}{\centering
            \includegraphics[width=.9\columnwidth]{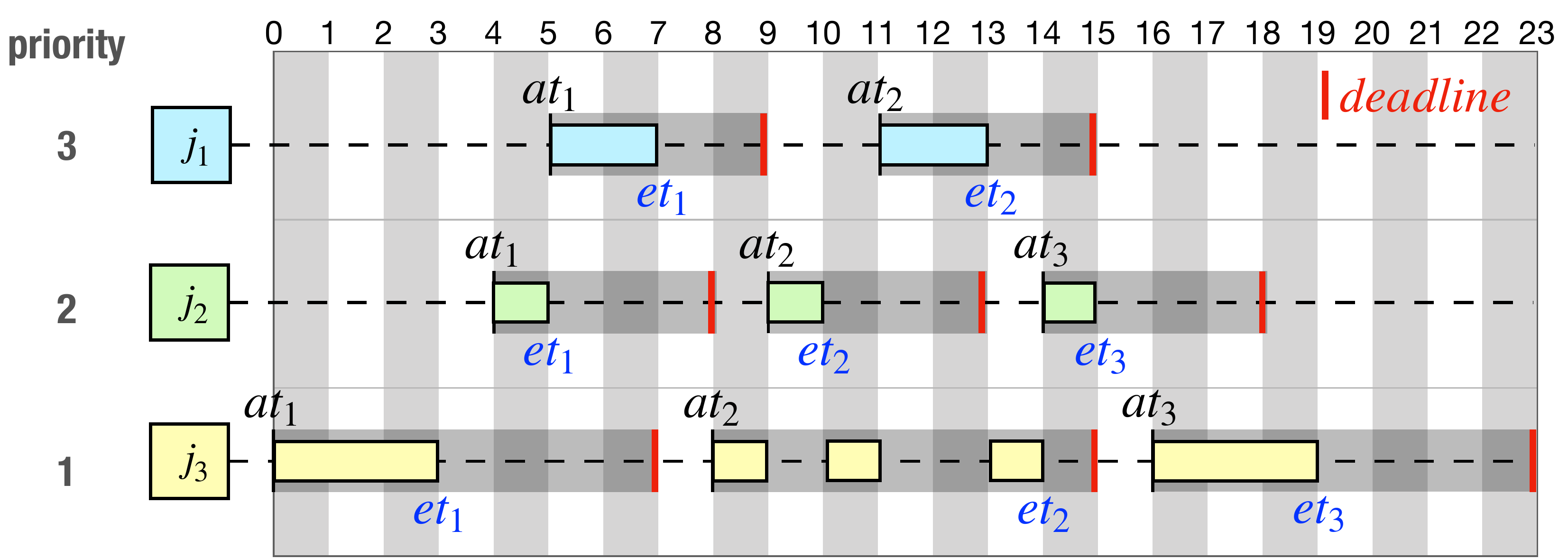}
            \label{fig:schedule 1}
        }
    }
    
    \subfloat[Schedule scenario $S^\prime$]{
        \parbox{1\columnwidth}{
            \centering
            \includegraphics[width=.9\columnwidth]{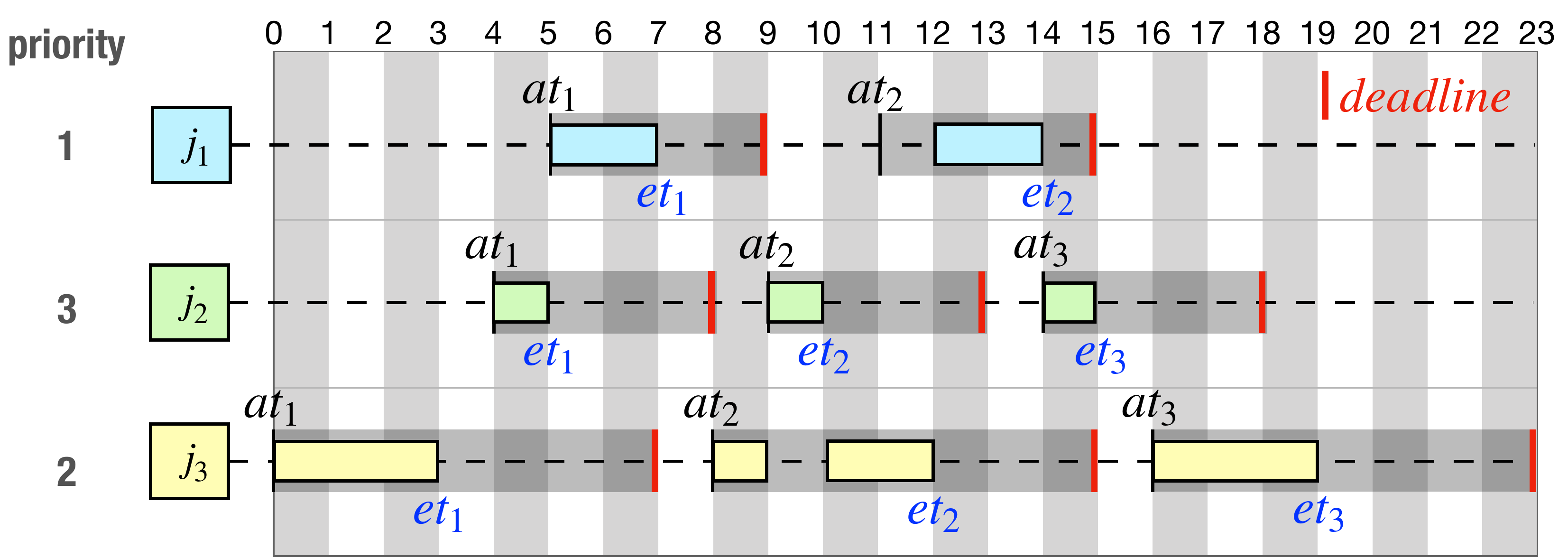}
            \label{fig:schedule 2}
        }
    }
    \caption{Example schedule scenarios $S$ and $S^\prime$ of three tasks: $j_1$, $j_2$, and $j_3$. (a)~The $S$ schedule scenario is produced when $\fun{pr}(j_1) = 3$, $\fun{pr}(j_2) = 2$, and $\fun{pr}(j_3) = 1$. (b)~The $S^\prime$ schedule scenario is produced when $\fun{pr}(j_1) = 1$, $\fun{pr}(j_2) = 3$, and $\fun{pr}(j_3) = 3$.}
    \label{fig:schedule}
\end{center}
\end{figure}

Figure~\ref{fig:schedule} shows two schedule scenarios $S$ (Figure~\ref{fig:schedule 1}) and $S^\prime$ (Figure~\ref{fig:schedule 2}) produced by a scheduler over the $[0,23]$ time period of a system run. Both $S$ and $S^\prime$ describe executions of three tasks, $j_1$, $j_2$, and $j_3$ arrived at the same time stamps (see $at_i$ in the figures). In both scenarios, the aperiodic task $j_1$ is characterized by: $\fun{pmin}(j_1) = 5$, $\fun{pmax}(j_1) = 13$, $\fun{dl}(j_1) = 4$, and $\fun{wcet}(j_1) = 2$. The aperiodic task $j_2$ is characterized by: $\fun{pmin}(j_2) = 3$, $\fun{pmax}(j_2) = 10$, $\fun{dl}(j_2) = 4$, and $\fun{wcet}(j_2) = 1$. The periodic task $j_3$ is characterised by: $\fun{pd}(j_3) = 8$, $\fun{dl}(j_3) = 7$, and $\fun{wcet}(j_3) = 3$. The priorities of the three tasks in $S$ (resp. $S^\prime$) satisfy the following: $pr(j_1) > pr(j_2) > pr(j_3)$ (resp. $pr(j_2) > pr(j_3) > pr(j_1)$). In both scenarios, task executions can be preempted depending on their priorities. Then, $S$ is defined by $S = \{(j_1, 5, 7)$, $\ldots$, $(j_2, 4, 5)$, $\ldots$, $(j_3, 8, 14)$, $(j_3, 16, 19))\}$; and $S^\prime$ is defined by $S^\prime = \{(j_1, 5, 7)$, $\ldots$, $(j_2, 4, 5)$, $\ldots$, $(j_3, 8, 12)$, $(j_3, 16, 19))\}$.

\textbf{Schedulability.} Given a schedule scenario $S$, a task $j$ is \emph{schedulable} if $j$ completes its execution before its deadline, i.e., for all $\fun{et}_k(j)$ observed in $S$, $\fun{et}_k(j) \le \fun{at}_k(j) + \fun{dl}(j)$. Let $J$ be a set of tasks to be scheduled by a scheduler. A set $J$ of tasks is then schedulable if for every schedule $S$ of $J$, we have no task $j \in J$ that misses its deadline.

As shown in schedule scenarios $S$ and $S^\prime$ presented in Figures~\ref{fig:schedule 1} and \ref{fig:schedule 2}, respectively, all three tasks, $j_1$, $j_2$, and $j_3$, are schedulable. However, we note that the overall amounts of remaining time, i.e., safety margins, from the tasks' completions to their deadlines observed in $S$ and $S^\prime$ are different (see the second completion times and deadlines of $j_1$, $j_2$, and $j_3$ in $S$ and $S^\prime$) because $S$ and $S^\prime$ are produced by using different priority assignments. Engineers typically desire to assign optimal priorities to real-time tasks that aim at maximizing such safety margins, as discussed below.

\textbf{Problem.} In real-time systems, fixed priorities are typically assigned to tasks~\citep{Davis2016,Lee2020Panda}. Finding an appropriate priority assignment is important not only for ensuring the schedulability of a system but also for maximizing the safety margins within which a system can tolerate unexpected execution time overheads. For example, if an unpredictable error occurs and triggers check-point mechanisms~\citep{Davis2007}, which re-execute part or all of a task $j$, then the execution time of $j$ unexpectedly overruns. Hence, engineers need an optimal priority assignment that maximizes the overall remaining times from task completion times to task deadlines, i.e., safety margins.

While assigning priorities to tasks, engineers also account for constraints, that are often but not always domain-specific. For example, aperiodic tasks' priorities should be lower than those of periodic tasks because periodic tasks are often more critical than aperiodic tasks. Hence, engineers develop a system that prioritizes executions of periodic tasks over aperiodic tasks. Recall from Section~\ref{sec:motivation}, this constraint is desirable by engineers. When needed, however, engineers can violate the constraint to some extent in order to ensure that aperiodic tasks complete within a reasonable amount of time while periodic tasks meet their deadlines. Constraints can be either \emph{hard} constraints, which must be satisfied, or \emph{soft} constraints, which are desired to be satisfied. In our study, hard constraints need to be assured while scheduling tasks, e.g., a running task's priority must be higher than a ready task's priority, which are enforced by a scheduler. In the context of optimizing priority assignments, we focus on maximizing the extent of satisfying soft constraints. We refer to a soft constraint as a constraint in this paper.

Our work aims at optimizing priority assignments that maximize the safety margins while satisfying such constraints. Specifically, for a set $J$ of tasks to be analyzed, we define three concepts as follows: (1)~a priority assignment for $J$ denoted by $\vv{P}$, (2)~the magnitude of safety margins for a priority assignment $\vv{P}$ denoted by $\fun{fs}(\vv{P})$, and (3)~the degree of constraint satisfaction denoted by $\fun{fc}(\vv{P})$. We note that Section~\ref{subsec:fitness} describes how we optimize $\vv{P}$, and compute $\fun{fs}(\vv{P})$ and $\fun{fc}(\vv{P})$ in detail. Our study aims at finding a set $\mathbf{B}$ of best possible priory assignments that are Pareto optimal~\citep{Knowles2000} such that a priority assignment $\vv{P} \in \mathbf{B}$ maximizes both $\fun{fs}(\vv{P})$ and $\fun{fc}(\vv{P})$, and any other priority assignments in $\mathbf{B}$ are equally viable.
\section{Related Work}
\label{sec:relatedwork}

This section discusses related research strands in the areas of priority assignments, real-time analysis using exhaustive techniques, search-based analysis in real-time systems, and coevolutionary analysis in software engineering.

\begin{table}[t]
\caption{Comparing our work, OPAM, with existing priority assignment techniques with respect to the properties captured in their underlying system models.}
\resizebox{\textwidth}{!}{
\begin{tabular}{l@{ }c@{ }c@{ }c@{ }c@{ }c@{ }c@{ }c@{ }c@{ }c@{ }c}
\toprule
\multicolumn{1}{c}{Properties} & OPAM & RMPO & DMPO & OPA & OPA-MLD & RPA & FNR-PA & PRPA & OPTA & EPAF \\
\midrule
\makecell[l]{Periodic\\ task} & $\circ$ & $\circ$ & $\circ$ & $\circ$ & $\circ$ & $\circ$ & $\circ$ & $\circ$ & $\circ$ & $\circ$ \\
\arrayrulecolor{lightgray}\hline\arrayrulecolor{black}
\makecell[l]{Aperiodic\\ task} & $\circ$ & & & $\circ$ & $\circ$ & $\circ$ & $\circ$ & $\circ$ & & \\
\arrayrulecolor{lightgray}\hline\arrayrulecolor{black}
\makecell[l]{Resource\\ dependency} & $\circ$ & & & & & & & & & \\
\arrayrulecolor{lightgray}\hline\arrayrulecolor{black}
\makecell[l]{Triggering\\ relationship} & $\circ$ & & & & & & & & & \\
\arrayrulecolor{lightgray}\hline\arrayrulecolor{black}
\makecell[l]{Multi-core\\ system} & $\circ$ & & & $\circ$ & $\circ$ & $\circ$ & $\circ$ & $\circ$ & & \\
\arrayrulecolor{lightgray}\hline\arrayrulecolor{black}
\makecell[l]{Safety\\ margin} & $\circ$ & & & & & $\circ$ & $\circ$ & $\circ$ & & \\
\arrayrulecolor{lightgray}\hline\arrayrulecolor{black}
\makecell[l]{Engineering\\ constraint} & $\circ$ & & & & $\circ$ & & & & & \\
\bottomrule
\end{tabular}
}
\label{tbl:comp-relatedwork}
\end{table}

\noindent\textbf{Priority assignment.} The problem of optimally assigning priorities to real-time tasks has been widely studied~\citep{Fineberg1967,Liu1973,Leung1982,Audsley1991,Tindell1994,George1996,Audsley2001,Davis2007,Chu2008,Davis2009,Davis2011,Davis2012,Davis2016,Zhao2017,Hatvani2018}.  \cite{Fineberg1967} reported early work that relies on a simple system model, assuming, for example, that all tasks arrive periodically, tasks run on a single processing core, tasks' deadlines are equal to their periods, and task executions are independent from one another. They proposed a priority assignment method, named rate-monotonic priority ordering (RMPO), that assigns higher priorities to the tasks with shorter periods. RMPO can find a feasible priority assignment that guarantees periodic tasks to be schedulable when such priority assignments exist~\citep{Liu1973}. \cite{Leung1982} extended RMPO to relax one of the underlying assumptions made in RMPO. Specifically, their priority assignment approach, known as deadline-monotonic priority ordering (DMPO), accounts for task deadlines that can be less than or equal to their periods. In contrast to our work, however, these methods are often not applicable to industrial systems that are not compatible with their simplified system models. Recall from Section~\ref{sec:problem} that a realistic system typically consists of both periodic and aperiodic tasks. Task executions depend on their relationships, i.e., resource dependencies and triggering relationships, with other tasks.

\cite{Audsley2001} designed a priority assignment method, named optimal priority assignment (OPA), that relies on an existing schedulability analysis method $M$. OPA guarantees to find a feasible priority assignment that is schedulable according to $M$ if such priority assignments exist. OPA is applicable to more complex systems than those supported by the methods mentioned above, i.e., RMPO and DMPO. Specifically, OPA can find a feasible priority assignment even in the following situations: (1)~First arrivals of periodic tasks occur after some offset time~\citep{Audsley1991}. (2)~Aperiodic tasks have arbitrary deadlines~\citep{Tindell1994}. (3)~Task executions are scheduled based on a non-preemptive scheduling policy~\citep{George1996}. (4)~Tasks run on multiple processing cores~\citep{Davis2011}. Unlike our approach that accounts for two objectives, safety margins and engineering constraints (see Section~\ref{sec:problem}), OPA attempts to find a feasible priority assignment whose only objective is to make all tasks schedulable. Note that such a feasible priority assignment does not necessarily maximize safety margins as discussed in Section~\ref{sec:problem}. Hence, a feasible priority assignment obtained by OPA is often fragile and sensitive any changes in task executions and unable to accommodate unexpected overheads in task execution times, which are commonly observed in industrial systems~\citep{Davis2007}. 

OPA has been extended by several works~\citep{Davis2007,Chu2008,Davis2009,Davis2012}. \cite{Davis2007} presented a robust priority assignment method (RPA) with a degree of tolerance for unexpected overruns of task execution times. \cite{Chu2008} introduced an extended OPA algorithm (OPA-MLD) that minimizes the lexicographical distance between the desired priority assignment and the one obtained by the algorithm. OPA-MLD enables important tasks to have higher priorities. \cite{Davis2012} proposed an RPA extension (FNR-PA) to make RPA work when a system allows task preemption to be deferred for some interval of time. \cite{Davis2009} developed a probabilistic robust priority assignment method (PRPA) for a real-time system to be less likely to violate its deadlines. Even though the prior works mentioned above improve OPA to some extent, they assume that task executions are independent of one another. In contrast to these existing approaches, OPAM accounts for dependencies among task executions, i.e., resource dependencies and triggering relationships (see our problem description in Section~\ref{sec:problem}).

Some recent priority assignment techniques address scalability. \cite{Hatvani2018} presented an optimal priority and preemption-threshold assignment algorithm (OPTA) that attempts to decrease the computation time for finding a feasible priority assignment. OPTA uses a heuristic to traverse a problem space while pruning infeasible paths to efficiently and effectively explore the problem space. \cite{Zhao2017} introduced an effective priority assignment framework (EPAF) that combines a commercial solver for integer linear programs and their problem-specific optimization algorithm. However, these methods rely on simple system models that assume, for example, task executions to be independent and running on a single processing core. Therefore, the applicability of these techniques is limited. In contrast, recall from Sections~\ref{sec:motivation} and \ref{sec:problem} that our approach aims at scaling to complex industrial systems while accounting for realistic system characteristics regarding task periods, inter-arrival times, resource dependencies, triggering relationships, and multiple processing cores. 

Table~\ref{tbl:comp-relatedwork} compares our work, OPAM, with the other priority assignment techniques mentioned above. As shown in the table, we note that prior works rely on system models that are very restrictive. In particular, existing work assumes that task executions are independent of one another. However, task dependencies such as resource dependencies and triggering relationships are commonly observed in industrial systems. In addition, we note that no existing solution simultaneously accounts for safety margins and engineering constraints. Hence, to our knowledge, OPAM is the first attempt to provide engineers with a set of equally viable priority assignments, allowing trade-off analysis with respect to the two objectives: maximizing safety margins and satisfying engineering constraints.

\noindent\textbf{Real-time analysis using exhaustive techniques.} Constraint programming and model checking have been applied to conclusively and exhaustively verify whether or not a system meets its deadlines~\citep{Kwiatkowska2011, Alesio2012, Nejati2012, Alesio2013}. Existing research on priority assignment based on OPA rely on such exhaustive techniques to prove the schedulability of a set of tasks for a given priority assignment. We note that schedulability analysis is, in general, an NP-hard problem~\citep{Davis2016} that cannot be solved in polynomial time. As a result, exhaustive techniques based on model checking and constraint solving are often not amenable to analyze large industrial systems such as ESAIL -- our motivating case study system -- described in Section~\ref{sec:motivation}. To assess if exhaustive techniques could scale to ESAIL, as discussed in Section~\ref{subsec:threats}, we performed a preliminary experiment using UPPAAL~\citep{Behrmann2004}, a model checker for real-time systems. We observed  that UPPAAL was not able to verify schedulability of ESAIL tasks for a fixed priority assignment even after letting it run for several days (see Section~\ref{subsec:threats} for more details). 

\noindent\textbf{Search-based analysis in real-time systems.}
In real-time systems, most of the existing works that use search-based techniques focus on testing~\citep{Wegener1997,Wegener1998,Briand2005,Lin2009,Arcuri2010}. \citeauthor{Wegener1997}~(\citeyear{Wegener1997}, \citeyear{Wegener1998}) introduced a testing approach based on a genetic algorithm that aims to check computation time, memory usage, and task synchronization by analyzing the control flow of a program. \cite{Briand2005} applied a genetic algorithm to find stress test scenarios for real-time systems. \cite{Lin2009} proposed a search-based approach to check whether a real-time system meets its timing and security constraints. \cite{Arcuri2010} presented a black-box system testing approach based on a genetic algorithm. Beyond testing real-time systems, \citeauthor{Nejati2013}~(\citeyear{Nejati2013}, \citeyear{Nejati2014}) developed a search-based trade-off analysis technique that helps engineers balance the satisfaction of temporal constraints and keeping the CPU time usage at an acceptable level. \cite{Lee2020} combined a search algorithm and machine learning to estimate safe ranges of worst-case task execution times within which tasks likely meet their deadlines. In contrast to these prior works, OPAM addresses the problem of optimally assigning priorities to real-time tasks while accounting for multiple objectives regarding safety margins and engineering constraints, thus enabling Pareto (trade-off) analysis. Further, OPAM uses a multi-objective, competitive coevolutionary search algorithm, which has been rarely applied to date in prior studies of real-time systems, as discussed next.

\noindent\textbf{Coevolutionary analysis in software engineering.}
Despite the success of search-based software engineering (SBSE) in many application domains including software testing~\citep{Wegener1997,Wegener1998,Lin2009,Arcuri2010,Shin2018}, program repair~\citep{Weimer2009,Tan2016,Abdessalem2020}, and self-adaptation~\citep{Andrade2013,Chen2018,Shin2020}, coevolutionary algorithms have been applied in only a few prior studies~\citep{Wilkerson2010,Wilkerson2012,Boussaa2013}. \citeauthor{Wilkerson2012}~(\citeyear{Wilkerson2010}, \citeyear{Wilkerson2012}) present a coevolution-based approach to automatically correct software. Their work introduced a program representation language to facilitate their automated corrections. \cite{Boussaa2013} developed a code-smells detection approach. The main idea is to evolve two competing populations of code-smell detection rules and artificial code-smells. Unlike these prior works, we study the problem of optimally assigning priorities to tasks in real-time systems. To our knowledge, we are the first to address the priority assignment problem using a multi-objective, competitive coevolutionary search algorithm. 

\section{Approach Overview}
\label{sec:overview}

Finding an optimal priority assignment is an inherently interactive process. In practice, once engineers assign priorities to the real-time tasks in a system, testers then stress the system to find a condition, i.e., a particular sequence of task arrivals, in which a task execution violates its deadline. Testers typically use a simulator or hardware equipment to stress the system by triggering plausible worst-case arrivals of tasks that maximize the likelihood of deadline misses. If testers find task arrivals that induce deadline misses, the task arrivals are reported to engineers in order to fix the problem by reassigning priorities. This interactive process of assigning priorities and testing schedulability continues until both engineers and testers ensure that the tasks meet their deadlines.

For such intrinsically interactive problem-solving domains, we conjecture that coevolutionary algorithms are potentially suitable solutions. A coevolutionary algorithm is a search algorithm that mutually adapts one of different species, e.g., in our study, two populations of priority assignments and task-arrival sequences, acting as foils against one another. Specifically, we apply multi-objective, two-population competitive coevolution~\citep{Luke2013} to address our problem of finding optimal priority assignments (see Section~\ref{sec:problem}). In our approach, the two populations of priority assignments and stress test scenarios, i.e., task-arrival sequences, evolve synchronously, competing with each other in order to search for optimal priority assignments that maximize the magnitude of safety margins from deadlines and the extent of constraint satisfaction. Note that better priority assignments enable a system to achieve larger safety margins. Hence, those priority assignments have a higher chance to pass stress test scenarios. This impacts the stress test scenarios because they need to evolve as well, aiming at inducing deadline misses in the system.

Recall from Section~\ref{sec:relatedwork} that most of the existing SBSE research relies on search algorithms using a single population~\citep{Chen2018,Abdessalem2020,Shin2020}. However, such algorithms do not fit the problem of priority assignments targeted here. When (1)~two competing traits between task arrivals and priority assignments are encoded together in an individual of a single population and (2)~two contradicting fitness functions regarding safety margins and deadline misses, which are exact opposites, assess such individuals, the notion of Pareto optimality is not applicable. In that case, maximizing the magnitude of safety margins necessarily entails minimizing the magnitude of deadline misses. Hence, a single population-based search algorithm cannot make Pareto improvements that maximize safety margins (resp. deadline misses) while not minimizing deadline misses (resp. safety margins). Specifically, the dominance relation over such individuals does not exist because if an individual $I$ is strictly better than another individual $I^\prime$ in one fitness value, $I$ is always worse than $I^\prime$ in the other fitness value. Hence, we are not able to obtain equally viable solutions with respect to the contradicting objectives using such a method.

\begin{figure}[t]
    \centering\includegraphics[width=.9\columnwidth]{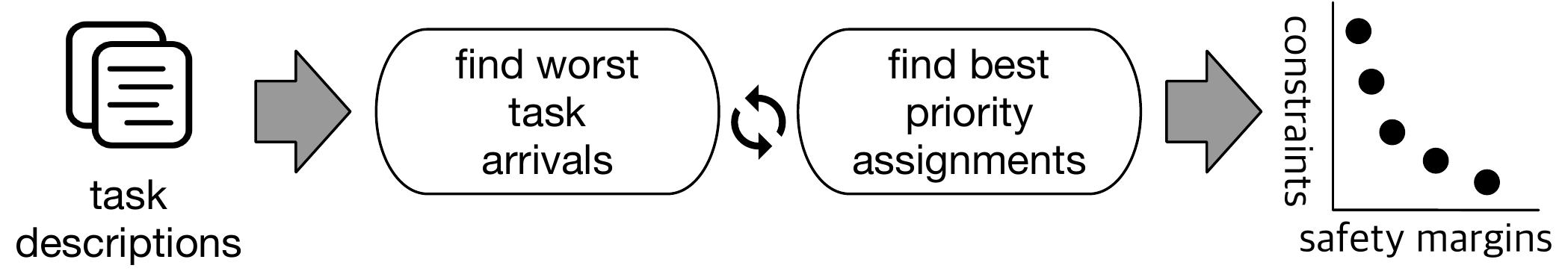}
    \caption{An overview of our \underline{O}ptimal \underline{P}riority \underline{A}ssignment \underline{M}ethod for real-time systems (OPAM).}
    \label{fig:overview}
\end{figure}

Figure~\ref{fig:overview} shows an overview of our proposed solution: \underline{O}ptimal \underline{P}riority \underline{A}ssignment \underline{M}ethod for real-time tasks (OPAM). OPAM requires as input task descriptions defined by engineers, which specify task characteristics and their relationships (see Section~\ref{sec:problem}). Given such input task descriptions, the ``find worst task arrivals' and ``find best priority assignments'' steps aim at generating worst-case sequences of task arrivals and best-case priority assignments, respectively. A worst-case sequence of task arrivals means that the magnitude of deadline misses, i.e., the amounts of time from task deadlines to task completion times, is maximized when tasks arrive as defined in the sequence. Note that if there is no deadline miss, a task-arrival sequence is considered worst-case if tasks complete their executions as close to their deadlines as possible. In contrast, a priority assignment is best-case when the magnitude of safety margins is maximized. Beyond maximizing safety margins, the ``find best priority assignments'' step accounts for satisfying engineering constraints in assigning priorities to tasks.
OPAM evolves two competing populations of task-arrival sequences and priority assignments synchronously generated from the two steps. OPAM then outputs a set of priority assignments that are Pareto optimal with regards to the magnitude of safety margins and the extent of satisfying constraints. Hence, OPAM allows engineers to perform domain-specific trade-off analysis among Pareto solutions and is useful in practice to support decision making with respect to their task design. For example, suppose engineers develop a weakly hard real-time systems~\citep{Bernat2001} that can tolerate occasional deadline misses. In that case, engineers may consider a few deadline misses as less important (as long as their consequences are negligible) than the overall magnitude of safety margins in their trade-off analysis. Section~\ref{sec:approach} describes OPAM in detail.
\section{Competitive Coevolution}
\label{sec:approach}

Figure~\ref{fig:coevolution} describes the OPAM algorithm for finding optimal priority assignments, which employs  multi-objective, two-population competitive coevolution. The algorithm first randomly initializes two populations $\mathbf{A}$ and $\mathbf{P}$ for task-arrival sequences and priority assignments, respectively (lines 13--15). For $\mathbf{A}$, OPAM randomly varies task arrivals of aperiodic tasks to create $\var{ps}_a$ task-arrival sequences, according to the input task descriptions $D$. Regarding $\mathbf{P}$, OPAM randomly creates $\var{ps}_p$ priority assignments that may include one defined by engineers if available.

\begin{figure}[t]
\begin{lstlisting}[style=Alg]
Algorithm Search optimal priority assignments
Input $D$: task descriptions
Input $n_c$: number of coevolution cycles //budget
Input $\var{ps}_a$: population size //task-arrival sequences
Input $\var{ps}_p$: population size //priority assignments
Input $\var{cp}_a$: crossover probability //task-arrival sequences
Input $\var{cp}_p$: crossover probability //priority assignments
Input $\var{mp}_a$: mutation probability //task-arrival sequences
Input $\var{mp}_p$: mutation probability //priority assignments
Input $\mathbf{E}$: set of task-arrival sequences //external evaluation
Output $\mathbf{B}$: best Pareto front

//initialize populations
$\mathbf{A} \leftarrow \fun{randomize\_arrivals}(D,\var{ps}_a)$
$\mathbf{P} \leftarrow \fun{randomize\_priorities}(D,\var{ps}_p)$

for $n_c$ times do
?\vrule?  //evolution: find worst-case sequences of task arrivals
?\vrule?  //objective: deadline misses
?\vrule?  $\fun{evaluate\_internal\_fitness\_arrivals}(\mathbf{A},\mathbf{P})$
?\vrule?  $\mathbf{A} \leftarrow \fun{bread\_arrivals}(\mathbf{A},\mathbf{P},\var{cp}_a,\var{mp}_a)$ //GA
?\vrule?  
?\vrule?  //evolution: find best-case priority assignments
?\vrule?  //objectives: safety margins and constraints
?\vrule?  $\fun{evaluate\_internal\_fitness\_priorities}(\mathbf{P},\mathbf{A})$
?\vrule?  $\mathbf{P} \leftarrow \fun{breed\_priorities}(\mathbf{P},\mathbf{A},\var{cp}_p,\var{mp}_p)$ //NSGAII
?\vrule?  
?\vrule?  //external fitness evaluation
?\vrule?  //objectives: safety margins and constraints
?\vrule?  $\fun{evaluate\_external\_fitness}(\mathbf{P},\mathbf{E})$
?\vrule?  $\mathbf{B} \leftarrow \fun{select\_best}(\mathbf{P} \cup \mathbf{B})$

return $\mathbf{B}$
\end{lstlisting}
\caption{Multi-objective two-population competitive coevolution for finding optimal priority assignments.}
\label{fig:coevolution}
\end{figure}

The two populations sequentially evolve during the allotted analysis budget (see line 17 in Figure~\ref{fig:coevolution}).
The best priority assignment is the one that makes tasks schedulable and maximizes the magnitude of safety margins, while satisfying engineering constraints for a given worst sequence of task arrivals. Hence, searching for the best priority assignments involves searching for the worst sequences of task arrivals. We create two populations $\mathbf{A}$ and $\mathbf{P}$ searching for the worst arrival sequences and the best priority assignments, respectively. The fitness values of task-arrival sequences in $\mathbf{A}$ are computed based on how well they challenge the priority assignments in $\mathbf{P}$, i.e., maximizing the magnitude of deadline misses (line 20). Likewise, the priority assignments in $\mathbf{P}$ are evaluated based on how well they perform against the task-arrival sequences in $\mathbf{A}$, i.e., maximizing the magnitude of safety margins while satisfying constraints (line 25). Once the two populations are assessed against each other, OPAM generates the next populations based on the computed fitness values (lines 21 and 26). OPAM tailors the breading mechanisms of steady-state genetic algorithms (GA)~\citep{Whitley1988} for $\mathbf{A}$ and NSGAII~\citep{Deb2002} for $\mathbf{P}$.

OPAM uses two types of fitness functions, namely internal and external fitness evaluations, which play a different and complementary role as described below. The two internal fitness evaluations in lines 20 and 25 of the listing in Figure~\ref{fig:coevolution} aim at selecting individuals -- task-arrival sequences and priority assignments -- for breeding the next $\mathbf{A}$ and $\mathbf{P}$ populations. OPAM evaluates the external fitness for the $\mathbf{P}$ population of priority assignments to find a best Pareto front (lines 28--31). As shown in lines 20 and 25, the internal fitness values of individuals in $\mathbf{A}$ (resp. $\mathbf{P}$) are computed based on how they perform with respect to individuals in $\mathbf{P}$ (resp. $\mathbf{A}$). Hence, an individual's internal fitness is assessed through interactions with competing individuals. For example, a priority assignment in the first generation may have acceptable fitness values regarding safety margins and constraint satisfaction with respect to the first generation of task-arrival sequences, which are likely far from worst-case sequences. However, priority assignment fitness may get worse in later generations as the task-arrival sequences evolve towards larger deadline misses. Thus, if OPAM simply monitors internal fitness, it cannot reliably detect coevolutionary progress as an individual's internal fitness changes according to competing individuals. The problem of monitoring progress in coevolution has been observed in many studies~\citep{Ficici2004,Popovici2012}. To address it, OPAM computes external fitness values of priority assignments in $\mathbf{P}$ based on a set $\mathbf{E}$ of task-arrival sequences generated independently from the coevolution process. By doing so, OPAM can observe the monotonic improvement of external fitness for priority assignments. We note that, in general, if interactions between two competing populations are finite and any interaction can be examined with non-zero probability at any time, monotonicity guarantees that a coevolutionary algorithm converges to a solution~\citep{Popovici2012}.

We note that our approach for evolving task-arrival sequences is based on past work~\citep{Briand2005}, where a specific genetic algorithm configuration was proposed to find worst-case task-arrival sequences. One significant modification is that OPAM accounts for task relationships -- resource-dependency and task triggering relationships -- and a multi-core scheduling policy based on simulations to evaluate the magnitude of deadline misses.

Following standard practice~\citep{Ralph2020}, the next sections describe OPAM in detail by defining the representations, the scheduler, the fitness functions, and the evolutionary algorithms for coevolving the task-arrival sequences and priority assignments. We then describe the external fitness evaluation of OPAM.

\subsection{Representations}
\label{subsec:representations}

OPAM coevolves two populations of task-arrival sequences and priority assignments. A task-arrival sequence is defined by their inter-arrival time characteristics (see Section~\ref{sec:problem}). A priority assignment is defined by a function that maps priorities to tasks.

\noindent\textbf{Task-arrival sequences.} Given a set $J$ of tasks to be scheduled, a feasible sequence of task arrivals is a set $A$ of tuples $(j, \fun{at}_k(j))$ where $j \in J$ and $\fun{at}_k(j)$ is the $k$th arrival time of a task $j$. Thus, a solution $A$ represents a valid sequence of task arrivals of $J$ (see valid $\fun{at}_k(j)$ computation in Section~\ref{sec:problem}). Let $\mathbb{T} = [0, \mathbf{T}]$ be the time period during which a scheduler receives task arrivals. The size of $A$ is equal to the number of task arrivals over the $\mathbb{T}$ time period. Due to the varying inter-arrival times of aperiodic tasks (Section~\ref{sec:problem}), the size of $A$ will vary across different sequences.

\noindent\textbf{Priority assignments.} Given a set $J$ of tasks to be scheduled, a feasible priority assignment is a list $\vv{P}$ of priority $\fun{pr}(j)$ for each task $j \in J$. OPAM assigns a non-negative integer to a priority $\fun{pr}(j)$ of $j$ such that priorities are comparable to one another. The size of $\vv{P}$ is equal to the number of tasks in $J$. Each task in $J$ has a unique priority. Hence, a priority assignment $\vv{P}$ is a permutation of all tasks' priorities. We note that these characteristics of priority assignments are common in many real-time analysis methods~\citep{Audsley2001,Davis2007,Zhao2017} and industrial systems (e.g., see our six industrial case study systems described in Section~\ref{subsec:industrial subjects}).

\subsection{Simulation}
\label{subsec:simulation}

OPAM relies on simulation for analyzing the schedulability of tasks in a scalable way. For instance, an inter-arrival time of a software update task in a satellite system is approximately at most three months. In such cases, conducting an analysis based on an actual scheduler is prohibitively expensive. Also, applying an exhaustive technique for schedulability analysis typically doesn't scale to an industrial system (e.g., see our experiment results using a model checker described in Section~\ref{subsec:threats}). Instead, OPAM uses a real-time task scheduling simulator, named OPAMScheduler, which applies a scheduling policy, i.e., single-queue multi-core scheduling policy~\citep{Arpaci2018}, based on discrete simulation time events. Note that we chose the single-queue multi-core scheduling policy for OPAMScheduler since our case study systems (described in Section~\ref{subsec:industrial subjects}) rely on this policy.

OPAMScheduler takes as input a feasible task-arrival sequence $A$ and a priority assignment $\vv{P}$ for scheduling a set $J$ of tasks. It then outputs a schedule scenario as a set $S$ of tuples $(j,\fun{at}_k(j),\fun{et}_k(j))$ where $\fun{at}_k(j)$ and $\fun{et}_k(j)$ are the $k$th arrival and end time values of a task $j$, respectively (see Section~\ref{sec:problem}). For each task $j$, OPAMScheduler computes $\fun{et}_k(j)$ based on its WCET and scheduling policy while accounting for task relationships (see the $\fun{dp}(j,j^\prime)$ resource-dependency relationship and the $\fun{tr}(j,j^\prime)$ task triggering relationship in Section~\ref{sec:problem}). To simulate the worst-case executions of tasks, OPAMScheduler assigns tasks' WCETs to their execution times.

\begin{sloppypar}
OPAMScheduler implements a single-queue multi-core scheduling policy~\citep{Arpaci2018}, which schedules a task $j$ with explicit priority $\fun{pr}(j)$ and deadline $\fun{dl}(j)$. When tasks arrive, OPAMScheduler puts them into a single queue that contains tasks to be scheduled. At any simulation time, if there are tasks in the queue and multiple cores are available to execute tasks, OPAMScheduler first fetches a task $j$ from the queue in which $j$ has the highest priority $\fun{pr}(j)$. OPAMScheduler then allocates task $j$ to any available core. Note that if task $j$ shares a resource with a running task $j^\prime$ in another core, i.e., the $\fun{dp}(j,j^\prime)$ resource-dependency relationship holds, $j$ will be blocked until $j^\prime$ releases the shared resource.
\end{sloppypar}

OPAMScheduler works under the assumption that context switching time is negligible, which is also a working assumption in many scheduling analysis methods~\citep{Liu1973,Audsley2001,Alesio2015}. Note that the assumption is practically valid and useful at an early development step in the context of real-time analysis. For instance, our collaborating partner, LuxSpace, accounts for the waiting time of tasks due to context switching between tasks through adding some extra time to WCET estimates at the task design stage. Note that OPAM can be applied with any scheduling policy, including those that account for context switching time and multiple queues.

\subsection{Fitness functions}
\label{subsec:fitness}

\noindent\textbf{Internal fitness: deadline misses.} Given a feasible task-arrival sequence $A$ and a priority assignment $\vv{P}$, we formulate a function, $\fun{fd}(A,\vv{P})$, to quantify the degree of deadline misses regarding a set $J$ of tasks to be scheduled. To compute $\fun{fd}(A,\vv{P})$, OPAM runs OPAMScheduler for $A$ and $\vv{P}$ and obtains a schedule scenario $S$. We denote by $\fun{dist}_k(j)$ the distance between the end time and the deadline of the $k$th arrival of task $j$ observed in $S$ and define $\fun{dist}_k(j) = \fun{et}_k(j) - \fun{at}_k(j) + \fun{dl}(j)$ (see Section~\ref{sec:problem} for the notation end time $\fun{et}_k(a)$, arrival time $\fun{at}_k(j)$, and deadline $\fun{dl}(j)$). We denote by $\fun{lk}(j)$ the last arrival index of a task $j$ in $A$. Given a set $J$ of tasks to be scheduled, the $\fun{fd}(A,\vv{P})$ function is defined as follows:
\begin{equation*}
\fun{fd}(A,\vv{P}) = \sum_{j \in J, k \in [1,\mathit{lk}(j)]}2^{\fun{dist}_k(j)}
\end{equation*}

Note that $\fun{fd}(A,\vv{P})$ is defined as an exponential equation. Hence, when all task executions observed in a schedule scenario $S$ meet their deadlines, $\fun{fd}(A,\vv{P})$ is a small value as any distance $\fun{dist}_k(j)$ between the task end time and the deadline of the $k$th arrival of task $j$ is a negative value. In contrast, deadline misses result in positive values for $\fun{dist}_k(j)$. In such cases, $\fun{fd}(A,\vv{P})$ is a large value. The exponential form of $\fun{fd}(A,\vv{P})$ was precisely selected for this reason, to assign large values for deadline misses but small values when deadlines are met. By doing so, $\fun{fd}(A,\vv{P})$ prevents an undesirable solution that would result into many task executions meeting deadlines obfuscating a smaller number of deadline misses.

Following the principles of competitive coevolution, individuals in a population $\mathbf{A}$ of task-arrival sequences need to be assessed by pitting them against individuals in the other population $\mathbf{P}$ of priority assignments. We denote by $\fun{fd}(A,\mathbf{P})$ the internal fitness function that quantifies the overall magnitude of deadline misses across all priority assignment $\vv{P} \in \mathbf{P}$, regarding a set $J$ of tasks to be scheduled. The $\fun{fd}(A,\mathbf{P})$ fitness is used for breeding the next population of task-arrival sequences. OPAM aims to maximize $\fun{fd}(A,\mathbf{P})$, defined as follows:
\begin{equation*}
\fun{fd}(A,\mathbf{P}) = \sum_{\vv{P} \in \mathbf{P}}\fun{fd}(A,\vv{P})/|\mathbf{P}|
\end{equation*}
 
\noindent\textbf{Internal fitness: safety margins.} Given a feasible priority assignment $\vv{P}$ and a task-arrival sequence $A$, we denote by $\fun{fs}(\vv{P},A)$ the magnitude of safety margins regarding a set $J$ of tasks to be scheduled. The computation of $\fun{fs}(\vv{P},A)$ is similar to the computation of $\fun{fd}(A,\vv{P})$ regarding the use of OPAMScheduler, which outputs a schedule scenario $S$. The difference is that OPAM reverses the sign of $\fun{fd}(A,\vv{P})$ as OPAM aims at maximizing the magnitude of safety margins. Given a set $J$ of tasks to be scheduled, the $\fun{fs}(\vv{P},A)$ function is defined as follows:
\begin{equation*}
\fun{fs}(\vv{P},A) = \sum_{j \in J, k \in [1,\mathit{lk}(j)]}-2^{\fun{dist}_k(j)} \text{\quad(i.e,}{-}\fun{fd}(A,\vv{P})\text{)}
\end{equation*}

Given two populations $\mathbf{P}$ and $\mathbf{A}$ of priority assignments and task-arrival sequences, similar to internal fitness $\fun{fd}(A,\mathbf{P})$, priority assignments in $\mathbf{P}$ need to be assessed against task-arrival sequences in $\mathbf{A}$. We formulate an internal fitness function, $\fun{fs}(\vv{P},\mathbf{A})$, to quantify the overall magnitude of safety margins across all task-arrival sequences $A \in \mathbf{A}$, regarding a set $J$ of tasks to be scheduled and a priority assignment $\vv{P}$. OPAM relies on the $\fun{fs}(\vv{P},\mathbf{A})$ function to breed the next population of priority assignments. OPAM aims to maximize $\fun{fs}(\vv{P},\mathbf{A})$, which is defined as follows:
\begin{equation*}
\fun{fs}(\vv{P},\mathbf{A}) = \sum_{A \in \mathbf{A}}\fun{fs}(\vv{P},\mathbf{A})/|\mathbf{A}|
\end{equation*}

\noindent\textbf{Internal fitness: constraints.}
Given a priority assignment $\vv{P}$, we formulate an internal fitness function, $\fun{fc}(\vv{P})$, to quantify the degree of satisfaction of soft constraints set by engineers. Such function is required as we recast the satisfaction of such constraints into an optimization problem, in order to minimize constraint violations. Specifically, OPAM accounts for the following constraint: aperiodic tasks should have lower priorities than those of periodic tasks. Recall from Section~\ref{sec:motivation} that engineers consider this constraint to be desirable. We denote by $\fun{lp}(\vv{P})$ the lowest priority of periodic tasks in $\vv{P}$. For a set $J$ of tasks to be scheduled, OPAM aims to maximize $\fun{fc}(\vv{P})$, which is defined as follows:
\begin{equation*}
\fun{fc}(\vv{P}) = \sum_{j \in J}
\begin{cases}
\fun{lp}(\vv{P}) - \fun{pr}(j) \text{, if $j$ is an aperiodic task}\\
0 \text{, otherwise}
\end{cases}
\end{equation*}

Greater $\fun{pr}(j)$ values denote higher priorities. Given a priority assignment $\vv{P}$, if $\fun{pr}(j)$ for an aperiodic task $j$ is lower than the priority of any of the periodic tasks, $\fun{lp}(\vv{P}) - \fun{pr}(j)$ is a positive value. OPAM measures the difference between priorities of aperiodic and periodic tasks. By doing so, $\fun{fc}(\vv{P})$ rewards aperiodic tasks that satisfy the above constraint and consistently penalizes those that violate it. Hence, OPAM aims at maximizing $\fun{fc}(\vv{P})$.

\noindent\textbf{External fitness: safety margins and constraints.}
To examine the quality of priority assignments and monitor the progress of coevolution, OPAM takes as input a set $\mathbf{E}$ of task-arrival sequences created independently from the coevolution process. Given a set $\mathbf{E}$ of task-arrival sequences and a priority assignment $\vv{P}$, OPAM utilizes $\fun{fs}(\vv{P},\mathbf{E})$ and $\fun{fc}(\vv{P})$ described above as external fitness functions for quantifying the magnitude of safety margins and the extent of constraint satisfaction, respectively. As $\mathbf{E}$ does not change over the coevolution process, $\fun{fs}(\vv{P},\mathbf{E})$ is used for evaluating a priority assignment $\vv{P}$ since it is not impacted by the evolution of task-arrival sequences. Hence, external fitness functions ensure that OPAM monitors the progress of coevolution in a stable manner. Given two populations $\mathbf{P}$ and $\mathbf{A}$ of priority assignments and task-arrival sequences, we recall that the $\fun{fd}(A,\mathbf{P})$ internal fitness function quantifies the overall magnitude of deadline misses across all priority assignments in $\mathbf{P}$ for the given sequence of task arrivals $A$. The $\fun{fs}(\vv{P},\mathbf{A})$ internal fitness function quantifies the overall magnitude of safety margins across all sequences of task arrivals in $\mathbf{A}$ for the given priority assignments $\vv{P}$. Hence, the internal fitness of $A$ (resp. $\vv{P}$) is assessed through interactions with competing individuals in $\mathbf{P}$ (resp. $\mathbf{A}$). Therefore, if OPAM relies only on the internal fitness functions, it cannot gauge the progress of coevolution in a stable manner as an individual's internal fitness depends on competing individuals.

We note that soft deadline tasks also require to execute within reasonable execution time, i.e., (soft) deadline. As the above fitness functions return quantified degrees of deadline misses and safety margins, OPAM uses the same fitness functions for both soft and hard deadline tasks.

\subsection{Evolution: Worst-case task arrivals}
\label{subsec:evolution arrivals}

\begin{figure}[t]
\begin{lstlisting}[style=Alg]
Algorithm Task-arrival sequences evolution
Input $\mathbf{A}$: population of task-arrival sequences
Input $\mathbf{P}$: population of priority assignments
Input $\var{cp}_a$: crossover probability //task-arrival sequences
Input $\var{mp}_a$: mutation probability //task-arrival sequences
Output $\mathbf{A}$: population of task-arrival sequences

//evaluate internal fitness values for ?$\color{javagreen}\mathbf{A}$?
for each $A_i \in \mathbf{A}$
?\vrule?  for each $\vv{P}_l \in \mathbf{P}$
?\vrule?  ?\vrule?  $S \leftarrow \fun{simulate}(A_i,\vv{P}_l)$ //OPAMScheduler
?\vrule?  ?\vrule?  //?$\color{javagreen}\fun{dist}_k(j)$? is computed based on ?$\color{javagreen}S$?
?\vrule?  ?\vrule?  $\fun{fd}(A_i,\vv{P}_l) = \sum_{j \in J, k \in [1,\mathit{lk}(j)]}2^{\fun{dist}_k(j)}$
?\vrule?  $\fun{fd}(A_i,\mathbf{P}) = \sum_{\vv{P}_l \in \mathbf{P}}\fun{fd}(A_i,\vv{P}_l)/|\mathbf{P}|$

//breed task-arrival sequences
$\var{parents} \leftarrow \fun{select\_arrivals}(\mathbf{A})$
$\var{offspring} \leftarrow \fun{crossover\_arrivals}(\var{parents},\var{cp}_a)$
$\var{offspring} \leftarrow \fun{mutate\_arrivals}(\var{offspring},\var{mp}_a)$
//evaluate internal fitness values for ?$\color{javagreen}\var{offspring}$?
for each $A_i \in \var{offspring}$
?\vrule?  for each $\vv{P}_l \in \mathbf{P}$
?\vrule?  ?\vrule?  $S \leftarrow \fun{simulate}(A_i,\vv{P}_l)$ //OPAMScheduler
?\vrule?  ?\vrule?  //?$\color{javagreen}\fun{dist}_k(j)$? is computed based on ?$\color{javagreen}S$?
?\vrule?  ?\vrule?  $\fun{fd}(A_i,\vv{P}_l) = \sum_{j \in J, k \in [1,\mathit{lk}(j)]}2^{\fun{dist}_k(j)}$
?\vrule?  $\fun{fd}(A_i,\mathbf{P}) = \sum_{\vv{P}_l \in \mathbf{P}}\fun{fd}(A_i,\vv{P}_l)/|\mathbf{P}|$
$\mathbf{A} \leftarrow \fun{replace\_arrivals}(\mathbf{A},\var{offspring})$

return $\mathbf{A}$
\end{lstlisting}
\caption{A steady-state GA-based algorithm for evolving task-arrival sequences.}
\label{fig:GA}
\end{figure}

The algorithm in Figure~\ref{fig:GA} describes in detail the evolution of task-arrival sequences in lines 18--21 of the listing in Figure~\ref{fig:coevolution}. OPAM adapts a steady-state Genetic Algorithm (GA)~\citep{Luke2013} for evolving task-arrival sequences. As shown in lines 8--14, OPAM first evaluates each task-arrival sequence in the $\mathbf{A}$ population against the $\mathbf{P}$ population of priority assignments. OPAM executes OPAMScheduler to obtain a schedule scenario $S$ for a task-arrival sequence $A_i \in \mathbf{A}$ and a priority assignment $\vv{P}_l \in \mathbf{P}$ (line 11). OPAM then computes the internal fitness $\fun{fd}(A_i,\mathbf{P})$ capturing the magnitude of deadline misses (lines 12--14). We note that a steady-state GA iteratively breeds offspring, assess their fitness, and then reintroduce them into a population. However, OPAM computes internal fitness of all task-arrival sequences in $\mathbf{A}$ at every generation. This is because internal fitness is computed in relation to  $\mathbf{P}$, which is coevolving with $\mathbf{A}$.

Breeding the next population is done by using the following genetic operators: (1)~\emph{Selection:} OPAM selects candidate task-arrival sequences using a tournament selection technique, with the tournament size equal to two which is the most common setting~\citep{Gendreau2010} (line 17 in Figure~\ref{fig:GA}). (2)~\emph{Crossover:} Selected candidate task-arrival sequences serve as parents to create offspring using a crossover operation (line 18). (3)~\emph{Mutation:} The offspring are then mutated (line 19). Below, we describe our crossover and mutation operators.

\emph{Crossover.} A crossover operator is used to produce offspring by mixing traits of parent solutions. OPAM modifies the standard one-point crossover operator~\citep{Luke2013} as two parent task-arrival sequences $A_p$ and $A_q$ may have different sizes, i.e., $|A_p| \neq |A_q|$. Let $J = \{j_1, j_2, \ldots, j_m\}$ be a set of tasks to be scheduled. Our crossover operator first randomly selects an aperiodic task $j_r \in J$. For all $i \in [1,r]$ and $j_i \in J$, OPAM then swaps all $j_i$ arrivals between the two task-arrival sequences $A_p$ and $A_q$. Since $J$ is fixed for all solutions, OPAM can cross over two solutions that may have different sizes.

\emph{Mutation operator} OPAM uses a heuristic mutation algorithm. For a task-arrival sequence $A$, OPAM mutates the $k$th task arrival time $\fun{at}_k(j)$ of an aperiodic task $j$ with a mutation probability. OPAM chooses a new arrival time value of $\fun{at}_k(j)$ based on the $[\mathit{pmin}(j), \mathit{pmax}(j)]$ inter-arrival time range of $j$. If such a mutation of the $k$th arrival time of $j$ does not affect the validity of the $k{+}1$th arrival time of $j$, the mutation operation ends. Specifically, let $d$ be a mutated value of $\fun{at}_k(j)$. In case $\fun{at}_{k+1}(j) \in [d + \mathit{pmin}(j), d + \mathit{pmax}(j)]$, OPAM returns the mutated $A$ task-arrival sequence.

After mutating the $k$th arrival time $\fun{at}_k(j)$ of a task $j$ in a solution $A$, if the $k{+}1$th arrival becomes invalid, OPAM corrects the remaining arrivals of $j$. Let $o$ and $d$ be, respectively, the original and mutated $k$th arrival time of $j$. For all the arrivals of $j$ after $d$, OPAM first updates their original arrival time values by adding the difference $d-o$. Let $\mathbb{T} = [0,\mathbf{T}]$ be the scheduling period. OPAM then removes some arrivals of $j$ if they are mutated to arrive after $\mathbf{T}$ or adds new arrivals of $j$ while ensuring that all tasks arrive within $\mathbb{T}$.

As shown in lines 20--26 in Figure~\ref{fig:GA}, the internal fitness of the generated offspring is computed based on the $\mathbf{P}$ population. OPAM then updates the $\mathbf{A}$ population of task-arrival sequences by comparing the offspring and individuals in $\mathbf{A}$ (line 27).

We note that when a system is only composed of periodic tasks, OPAM will skip evolving for worst-case arrival sequences as arrivals of periodic tasks are deterministic (see Section~\ref{sec:problem}). Nevertheless, OPAM will optimize priority assignments based on given arrivals of periodic tasks. When needed, OPAM can be easily extended to manipulate offset and period values for periodic tasks, in a way identical to how we currently handle inter-arrival times for aperiodic tasks.

\subsection{Evolution: Best-case priority assignments}
\label{subsec:evolution priorities}

\begin{figure}[t]
\begin{lstlisting}[style=Alg]
Algorithm Priority assignments evolution
Input $\mathbf{A}$: population of task-arrival sequences
Input $\mathbf{P}$: population of priority assignments
Input $\var{ps}_p$: population size //priority assignments
Input $\var{cp}_p$: crossover probability //priority assignments
Input $\var{mp}_p$: mutation probability //priority assignments
Output $\mathbf{P}$: population of priority assignments

//evaluate internal fitness values for ?$\color{javagreen}\mathbf{P}$?
for each $\vv{P}_i \in \mathbf{P}$
?\vrule?  for each $A_l \in \mathbf{A}$
?\vrule?  ?\vrule?  $S \leftarrow \fun{simulate}(A_l,\vv{P}_i)$ //OPAMScheduler
?\vrule?  ?\vrule?  //?$\color{javagreen}\fun{dist}_k(j)$? is computed based on ?$\color{javagreen}S$?
?\vrule?  ?\vrule?  $\fun{fs}(\vv{P}_i,A_l) = \sum_{j \in J, k \in [1,\mathit{lk}(j)]}-2^{\fun{dist}_k(j)}$
?\vrule?  $\fun{fs}(\vv{P}_i,\mathbf{A}) = \sum_{A_l \in \mathbf{A}}\fun{fs}(\vv{P}_i,\mathbf{A})/|\mathbf{A}|$
?\vrule?  ?$\fun{fc}(\vv{P}_i) = \sum_{j \in J}
\begin{cases}
\fun{lp}(\vv{P}_i) - \fun{pr}(j) \text{, if $j$ is an aperiodic task}\\
0 \text{, otherwise}
\end{cases}$?

//breed priority assignments
$\vv{R} \leftarrow \fun{sort\_non\_dominated\_fronts}(\mathbf{P})$
$\fun{assign\_crowding\_distance}(\vv{R})$
$\mathbf{P}_\alpha \leftarrow \fun{NSGAII\_breed}(\vv{R},\var{ps}_p,\var{cp}_p,\var{mp}_p)$
//evaluate internal fitness values for ?$\color{javagreen}\mathbf{P}_\alpha$?
for each $\vv{P}_i \in \mathbf{P}_\alpha$
?\vrule?  for each $A_l \in \mathbf{A}$
?\vrule?  ?\vrule?  $S \leftarrow \fun{simulate}(A_l,\vv{P}_i)$ //OPAMScheduler
?\vrule?  ?\vrule?  //?$\color{javagreen}\fun{dist}_k(j)$? is computed based on ?$\color{javagreen}S$?
?\vrule?  ?\vrule?  $\fun{fs}(\vv{P}_i,A_l) = \sum_{j \in J, k \in [1,\mathit{lk}(j)]}-2^{\fun{dist}_k(j)}$
?\vrule?  $\fun{fs}(\vv{P}_i,\mathbf{A}) = \sum_{A_l \in \mathbf{A}}\fun{fs}(\vv{P}_i,\mathbf{A})/|\mathbf{A}|$
?\vrule?  ?$\fun{fc}(\vv{P}_i) = \sum_{j \in J}
\begin{cases}
\fun{lp}(\vv{P}_i) - \fun{pr}(j) \text{, if $j$ is an aperiodic task}\\
0 \text{, otherwise}
\end{cases}$?
$\vv{R} \leftarrow \fun{sort\_non\_dominated\_fronts}(\mathbf{P} \cup \mathbf{P}_\alpha)$
$\fun{assign\_crowding\_distance}(\vv{R})$
$\mathbf{P} \leftarrow \fun{select\_archive}(\vv{R},\var{ps}_p)$

return $\mathbf{P}$
\end{lstlisting}
\caption{An NSGAII-based algorithm for evolving priority assignments.}
\label{fig:NSGAII}
\end{figure}

Figure~\ref{fig:NSGAII} shows the evolution procedure of priority assignments, which refines lines 23--26 in Figure~\ref{fig:coevolution}. OPAM tailors the Non-dominated Sorting Genetic Algorithm version 2 (NSGAII)~\citep{Deb2002} to generate a non-dominating (equally viable) set of priority assignments, representing the best trade-offs found among the given internal fitness functions. This is referred to as a Pareto nondominated front~\citep{Knowles2000}, where the dominance relation over priority assignments is defined as follows: A priority assignment $\vv{P}$ dominates another priority assignment $\vv{P}^\prime$ if $\vv{P}$ is not worse than $\vv{P}^\prime$ in all fitness values, and $\vv{P}$ is strictly better than $\vv{P}^\prime$ in at least one fitness value. NSGAII has been applied to many multi-objective optimization problems~\citep{Langdon2010,Shin2018,Wang2020}.

OPAM maintains a population $\mathbf{P}$ of priority assignments as an archive that contains the best priority assignments discovered during coevolution. Unlike a standard application of NSGAII, in our study, we need to reevaluate the internal fitness values for priority assignments in $\mathbf{P}$ at every generation as the internal fitness values are computed based on the $\mathbf{A}$ population of task-arrival sequences, which coevolves. As shown in lines 9--16 in Figure~\ref{fig:NSGAII}, OPAM first computes the internal fitness functions that measure the magnitude of safety margins and the extent of constraint satisfaction. OPAM then sorts non-dominated Pareto fronts (line 19) and assigns crowding distance (line 20) to introduce diversity among non-dominated priority assignments~\citep{Deb2002}. 

For breeding the next population of priority assignments (line 21 in Figure~\ref{fig:NSGAII}, OPAM applies the following standard genetic operators~\citep{Sivan2008} that have been applied to many similar problems~\citep{Islam2012,Marchetto2016,Shin2018}: (1)~\emph{Selection.} OPAM uses a binary tournament selection based on non-domination ranking and crowding distance. The binary tournament selection has been used in the original implementation of NSGAII~\citep{Deb2002}. (2)~\emph{Crossover.} OPAM applies a partially mapped crossover (PMX)~\citep{Goldberg1985}. PMX ensures that the generated offspring are valid permutations of priorities. (3)~\emph{Mutation.} OPAM uses a permutation swap method for mutating a priority assignment. This mutation method interchanges two randomly-selected priorities in a priority assignment according to a given mutation probability.

For the generated population $\mathbf{P}_\alpha$ of priority assignments, OPAM computes the two internal fitness functions (lines 22--29 in Figure~\ref{fig:NSGAII}). OPAM then sorts non-dominated Pareto fronts for the union of the current $\mathbf{P}$ and next $\mathbf{P}_\alpha$ populations (line 30), assign crowding distance (line 31), and select the best archive by accounting for the computed non-domination ranking and crowding distance (line 32). 
 
\subsection{External fitness evaluation}
\label{subsec:external}

\begin{figure}[t]
\begin{lstlisting}[style=Alg]
Algorithm Priority assignments evolution
Input $\mathbf{E}$: set of task-arrival sequences //external evaluation
Input $\mathbf{P}$: population of priority assignments
Input $\var{ps}_p$: population size //priority assignments
Input $\var{cp}_p$: crossover probability //priority assignments
Input $\var{mp}_p$: mutation probability //priority assignments
Output $\mathbf{P}$: population of priority assignments

//evaluate external fitness values for ?$\color{javagreen}\mathbf{P}$?
for each $\vv{P}_i \in \mathbf{P}$
?\vrule?  for each $E_l \in \mathbf{E}$
?\vrule?  ?\vrule?  $S \leftarrow \fun{simulate}(E_l,\vv{P}_i)$ //OPAMScheduler
?\vrule?  ?\vrule?  //?$\color{javagreen}\fun{dist}_k(j)$? is computed based on ?$\color{javagreen}S$?
?\vrule?  ?\vrule?  $\fun{fs}(\vv{P}_i,E_l) = \sum_{j \in J, k \in [1,\mathit{lk}(j)]}-2^{\fun{dist}_k(j)}$
?\vrule?  $\fun{fs}(\vv{P}_i,\mathbf{E}) = \sum_{E_l \in \mathbf{E}}\fun{fs}(\vv{P}_i,\mathbf{E})/|\mathbf{E}|$
?\vrule?  ?$\fun{fc}(\vv{P}_i) = \sum_{j \in J}
\begin{cases}
\fun{lp}(\vv{P}_i) - \fun{pr}(j) \text{, if $j$ is an aperiodic task}\\
0 \text{, otherwise}
\end{cases}$?
$\vv{R} \leftarrow \fun{sort\_non\_dominated\_fronts}(\mathbf{P} \cup \mathbf{B})$
$\fun{assign\_crowding\_distance}(\vv{R})$
$\mathbf{B} \leftarrow \fun{select\_best\_front}(\vv{R})$ //?$\color{javagreen}|\mathbf{B}| \le |\mathbf{P}|$? 

return $\mathbf{B}$
\end{lstlisting}
\caption{An algorithm for evaluating external fitness and finding the best Pareto front.}
\label{fig:external}
\end{figure}

Figure~\ref{fig:external} shows an algorithm that computes the external fitness functions and finds the best Pareto front, which refines lines 28--31 in Figure~\ref{fig:coevolution}. To monitor the coevolution progress in a stable manner, OPAM takes as input a set $\mathbf{E}$ of task-arrival sequences that are generated independently from the coevolution process. We use an adaptive random search technique~\citep{Chen2010} to sample task-arrival sequences in order to create $\mathbf{E}$. The adaptive random search extends the naive random search by maximizing the Euclidean distance between the sampled points such that it maximizes the diversity of task-arrival sequences in $\mathbf{E}$.

As shown in lines 9--16 in Figure~\ref{fig:external}, OPAM computes the two external fitness values for each priority assignment in the $\mathbf{P}$ population based on a given set $\mathbf{E}$ of task-arrival sequences. OPAM then sorts non-dominated Pareto fronts for the union of the $\mathbf{P}$ population and the current best Pareto front (line 17), assigns crowding distance (line 18), and selects the best Pareto front by accounting for the computed non-domination ranking and crowding distance (line 32). OPAM adopts NSGAII in order to maximize the diversity of priority assignments in the best Pareto front.

\section{Evaluation}
\label{sec:eval}

This section describes our evaluation of OPAM through six industrial case studies from different domains and several synthetic subjects. Our full evaluation package is available online~\citep{Artifacts}.

\subsection{Research questions}
\label{subsec:RQs}

\noindent\textbf{RQ1 (Sanity check):} \textit{How does OPAM perform compared with Random Search?}
For search-based solutions, this RQ is an important \emph{sanity check} to ensure that success is not due to the search problem being easy~\citep{Arcuri2014}. Our conjecture is that a search-based algorithm, although expensive, will significantly outperform naive random search (RS).

\noindent\textbf{RQ2 (Coevolution):} \textit{Is competitive coevolution suitable to find best-case priority assignments?}
We conjecture that a coevolutionary algorithm is a suitable solution to address the priority assignment problem since it is solved, in practice, through a competing interactive process between the development and testing teams. To answer this RQ, we compare OPAM with a sequential approach that first looks for worst-case sequences of task arrivals and then tries to find best-case priority assignments.

\noindent\textbf{RQ3 (Scalability):} \textit{Can OPAM find (near-)optimal solutions for large-scale systems in a reasonable time budget?}
In this RQ, we investigate the scalability of OPAM by conducting some experiments with systems of various sizes, including six industrial and several synthetic subjects. We study the relationship between OPAM's performance measures and the characteristics of study subjects.

\noindent\textbf{RQ4 (Usefulness):} \textit{How do priority assignments generated by OPAM compare with priority assignments defined by engineers?}
OPAM can be considered useful only when it finds priority assignments that show benefits over those defined (manually) by engineers with domain expertise. This RQ therefore compares the quality of priority assignments generated by OPAM with those defined by engineers. We further discuss the usefulness of OPAM from a practical perspective, based on the feedback received from engineers in LuxSpace.
\subsection{Industrial study subjects}
\label{subsec:industrial subjects}

\begin{table*}[t]
\caption{Description of the six industrial subject systems: number of periodic and aperiodic tasks, resource dependencies, triggering relations, and platform cores.}
\label{tbl:subjects}
\begin{center}
\begin{tabular}{lccccc}
\toprule
& \multicolumn{2}{c}{Task types} & \multicolumn{2}{c}{Relationships} & Platform \\
\cmidrule(lr){2-3}\cmidrule(lr){4-5}\cmidrule(lr){6-6}
\multicolumn{1}{c}{System} & Periodic & Aperiodic & Dependencies & Triggering & Cores \\
\midrule
ICS    &  3 &  3 & 3 & 0 & 3 \\
CCS    &  8 &  3 & 3 & 6 & 2 \\
UAV    & 12 &  4 & 4 & 0 & 3 \\
GAP    & 15 &  8 & 6 & 5 & 2 \\
HPSS   & 23 &  9 & 5 & 0 & 1 \\
ESAIL  & 11 & 14 & 0 & 0 & 1 \\
\bottomrule
\end{tabular}
\end{center}
\end{table*}

\textcolor{rev3}{To evaluate RQs in realistic and diverse settings, we apply OPAM to six industrial study subjects from different domains such as aerospace, automotive, and avionics domains. Specifically, we obtained one case study subject from our industry partner, LuxSpace. We found the other five industrial study subjects in the literature~\citep{Alesio2015}, which, consistent with the LuxSpace system, all assume a single-queue, multi-core, fixed-priority scheduling policy. Note that OPAM uses the same scheduling policy (described in Section~\ref{subsec:simulation}) as in \citeauthor{Alesio2015}'s work. This policy uses fixed priorities that are determined offline and therefore do not change dynamically.}
Table~\ref{tbl:subjects} summarizes the relevant attributes of these subjects, presenting the number of periodic and aperiodic tasks, resource dependencies, triggering relations, and platform cores. \textcolor{rev3}{The subjects are characterized by real-time parameters, e.g., periods, deadlines, and priorities, described in Section~\ref{sec:problem}.  We note that all the study subjects are deadlock-free systems as they do not have circular resource dependencies.} Regarding task priorities, all tasks in the six subjects have fixed priorities, which are defined by experts in their domains. The full task descriptions (including WCET, inter-arrival times, periods, deadlines, priorities, and relationship details) of the subjects are available online~\citep{Artifacts}. The main missions of the six subjects are described as follows: 

\begin{sloppypar}
\begin{itemize}[leftmargin=1em]
\item ICS is an ignition control system that checks the status of an automotive engine and corrects any errors of the engine~\citep{Frati2008}. The system was developed by Bosch GmbH\footnote{Bosch {GmbH}: https://www.bosch.com/}.
\item CCS is a cruise control system that acquires data from vehicle sensors and maintains the specified vehicle speed~\citep{Anssi2011}. Continental AG\footnote{Continental {AG}: https://www.continental.com} developed the system.
\item UAV is a mini unmanned air vehicle that follows dynamically defined way-points and communicates with a ground station to receive instructions~\citep{Traore2006}. The system was developed in collaboration with the University of Poitiers France and ENSMA\footnote{ENSMA: https://www.ensma.fr/}.
\item GAP is a generic avionics platform for a military aircraft~\citep{Locke1990}. The system was designed in a joint project with Carnegie Mellon University, the US Navy, and IBM\footnote{IBM: https://www.ibm.com/}, aiming at supporting several missions regarding air-to-surface attacks. 
\item HPSS is a satellite system for two satellites, named Herschel and Planck~\citep{Marius2010}. The two satellites share the same computational architecture, although they have different scientific missions. Herschel aims at studying the origin and evolution of stars and galaxies. Planck's primary mission is the study of the relic radiation from the Big Bang. ESA\footnote{ESA: https://www.esa.int/} carried out the HPSS project.
\item ESAIL is a microsatellite for tracking ships worldwide by detecting messages that ships radio-broadcast (see Section~\ref{sec:motivation}). Luxspace, our industry partner, developed ESAIL in an ESA project.
\end{itemize}
\end{sloppypar}
\subsection{Synthetic study subjects}
\label{subsec:synthetic subjects}

To investigate RQ3, we use synthetic subjects in order to freely control key parameters in real-time systems. We create a set of tasks by adopting a well-known procedure~\citep{Emberson2010} for synthesizing real-time tasks, which has been applied in many schedulability analysis studies~\citep{Davis2008,Zhang2009,Davis2011,Grass2018,Durr2019}.

\lstset{morekeywords={continue,not,is}, escapeinside={\and,\times}{and,times}}
\begin{figure}[t]
\begin{lstlisting}[style=Alg]
Algorithm Synthetic task generation
Input ${n}$: number of tasks
Input $u_t$: target utilization
Input $\var{pd}_\var{min}$: minimum task period
Input $\var{pd}_\var{max}$: maximum task period
Input $g$: granularity of task periods
Input $\gamma$: ratio of aperiodic tasks
Input $\mu$: range factor to determine maximum inter-arrival ?times?
Output $\mathbf{S}$: set of tasks

$\mathbf{S} \leftarrow$ $\{\}$, $\mathbf{C} \leftarrow$ $\{\}$
// synthesize a set of periodic tasks
$\mathbf{U} \leftarrow \fun{UUniFast\_discard}(n, u_t)$ // task utilizations
$\mathbf{I} \leftarrow \fun{generate\_task\_periods}(n, \var{pd}_\var{min}, \var{pd}_\var{max}, g)$ // task periods
for each $j \in [1,n]$
?\vrule?  $\mathbf{C}$ $\leftarrow$ $\mathbf{C} \cup \{U_j {\cdot} I_j\}$, where $U_j \in \mathbf{U}$ and $I_j \in \mathbf{I}$ // WCETs
$ \mathbf{S} \leftarrow \fun{generate\_task\_set}(\mathbf{I}, \mathbf{C})$ // set of tasks
// convert some periodic tasks to aperiodic tasks
$ \mathbf{S} \leftarrow \fun{convert\_aperiodic\_tasks}(\mathbf{S}, \gamma, \mu)$

return $\mathbf{S}$
\end{lstlisting}
\caption{An algorithm for synthesizing a set of tasks.}
\label{fig:synthetic}
\end{figure}

Figure~\ref{fig:synthetic} describes a procedure that synthesizes a set of real-time tasks. For a given number $n$ of tasks and a target utilization $u_t$, the procedure first generates a set $\mathbf{U}$ of task utilization values by using the UUniFast-Discard algorithm~\citep{Davis2011} (line 13). The UUniFast-Discard algorithm is devised to give an unbiased distribution of utilization values, where a utilization $U_j \in \mathbf{U}$ is a positive value and $\sum_{U_j \in \mathbf{U}} U_j = u_t$.

The procedure then generates a set $\mathbf{I}$ of $n$ task periods according to a log-uniform distribution within a range  $[\var{pd}_\var{min}, \var{pd}_\var{max}]$, i.e., given a task period (random variable) $I_j \in \mathbf{I}$, $\log{I_j}$ follows a uniform distribution (line 14 in Figure~\ref{fig:synthetic}). For example, when the minimum and maximum task periods are $\var{pd}_\var{min} = 10\text{ms}$ and $\var{pd}_\var{max} = 1000\text{ms}$, respectively, the procedure generates (approximately) an equal number of tasks in time intervals  [10ms, 100ms] and [100ms, 1000ms]. The parameter $g$ is used to choose the granularity of the periods, i.e., task periods are multiples of $g$. Such a distribution of task periods provides a reasonable degree of realism with respect to what is usually observed in real systems~\citep{Baruah2011}.

As shown in lines 15--16 of the procedure in Figure~\ref{fig:synthetic}, a set $\mathbf{C}$ of task WCETs are computed based on the set $\mathbf{U}$ of task utilization values and the set $\mathbf{I}$ of task periods. Specifically, a task WCET $C_j \in \mathbf{C}$ is computed as $C_j = U_j \cdot I_j$.

As per line 17 of the listing in Figure~\ref{fig:synthetic}, the procedure synthesizes a set $\mathbf{S}$ of tasks. A task $j$ is characterized by a period $I_j$ and a WCET $C_j$ and it is associated with a deadline $\fun{dl}(j)$ and a priority $\fun{pr}(j)$. According to the rate-monotonic scheduling policy~\citep{Liu1973}, tasks' deadlines are equal to their periods and tasks with shorter periods are given higher priorities.

To synthesize aperiodic tasks, the procedure converts some periodic tasks to aperiodic tasks according to a given ratio $\gamma$ of aperiodic tasks among all tasks (see line 19 in Figure~\ref{fig:synthetic}). A range factor $\mu$ is used to determine maximum inter-arrival times of aperiodic tasks. Specifically, for a task $j$ to be converted, the procedure sets the minimum inter-arrival time $\fun{pmin}(j)$ as $\fun{pmin}(j) = I_j$. The procedure then selects a uniformly distributed value $x$ from the range $(1, \mu]$ and computes the maximum inter-arrival time $\fun{pmax}(j)$ as $\fun{pmax}(j) = x \cdot I_j$.
\subsection{Experimental Design}
\label{subsec:design}

This section describes how we design experiments to answer the RQs described in Section~\ref{subsec:RQs}. We conducted four experiments, EXP1, EXP2, EXP3, and EXP4, as described below.

\noindent\textbf{EXP1.}
To answer RQ1, EXP1 compares OPAM with our baseline, which relies on random search, to ensure that the effectiveness of OPAM is not due to the search problem being simple. Our baseline, named RS, replaces GA with a random search for finding worst-case sequences of task arrivals and NSGAII with a random search for finding best-case priority assignments. Note that RS uses the same internal and external fitness functions (see Section~\ref{subsec:fitness}) and also maintains the best populations during search; however, it does not employ any genetic operators, i.e., crossover and mutation. In EXP1, we applied OPAM and RS to the six industrial subjects described in Section~\ref{subsec:industrial subjects}.

Recall from Section~\ref{subsec:fitness} that OPAM uses a set $\mathbf{E}$ of task-arrival sequences that are generated independently from the coevolution process in order to monitor the coevolution progress in a stable manner. As OPAM and RS use the same set $\mathbf{E}$ of task-arrival sequences, EXP1 first compares OPAM and RS based on $\mathbf{E}$. In addition, EXP1 examines how well the solutions, i.e., priority assignments, found by OPAM and RS perform with other sequences of task arrivals. To do so, we create six sets of sequences of task arrivals for each study subject by varying the method to generate task-arrival sequences and the number of task-arrival sequences. Note that task-arrival sequences generated by different methods are valid with respect to the inter-arrival times defined in each study subject. Below we describe the six sets of task-arrival sequences generated for each subject.

\begin{itemize}[leftmargin=1em]
\item $\mathbf{T}_{a}^{10}$: A set of task-arrival sequences generated by using an adaptive random search technique~\citep{Chen2010} that aims at maximizing the diversity of task-arrival sequences. The $\mathbf{T}_{a}^{10}$ set contains 10 sequences of task arrivals.
\item $\mathbf{T}_{w}^{10}$: A set of task-arrival sequences generated by using a stress test case generation method that aims at maximizing the chances of deadline misses in task executions. The stress test case generation method extends prior work~\citep{Briand2005}. The extended method uses the fitness function regarding deadline misses and genetic operators that OPAM introduces for evolving worst-case task-arrival sequences (see Section~\ref{sec:approach}). The $\mathbf{T}_{w}^{10}$ set contains 10 sequences of task arrivals.
\item $\mathbf{T}_{r}^{10}$: A set of task-arrival sequences generated randomly. The $\mathbf{T}_{r}^{10}$ set has 10 sequences of task arrivals.
\item $\mathbf{T}_{a}^{500}$: A set of task-arrival sequences generated by using the adaptive random search technique. The $\mathbf{T}_{a}^{500}$ set contains 500 sequences of task arrivals.
\item $\mathbf{T}_{w}^{500}$: A set of task-arrival sequences generated by using the stress test case generation method. The $\mathbf{T}_{w}^{500}$ set contains 500 sequences of task arrivals.
\item $\mathbf{T}_{r}^{500}$: A set of task-arrival sequences generated randomly. The $\mathbf{T}_{r}^{500}$ set has 500 sequences of task arrivals.
\end{itemize}

\noindent\textbf{EXP2.}
To answer RQ2, EXP2 compares OPAM with a priority assignment method, named SEQ, that relies on one-population search algorithms. SEQ first finds a set of worst-case sequences of task arrivals using GA with the fitness function that measures the magnitude of deadline misses (see $\fun{fd}()$ in Section~\ref{subsec:fitness}) and the genetic operators described in Section~\ref{subsec:evolution arrivals}. Given a set of worst-case task-arrival sequences obtained from GA, SEQ then aims at finding best-case priority assignments using NSGAII with the fitness functions that quantify the magnitude of safety margins and the degree of constraint satisfaction (see $\fun{fs}()$ and $\fun{fc}()$, respectively, in Section~\ref{subsec:fitness}) and the genetic operators described in Section~\ref{subsec:evolution priorities}.

We note that SEQ does not use the external fitness functions as it does not coevolve task-arrival sequences and priority assignments. Hence, the numbers of fitness evaluations of the two methods are not comparable. To fairly compare OPAM and SEQ, we set the same time budget for the two methods. Specifically, we first measure the execution time of OPAM for analyzing each subject. We then split the execution time in half and set each half time as the execution budget of the GA and NSGAII steps in SEQ for the corresponding subject. In order to assess the quality of priority assignments obtained from OPAM and SEQ, we use the sets of task-arrival sequences described in EXP1, i.e., $\mathbf{T}_{a}^{10}$, $\mathbf{T}_{w}^{10}$, $\mathbf{T}_{r}^{10}$, $\mathbf{T}_{a}^{500}$, $\mathbf{T}_{w}^{500}$, and $\mathbf{T}_{r}^{500}$, which are created independently from the two methods.

\noindent\textbf{EXP3.}
To answer RQ3, EXP3 examines not only the six industrial subjects but also 370 synthetic subjects. We create the synthetic subjects to study correlations between the execution time and memory usage of OPAM and the following parameters: the number of tasks ($n$), a (part-to-whole) ratio of aperiodic tasks ($\gamma$), a range factor for maximum inter-arrival times ($\mu$), and simulation time ($T$), as described in Sections~\ref{subsec:synthetic subjects} and~\ref{sec:approach}. We note that we chose to control parameters $n$, $\gamma$, and $\mu$ because they are the main parameters on which engineers have control to define tasks in real-time systems. Simulation time $T$ obviously impacts the execution time of OPAM as well. But EXP3 aims at modeling such correlations precisely and providing experimental results. Regarding the other factors that define, for example, task relationships and platform cores, we note significant diversity across the six industrial subjects.

Recall from Section~\ref{subsec:synthetic subjects} that we use the task generation procedure presented in Figure~\ref{fig:synthetic} to synthesize tasks. For EXP3, we set some parameter values of the procedure as follows: (1)~Target utilization $u_t = 0.7$, which is a common objective in the development of a real-time system in order to guarantee the schedulability of tasks~\citep{Fineberg1967,Durr2019}. (2)~The range of task periods  $[\var{pd}_\var{min}, \var{pd}_\var{max}] = [10\text{ms}, 1\text{s}]$, which are common values in many real-time systems~\citep{Emberson2010,Baruah2011}. (3)~The granularity of task periods $g = 10\text{ms}$ in order to increase realism as most of the task periods in our industrial subjects are multiples of 10ms. Because of some degree of randomness in the procedure of Figure~\ref{fig:synthetic}, we create ten synthetic subjects per configuration. Below we further describe how synthetic subjects are created for each controlled experiment.

\emph{EXP3.1.} To study the correlations between the execution time and memory usage of OPAM with the number of tasks $n$, we create nine sets of ten synthetic subjects such that no two sets have the same number of tasks. Specifically, we create sets with 10, 15, ..., 50 tasks, respectively. Regarding the ratio of aperiodic tasks, $\gamma = 0.4$ as, on average, the ratio of aperiodic tasks to periodic tasks in our industrial subjects is 2/3. For the range factor, $\mu = 2$, which is determined based on the inter-arrival times of aperiodic tasks in our industry subjects. We set the simulation time $T$ to 2s in order to ensure that any aperiodic task arrives at least once during that time. We note that, given the maximum task period $\var{pd}_\var{max} = 1\text{s}$ and the range factor $\mu = 2$, the maximum inter-arrival time of an aperiodic task is at most 2s (see Section~\ref{subsec:synthetic subjects}).

\emph{EXP3.2.} To study the correlations between the execution time and memory usage of OPAM with the ratio of aperiodic tasks $\gamma$, we create ten sets of synthetic subjects by setting this ratio to the following values: 0.05, 0.10, ..., 0.50. We set the number of tasks to 20 ($n = 20$), which is the average number of tasks in our six industrial subjects. Regarding the other parameters, range factor and simulation time, $\mu = 2$ and $T = 2\text{s}$ are set as discussed in EXP3.1.

\emph{EXP3.3.} To study the correlations between the execution time and memory usage of OPAM with the range factor $\mu$ that is used to determine the maximum inter-arrival times, we create nine sets of synthetic subjects by setting $\mu$ to 2, 3, ..., 10. We set the simulation time as follows: $T = 10\text{s}$. This ensures that any aperiodic task arrives at least once during the simulation time when $\mu$ is at most 10 (see Section~\ref{subsec:synthetic subjects}). The other parameters, the number of tasks and ratio of aperiodic tasks, $n = 20$ and $\gamma = 0.4$ are set as discussed in EXP3.1 and EXP3.2.

\emph{EXP3.4.} To study the correlations between the execution time and memory usage of OPAM with the simulation time $T$, we create nine sets of synthetic subjects by setting $T$ to 2s, 3s, ..., 10s. The other parameters, e.g., the number of tasks, the ratio of aperiodic tasks, and the range factor, $n = 20$, $\gamma = 0.4$, and $\mu = 2$, are set as discussed in EXP3.1 and EXP3.2.

\noindent\textbf{EXP4.}
To answer RQ4, EXP4 compares priority assignments optimized by OPAM and those defined by engineers. We apply OPAM to the six industrial subjects (see Section~\ref{subsec:industrial subjects}) which include priority assignments defined by practitioners. Note that we focus here on the ESAIL subject in collaboration with our industry partner, LuxSpace; The other five subjects are from the literature~\citep{Alesio2015} and hence we can only collect feedback from practitioners for ESAIL.
\subsection{Evaluation metrics}
\label{subsec:metrics}

\noindent\textbf{Multi-objective evaluation metrics.}
In order to fairly compare the results of search algorithms, based on existing guidelines~\citep{Chen2020} for assessing multi-objective search algorithms, we use complementary quality indicators: \emph{Hypervolume} (HV)~\citep{Zitzler1999}, \emph{Pareto Compliant Generational Distance} (GD+)~\citep{Ishibuchi2015}, and \emph{Spread} ($\Delta$)~\citep{Deb2002}. To compute the GD+ and $\Delta$ quality indicators, following the usual procedure~\citep{Chen2020}, we create a reference Pareto front as the union of all the non-dominated solutions obtained from all runs of the algorithms being compared. Identifying the optimal (ideal) Pareto front is typically infeasible for a complex optimization problem~\citep{Chen2020}. Key features of the three quality indicators are described below.

\begin{itemize}[leftmargin=1em]
\item HV is defined to measure the volume in the objective space that is covered by members of a Pareto front generated by a search algorithm~\citep{Zitzler1999}. The higher the HV values, the more optimal the search outputs.
\item GD+ is defined to measure the distance between the points on a Pareto front obtained from a search algorithm and the nearest points on a reference Pareto front~\citep{Ishibuchi2015}. GD+ modifies General Distance (GD)~\citep{Veldhuizen1998} to account for the dominance relations when computing the distances. The lower the GD+ values, the more optimal the search outputs.
\item $\Delta$ is defined to measure the extent of spread among the points on a Pareto front computed by a search algorithm~\citep{Deb2002}. We note that OPAM aims at obtaining a wide variety of equally-viable priority assignments on a Pareto front (see Section~\ref{sec:approach}). The lower the Spread values, the more spread out the search outputs. 
\end{itemize}

\noindent\textbf{Interpretable metrics.}
The two external fitness functions described in Section~\ref{sec:approach} mainly aim at effectively guiding search. It is, however, difficult for practitioners to interpret the computed fitness values. Since they are not intuitive to practitioners, to assess the usefulness of OPAM from a practitioner perspective, we measure (1)~the safety margins from tasks' completion times to their deadlines across our experiments and (2)~the number of constraint violations in a priority assignment. In addition, we measure the execution time and memory usage of OPAM. 

\noindent\textbf{Statistical comparison metrics.}
To statistically compare our experiment results, we use the Mann-Whitney U-test~\citep{Mann1947} and Vargha and Delaney's $\hat{A}_{12}$ effect size~\citep{Vargha2000}, which have been frequently applied for evaluating search-based algorithms~\citep{Arcuri2010, Hemmati2013, Shin2018}. Mann-Whitney U-test determines whether two independent samples are likely or not to belong to the same distribution. We set the level of significance, $\alpha$, to 0.05. Vargha and Delaney’s $\hat{A}_{12}$ measures probabilistic superiority -- effect size -- between search algorithms. Two algorithms are considered to be equivalent when the value of $\hat{A}_{12}$ is 0.5.
\subsection{Parameter tuning and implementation}
\label{subsec:param}

\noindent\textbf{Parameters for coevolutionary search.}
For the coevolutionary search parameters, we set the population size to 10, the crossover rate to 0.8, and the mutation rate to $1/|J|$, where $|J|$ denotes the number of tasks. We apply these parameter values for both the evolution of task-arrival sequences and priority assignments (see Section~\ref{sec:approach}). These values are determined based on existing guidelines~\citep{Arcuri2011,Sayyad2013} and previous work~\citep{Lee2020}. 

We determine the number of coevolution cycles (see Section~\ref{sec:approach}) based on an initial experiment. We applied OPAM to the six industrial subjects and ran OPAM 50 times for each subject. From the experiment results, we observed that there is no notable difference in Pareto fronts generated after 1000 cycles. Hence, we set the number of coevolution cycles to 1000 in our experiments, i.e., EXP1, EXP2, and EXP3 described in Section~\ref{subsec:design}.

\noindent\textbf{Parameters for evaluating fitness functions.}
To evaluate external fitness functions, we use a set of task-arrival sequences that are generated independently from the coevolution process (see Section~\ref{subsec:external}). We use an adaptive random search~\citep{Chen2010} to generate a set $\mathbf{E}$ of task-arrival sequences, which varies task arrival times within the specified inter-arrival time ranges of aperiodic tasks. We set the size of $\mathbf{E}$ to 10. From our initial experiment, we observed that this is sufficient to compute the external fitness functions of OPAM under a reasonable time, i.e., less than 15s. We note that $\mathbf{E}$ contains two default sequences of task arrivals as follows: (seq.~1)~aperiodic tasks always arrive at their maximum inter-arrival times and (seq.~2)~aperiodic tasks always arrive at their minimum inter-arrival times. By having those two sequences of task arrivals as initial elements in $\mathbf{E}$, the adaptive random search finds other sequences of task arrivals to maximize the diversity of elements in $\mathbf{E}$. 

If a system contains only periodic tasks, the simulation time is often set as the least common multiple (LCM) of their periods to account for all possible arrivals~\citep{Peng1997}. However, as the six industrial subjects include aperiodic tasks, this is not applicable. For the experiments with the six industrial subjects, we set the simulation time to the maximum time between the LCM of periodic tasks' periods and the maximum inter-arrival time among aperiodic tasks. By doing so, all possible arrival patterns of periodic tasks are examined and any aperiodic task arrives at least once during  simulation. Recall from Section~\ref{subsec:evolution arrivals} that OPAM varies arrival times of aperiodic tasks to find worst-case sequences of task arrivals.

We note that the parameters mentioned above can probably be further tuned to improve the performance of our approach. However, since with our current setting, we were able to convincingly and clearly support our conclusions, we do not report further experiments on tuning those values.

\textbf{Implementation.}
We implemented OPAM by extending jMetal~\citep{Durillo2011}, which is a metaheuristic optimization framework supporting NSGAII and GA. We conducted our experiments using the high-performance computing cluster~\citep{Varrette2014} at the University of Luxembourg. To account for randomness, we repeated each run of OPAM 50 times for all experiments. Each run of OPAM was executed on a different node (equipped with five 2.5GHz cores and 20GB memory) of the cluster, and took less than 16 hours.
\subsection{Results}
\label{subsec:results}

\begin{figure}[t]
\begin{center}
    \subfloat[ICS]{
        \includegraphics[width=0.45\columnwidth]{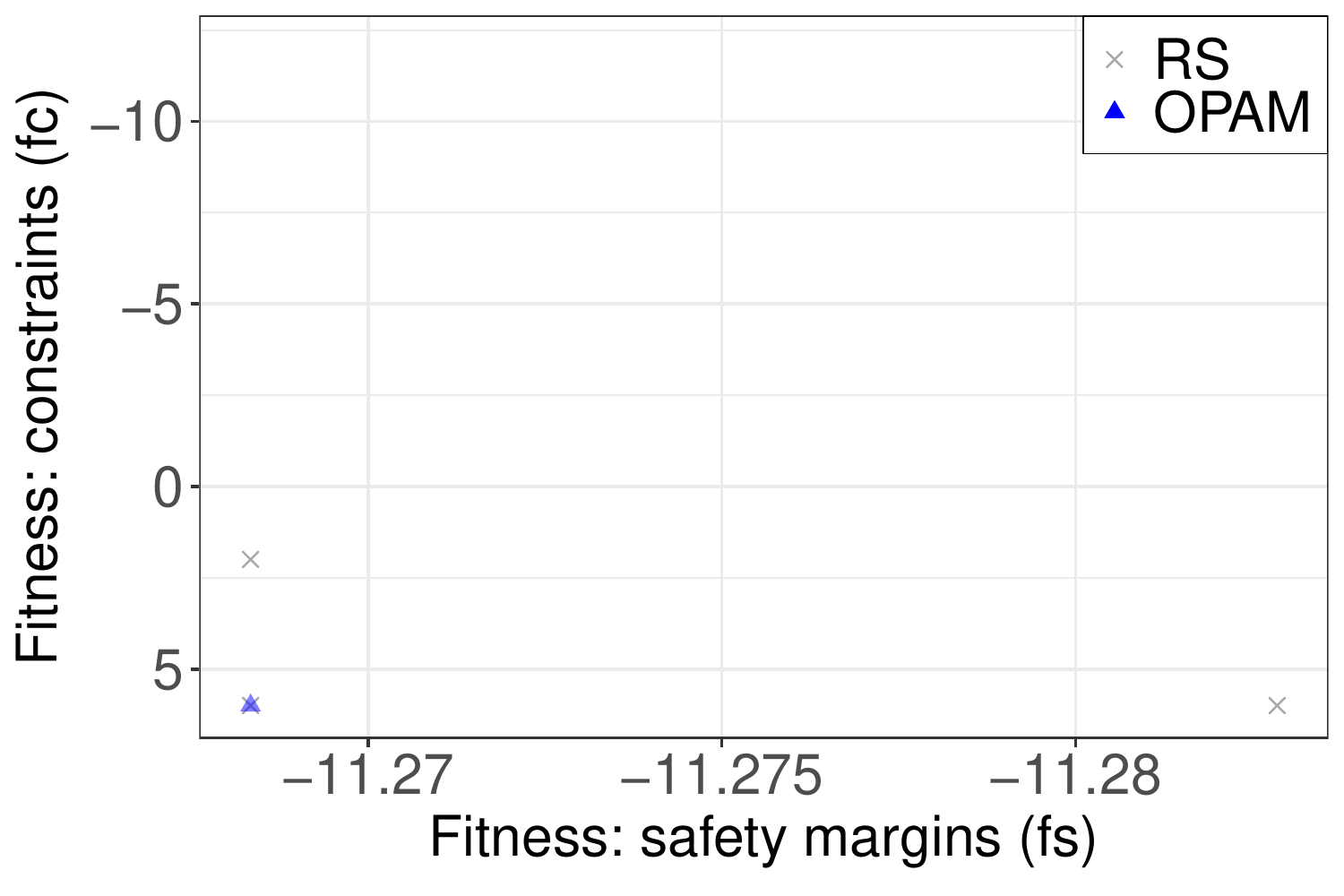}
        \label{fig:rq1scatter ics}
    }
    \hspace*{\fill}
    \subfloat[CCS]{
        \includegraphics[width=.45\columnwidth]{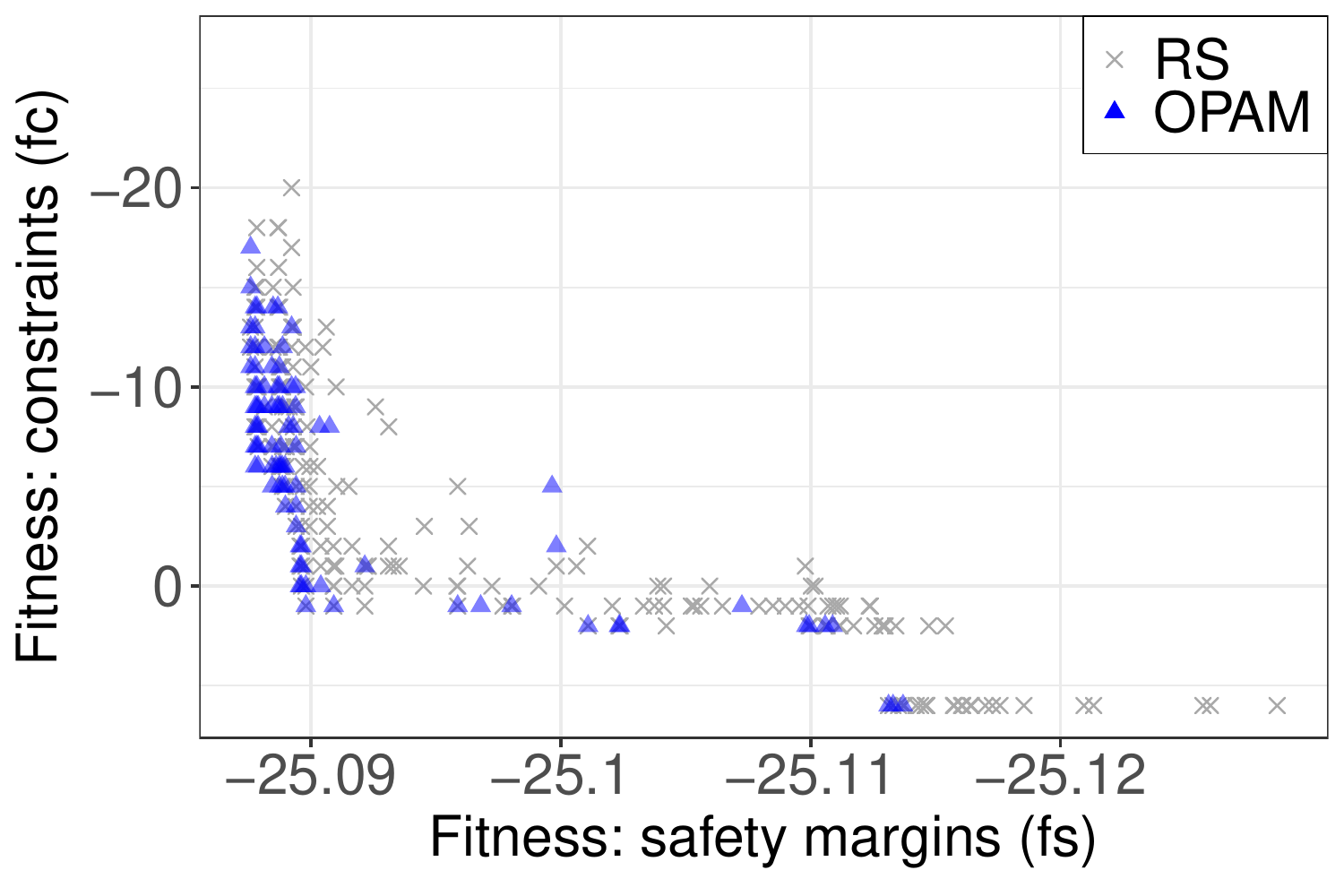}
        \label{fig:rq1scatter ccs}
    }
    \hfill
    \subfloat[UAV]{
        \includegraphics[width=.45\columnwidth]{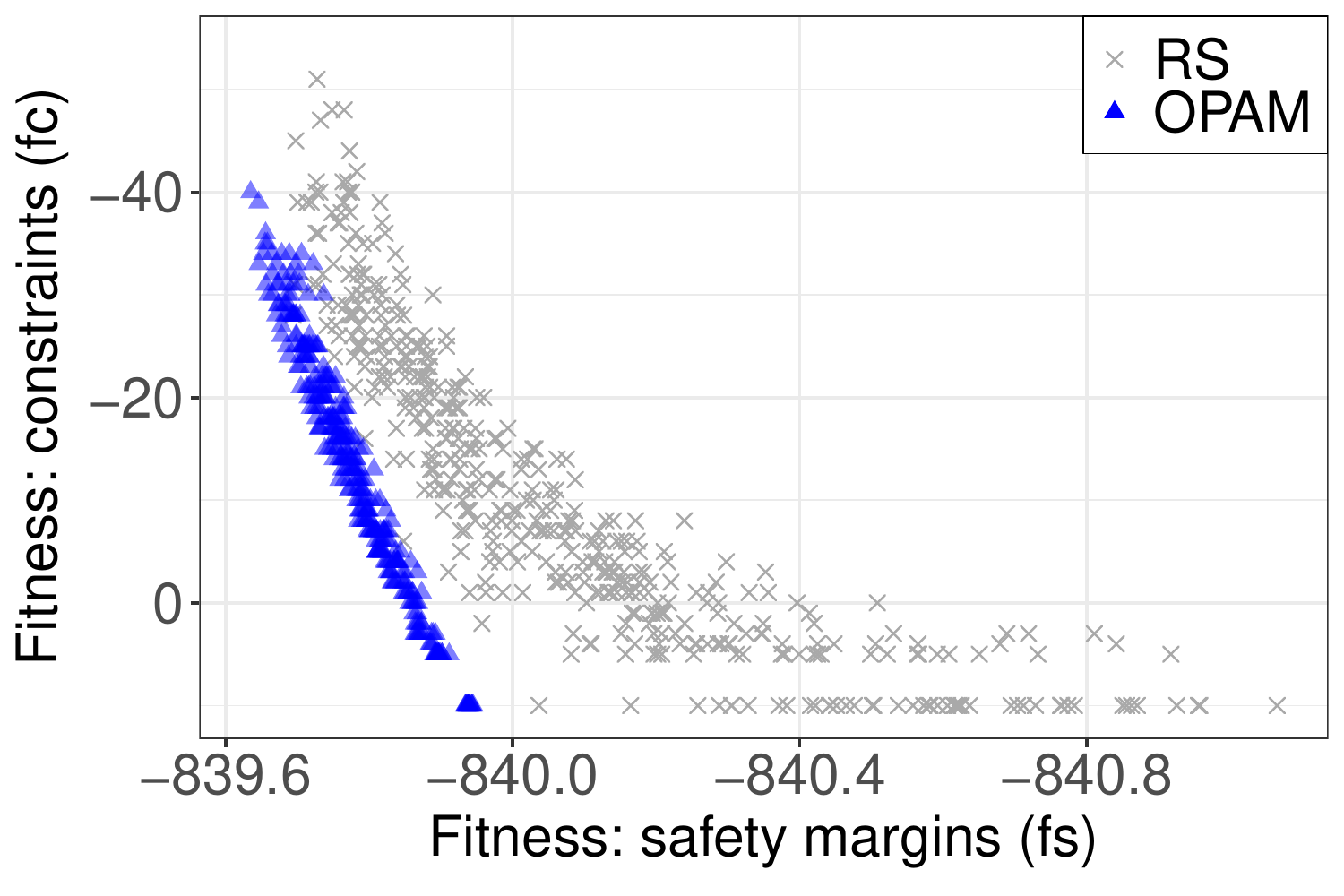}
        \label{fig:rq1scatter uav}
    }
    \hspace*{\fill}
    \subfloat[GAP]{
        \includegraphics[width=.45\columnwidth]{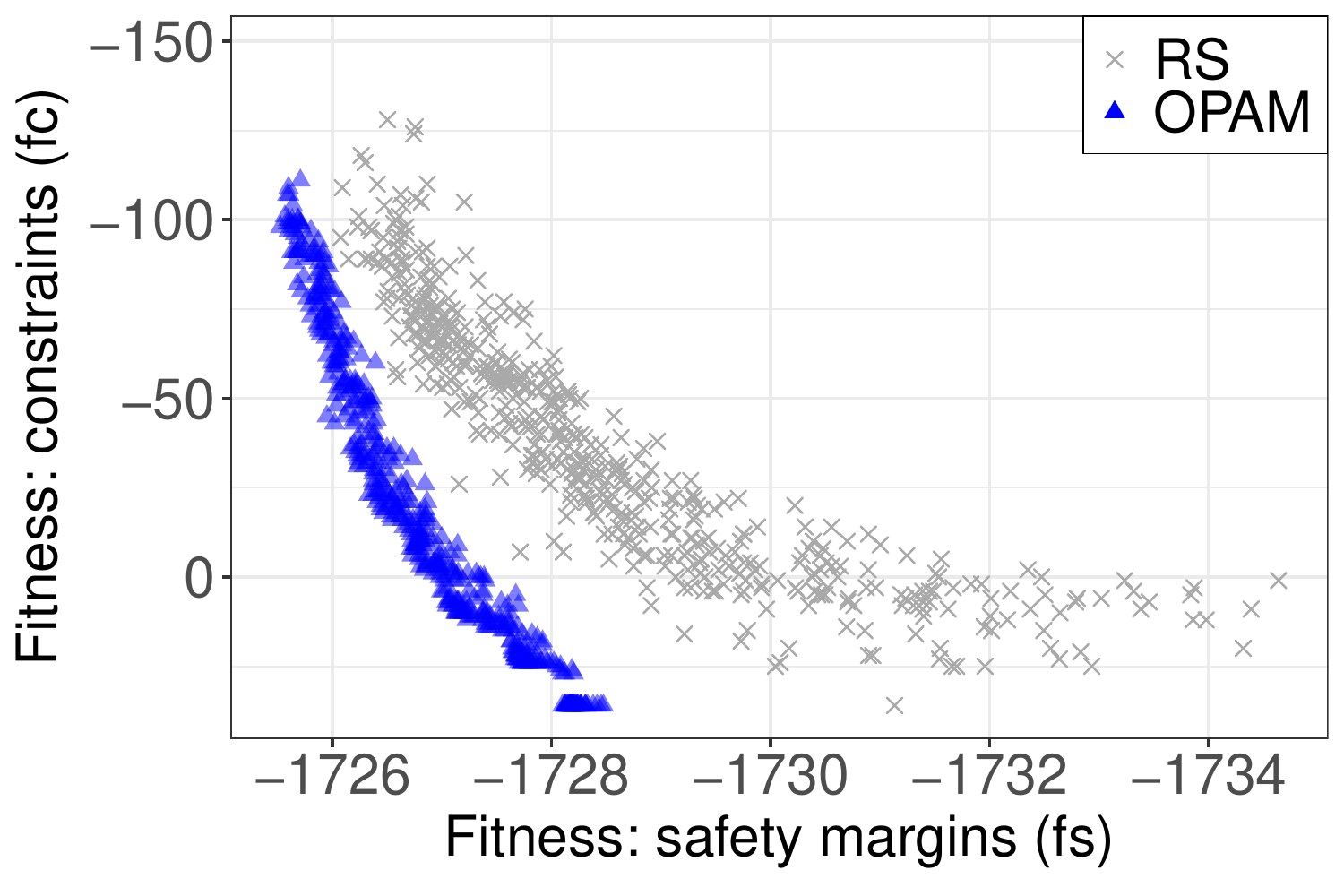}
        \label{fig:rq1scatter gap}
    }
    \hfill
    \subfloat[HPSS]{
        \includegraphics[width=.45\columnwidth]{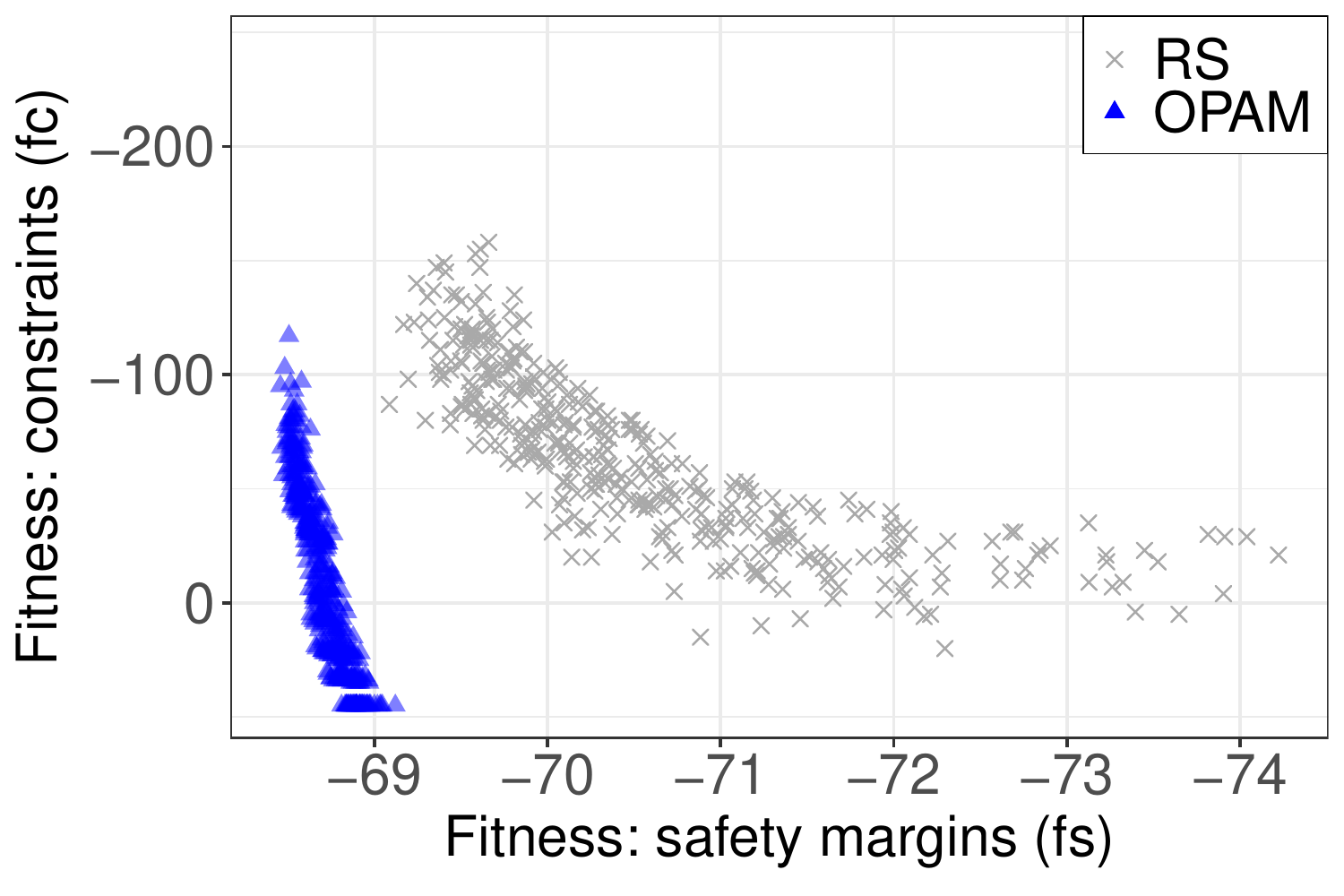}
        \label{fig:rq1scatter hpss}
    }
    \hspace*{\fill}
    \subfloat[ESAIL]{
        \includegraphics[width=0.45\columnwidth]{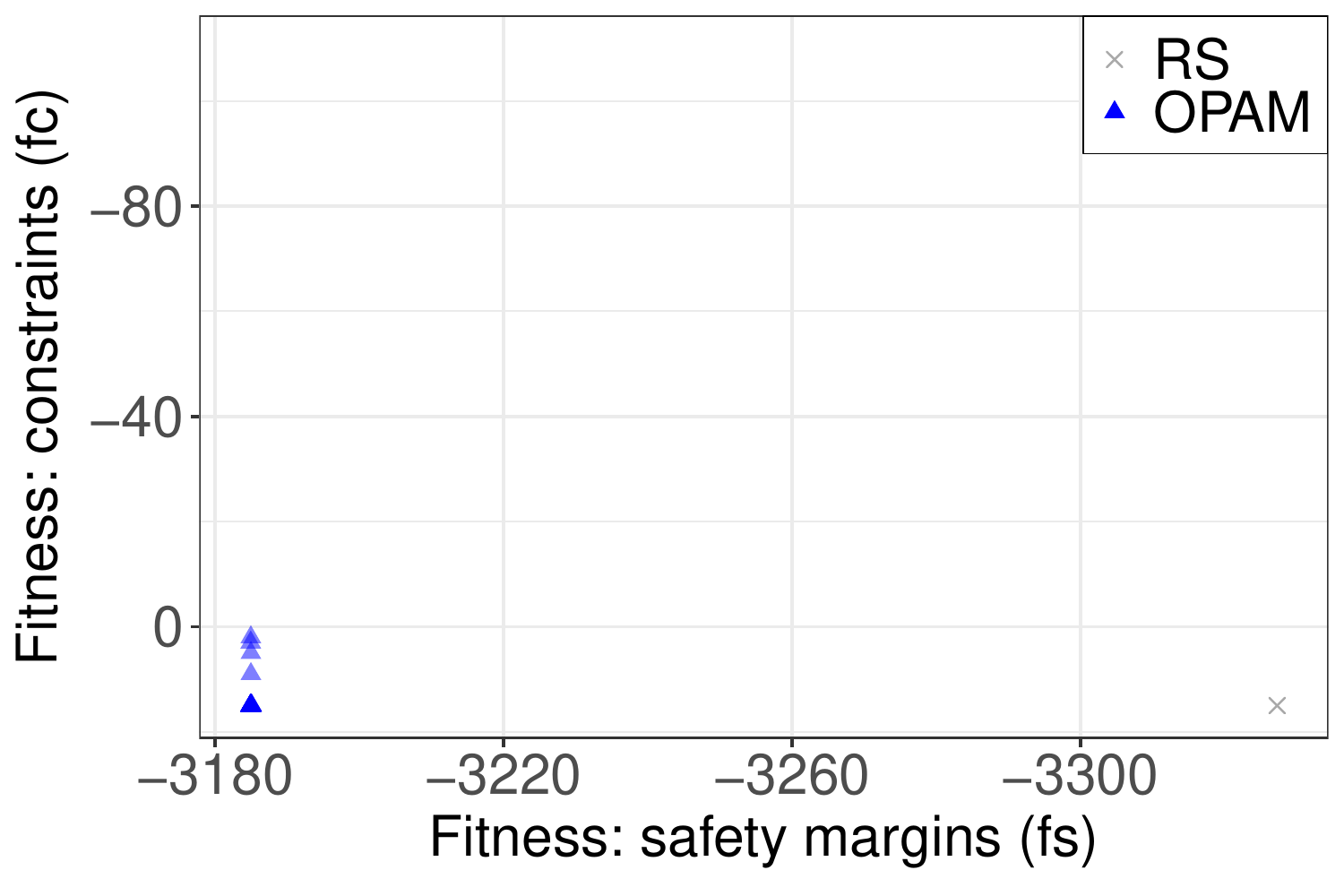}
        \label{fig:rq1scatter esail}
    }
\end{center}
\caption{Pareto fronts obtained by OPAM and RS for the six industrial subjects: (a)~ICS, (b)~CCS, (c)~UAV, (d)~GAP, (e)~HPSS, and (f)~ESAIL. The fitness values are computed based on each subject's set $\mathbf{E}$ of task-arrival sequences (see Section~\ref{subsec:param}). The points located closer to the bottom left of each plot are considered to be better priority assignments when compared to points closer to the top right.}
\label{fig:rq1scatter}
\end{figure}

\noindent\textbf{RQ1.}
Figure~\ref{fig:rq1scatter} shows the best Pareto fronts obtained with 50 runs of OPAM and RS, for the six industrial study subjects described in Section~\ref{subsec:industrial subjects}. The fitness values presented in the figures are computed based on each subject's set $\mathbf{E}$ of task-arrival sequences (see Section~\ref{subsec:param}), which is created independently from OPAM and RS. Figures~\ref{fig:rq1scatter ics}, \ref{fig:rq1scatter uav}, \ref{fig:rq1scatter gap}, \ref{fig:rq1scatter hpss}, and \ref{fig:rq1scatter esail} indicate that OPAM finds significantly better solutions than RS for ICS, UAV, GAP, HPSS, and ESAIL. 
Regarding CCS (see Figure~\ref{fig:rq1scatter ccs}), it is difficult to conclude anything based only on visual inspection. Hence, we compared Pareto fronts obtained by OPAM and RS using the three quality indicators HV, GD+, and $\Delta$, described in Section~\ref{subsec:metrics}.

\begin{figure*}[t]
\begin{center}
    \subfloat[HV]{
        \includegraphics[width=.49\columnwidth]{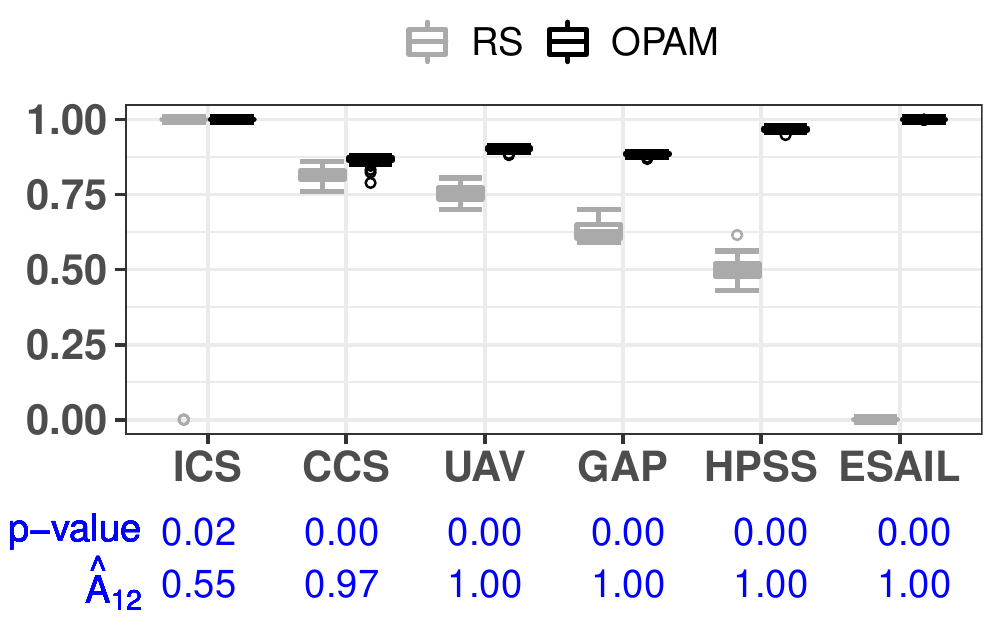}
        \label{fig:rq1 hv}
    }
    \hfill
    \subfloat[GD+]{
        \includegraphics[width=0.49\columnwidth]{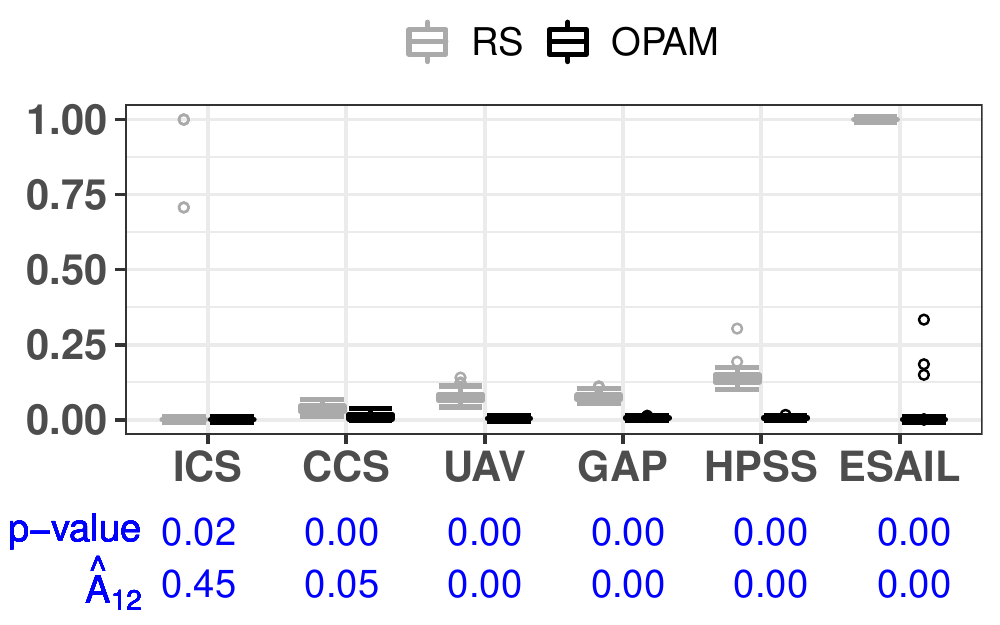}
        \label{fig:rq1 gd}
    }
    \hfill
    \subfloat[$\Delta$]{
        \includegraphics[width=0.49\columnwidth]{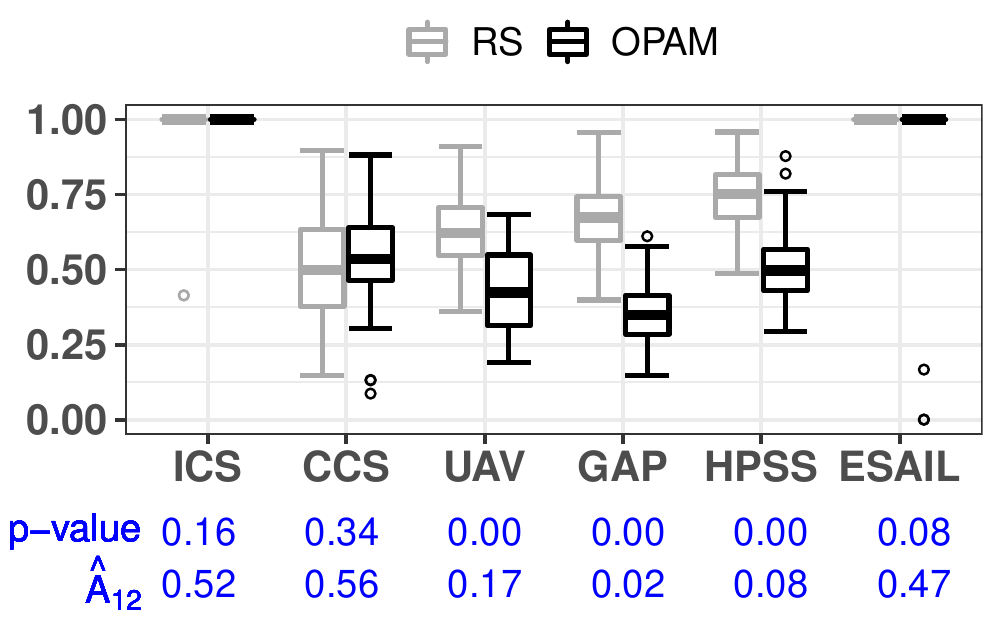}
        \label{fig:rq1 sp}
    }
\end{center}
\caption{Comparing OPAM and RS using the three quality indicators: (a)~HV, (b)~GD+, and (c)~$\Delta$. The boxplots (25\%-50\%-75\%) show the quality values obtained from 50 runs of OPAM and RS. The quality values are computed based on the Pareto fronts obtained by the algorithms and each subject's set $\mathbf{E}$ of task-arrival sequences (see Section~\ref{subsec:param}).}
\label{fig:rq1}
\end{figure*}

Figure~\ref{fig:rq1} depicts distributions of HV (Figure~\ref{fig:rq1 hv}), GD+ (Figure~\ref{fig:rq1 gd}), and $\Delta$ (Figure~\ref{fig:rq1 sp}) for the six industrial subjects. The boxplots in the figures present the distributions (25\%-50\%-75\%) of the quality values obtained from 50 runs of OPAM and RS. The quality values are computed based on the Pareto fronts obtained by the algorithms and each subject's set $\mathbf{E}$ of task-arrival sequences (see Section~\ref{subsec:param}). In the figures, statistical comparisons of the two corresponding distributions are summarized using p-values and $\hat{A}_{12}$ values, as described in Section~\ref{subsec:metrics}, under each subject name.

As shown in Figures~\ref{fig:rq1 hv} and \ref{fig:rq1 gd}, OPAM obtains better distributions of HV and GD+ compared to RS for all six subjects. All the differences are statistically significant as the p-values are below 0.05. Regarding $\Delta$, as depicted in Figure~\ref{fig:rq1 sp}, OPAM yields higher diversity in Pareto front solutions than RS for the following subjects: UAV, GAP, and HPSS. For ICS, CCS, and ESAIL, OPAM and RS obtain similar $\Delta$ values. From Figures~\ref{fig:rq1 hv} and \ref{fig:rq1 gd}, and Table~\ref{tbl:subjects}, we also observe that the higher the number of aperiodic tasks in a subject, the larger the differences in HV and GD+ between OPAM and RS. Hence, for these two quality indicators, OPAM outperforms RS more significantly for more complex search problems. Note that the number of aperiodic tasks is one of the main factors that drives the degree of uncertainty in task arrivals.

\begin{table}[!p]
	\caption{Comparing OPAM and RS using the three quality indicators: HV, GD+, and $\Delta$. Average quality values computed based on 50 runs of OPAM and RS using the different sets of task-arrival sequences (see Section~\ref{subsec:design}).}
	\vspace{-1.2em}
	\fontsize{8}{8}\selectfont
	\def\arraystretch{0.5}
\begin{center}
\begin{tabularx}{\columnwidth}{m{2.5em}@{}c@{\hspace{0.3em}}r@{\hspace{1em}} RRRRRR}
		\toprule
		\addlinespace[0.5em]
		\multicolumn{3}{c}{} & \multicolumn{1}{c}{\textbf{ICS}} & \multicolumn{1}{c}{\textbf{CCS}}
		 & \multicolumn{1}{c}{\textbf{UAV}} & \multicolumn{1}{c}{\textbf{GAP}} & \multicolumn{1}{c}{\textbf{HPSS}} & \multicolumn{1}{c}{\textbf{ESAIL}} \\
		\addlinespace[0.2em]
\midrule 
\multirow{9}{*}{\rotatebox{90}{
        \parbox{8em}{\centering{$\mathbf{T}^{10}_{a}$ \\ (adaptive, size 10)}}
}}
		& \multirow{3}{*}{\textbf{HV}}	& \textbf{OPAM} & \cellcolor{blue!30}\textbf{1.0000} & \cellcolor{blue!30}\textbf{0.7168} & \cellcolor{blue!30}\textbf{0.8923} & \cellcolor{blue!30}\textbf{0.8864} & \cellcolor{blue!30}\textbf{0.9629} & \cellcolor{blue!30}\textbf{0.9998} \\
		&						& \textbf{RS} & 0.9000 & 0.6633 & 0.7488 & 0.6278 & 0.5120 & 0.0000 \\
		&						& $p\vert\hat{A}_{12}$ & 0.02$\vert$0.55 & 0.00$\vert$0.80 & 0.00$\vert$1.00 & 0.00$\vert$1.00 & 0.00$\vert$1.00 & 0.00$\vert$1.00 \\
		\addlinespace[0.5em] 
		& \multirow{3}{*}{\textbf{GD+}}	& \textbf{OPAM} & \cellcolor{blue!30}\textbf{0.0000} & \cellcolor{blue!30}\textbf{0.0203} & \cellcolor{blue!30}\textbf{0.0068} & \cellcolor{blue!30}\textbf{0.0067} & \cellcolor{blue!30}\textbf{0.0073} & \cellcolor{blue!30}\textbf{0.0135} \\
		&						& \textbf{RS} & 0.0883 & 0.0472 & 0.0745 & 0.0780 & 0.1380 & 1.0000 \\
		&						& $p\vert\hat{A}_{12}$ & 0.02$\vert$0.45 & 0.00$\vert$0.04 & 0.00$\vert$0.00 & 0.00$\vert$0.00 & 0.00$\vert$0.00 & 0.00$\vert$0.00 \\
		\addlinespace[0.5em] 
		& \multirow{3}{*}{\textbf{$\Delta$}}	& \textbf{OPAM} & 1.0000 & 0.7650 & \cellcolor{blue!30}\textbf{0.4256} & \cellcolor{blue!30}\textbf{0.3631} & \cellcolor{blue!30}\textbf{0.5355} & 0.9433 \\
		&						& \textbf{RS} & 0.9766 & \cellcolor{gray!20}\textbf{0.5879} & 0.6112 & 0.6605 & 0.7508 & 1.0000 \\
		&						& $p\vert\hat{A}_{12}$ & 0.16$\vert$0.52 & 0.00$\vert$0.76 & 0.00$\vert$0.15 & 0.00$\vert$0.03 & 0.00$\vert$0.12 & 0.08$\vert$0.47 \\
\midrule 
\multirow{9}{*}{\rotatebox{90}{
        \parbox{8em}{\centering{$\mathbf{T}^{10}_{w}$ \\ (worst, size 10)}}
}}
		& \multirow{3}{*}{\textbf{HV}}	& \textbf{OPAM} & 0.0000 & \cellcolor{blue!30}\textbf{0.7878} & \cellcolor{blue!30}\textbf{0.9152} & \cellcolor{blue!30}\textbf{0.9280} & \cellcolor{blue!30}\textbf{0.9652} & \cellcolor{blue!30}\textbf{0.9997} \\
		&						& \textbf{RS} & 0.0000 & 0.7591 & 0.7782 & 0.6743 & 0.5180 & 0.0000 \\
		&						& $p\vert\hat{A}_{12}$ & 1.00$\vert$0.50 & 0.01$\vert$0.65 & 0.00$\vert$1.00 & 0.00$\vert$1.00 & 0.00$\vert$1.00 & 0.00$\vert$1.00 \\
		\addlinespace[0.5em] 
		& \multirow{3}{*}{\textbf{GD+}}	& \textbf{OPAM} & 0.0000 & 0.0809 & \cellcolor{blue!30}\textbf{0.0053} & \cellcolor{blue!30}\textbf{0.0042} & \cellcolor{blue!30}\textbf{0.0108} & \cellcolor{blue!30}\textbf{0.0135} \\
		&						& \textbf{RS} & 0.0200 & 0.0866 & 0.0740 & 0.0760 & 0.1405 & 1.0000 \\
		&						& $p\vert\hat{A}_{12}$ & 0.16$\vert$0.48 & 0.75$\vert$0.52 & 0.00$\vert$0.00 & 0.00$\vert$0.00 & 0.00$\vert$0.00 & 0.00$\vert$0.00 \\
		\addlinespace[0.5em] 
		& \multirow{3}{*}{\textbf{$\Delta$}}	& \textbf{OPAM} & 1.0000 & 0.7012 & \cellcolor{blue!30}\textbf{0.4508} & \cellcolor{blue!30}\textbf{0.4009} & \cellcolor{blue!30}\textbf{0.4872} & 0.9433 \\
		&						& \textbf{RS} & 0.9600 & \cellcolor{gray!20}\textbf{0.4764} & 0.6032 & 0.7002 & 0.7328 & 1.0000 \\
		&						& $p\vert\hat{A}_{12}$ & 0.16$\vert$0.52 & 0.00$\vert$0.79 & 0.00$\vert$0.22 & 0.00$\vert$0.03 & 0.00$\vert$0.11 & 0.08$\vert$0.47 \\
\midrule 
\multirow{9}{*}{\rotatebox{90}{
        \parbox{8em}{\centering{$\mathbf{T}^{10}_{r}$ \\ (random, size 10)}}
}}
		& \multirow{3}{*}{\textbf{HV}}	& \textbf{OPAM} & 0.0000 & \cellcolor{blue!30}\textbf{0.8976} & \cellcolor{blue!30}\textbf{0.9792} & \cellcolor{blue!30}\textbf{0.9449} & \cellcolor{blue!30}\textbf{0.9837} & \cellcolor{blue!30}\textbf{0.9999} \\
		&						& \textbf{RS} & 0.0000 & 0.8517 & 0.8191 & 0.6879 & 0.5183 & 0.0000 \\
		&						& $p\vert\hat{A}_{12}$ & 1.00$\vert$0.50 & 0.00$\vert$0.90 & 0.00$\vert$1.00 & 0.00$\vert$1.00 & 0.00$\vert$1.00 & 0.00$\vert$1.00 \\
		\addlinespace[0.5em] 
		& \multirow{3}{*}{\textbf{GD+}}	& \textbf{OPAM} & 0.0000 & \cellcolor{blue!30}\textbf{0.0806} & \cellcolor{blue!30}\textbf{0.0035} & \cellcolor{blue!30}\textbf{0.0043} & \cellcolor{blue!30}\textbf{0.0211} & \cellcolor{blue!30}\textbf{0.0134} \\
		&						& \textbf{RS} & 0.0200 & 0.1252 & 0.0912 & 0.0789 & 0.1580 & 1.0000 \\
		&						& $p\vert\hat{A}_{12}$ & 0.16$\vert$0.48 & 0.00$\vert$0.09 & 0.00$\vert$0.00 & 0.00$\vert$0.00 & 0.00$\vert$0.00 & 0.00$\vert$0.00 \\
		\addlinespace[0.5em] 
		& \multirow{3}{*}{\textbf{$\Delta$}}	& \textbf{OPAM} & 1.0000 & 0.8662 & \cellcolor{blue!30}\textbf{0.4603} & \cellcolor{blue!30}\textbf{0.3951} & \cellcolor{blue!30}\textbf{0.4728} & 0.9433 \\
		&						& \textbf{RS} & 0.9600 & \cellcolor{gray!20}\textbf{0.6579} & 0.6331 & 0.7035 & 0.7617 & 1.0000 \\
		&						& $p\vert\hat{A}_{12}$ & 0.16$\vert$0.52 & 0.00$\vert$0.73 & 0.00$\vert$0.20 & 0.00$\vert$0.02 & 0.00$\vert$0.05 & 0.08$\vert$0.47 \\
\midrule 
\multirow{9}{*}{\rotatebox{90}{
        \parbox{8.5em}{\centering{$\mathbf{T}^{500}_{a}$ \\ (adaptive, size 500)}}\hspace{-0.3em}
}}
		& \multirow{3}{*}{\textbf{HV}}	& \textbf{OPAM} & \cellcolor{blue!30}\textbf{1.0000} & \cellcolor{blue!30}\textbf{0.7032} & \cellcolor{blue!30}\textbf{0.9424} & \cellcolor{blue!30}\textbf{0.9089} & \cellcolor{blue!30}\textbf{0.9803} & \cellcolor{blue!30}\textbf{0.9999} \\
		&						& \textbf{RS} & 0.9000 & 0.6518 & 0.7893 & 0.6561 & 0.5167 & 0.0000 \\
		&						& $p\vert\hat{A}_{12}$ & 0.02$\vert$0.55 & 0.00$\vert$0.86 & 0.00$\vert$1.00 & 0.00$\vert$1.00 & 0.00$\vert$1.00 & 0.00$\vert$1.00 \\
		\addlinespace[0.5em] 
		& \multirow{3}{*}{\textbf{GD+}}	& \textbf{OPAM} & \cellcolor{blue!30}\textbf{0.0000} & \cellcolor{blue!30}\textbf{0.0159} & \cellcolor{blue!30}\textbf{0.0035} & \cellcolor{blue!30}\textbf{0.0051} & \cellcolor{blue!30}\textbf{0.0064} & \cellcolor{blue!30}\textbf{0.0134} \\
		&						& \textbf{RS} & 0.0883 & 0.0393 & 0.0850 & 0.0746 & 0.1422 & 1.0000 \\
		&						& $p\vert\hat{A}_{12}$ & 0.02$\vert$0.45 & 0.00$\vert$0.03 & 0.00$\vert$0.00 & 0.00$\vert$0.00 & 0.00$\vert$0.00 & 0.00$\vert$0.00 \\
		\addlinespace[0.5em] 
		& \multirow{3}{*}{\textbf{$\Delta$}}	& \textbf{OPAM} & 1.0000 & 0.7842 & \cellcolor{blue!30}\textbf{0.4715} & \cellcolor{blue!30}\textbf{0.3680} & \cellcolor{blue!30}\textbf{0.4850} & 0.9433 \\
		&						& \textbf{RS} & 0.9766 & \cellcolor{gray!20}\textbf{0.5354} & 0.6357 & 0.6850 & 0.7565 & 1.0000 \\
		&						& $p\vert\hat{A}_{12}$ & 0.16$\vert$0.52 & 0.00$\vert$0.84 & 0.00$\vert$0.21 & 0.00$\vert$0.01 & 0.00$\vert$0.09 & 0.08$\vert$0.47 \\
\midrule 
\multirow{9}{*}{\rotatebox{90}{
        \parbox{8em}{\centering{$\mathbf{T}^{500}_{w}$ \\ (worst, size 500)}}
}}
		& \multirow{3}{*}{\textbf{HV}}	& \textbf{OPAM} & \cellcolor{blue!30}\textbf{1.0000} & \cellcolor{blue!30}\textbf{0.6535} & \cellcolor{blue!30}\textbf{0.9223} & \cellcolor{blue!30}\textbf{0.9307} & \cellcolor{blue!30}\textbf{0.9635} & \cellcolor{blue!30}\textbf{0.9997} \\
		&						& \textbf{RS} & 0.9000 & 0.6050 & 0.7791 & 0.6770 & 0.5032 & 0.0000 \\
		&						& $p\vert\hat{A}_{12}$ & 0.02$\vert$0.55 & 0.00$\vert$0.77 & 0.00$\vert$1.00 & 0.00$\vert$1.00 & 0.00$\vert$1.00 & 0.00$\vert$1.00 \\
		\addlinespace[0.5em] 
		& \multirow{3}{*}{\textbf{GD+}}	& \textbf{OPAM} & \cellcolor{blue!30}\textbf{0.0000} & \cellcolor{blue!30}\textbf{0.0302} & \cellcolor{blue!30}\textbf{0.0037} & \cellcolor{blue!30}\textbf{0.0040} & \cellcolor{blue!30}\textbf{0.0054} & \cellcolor{blue!30}\textbf{0.0136} \\
		&						& \textbf{RS} & 0.0883 & 0.0545 & 0.0768 & 0.0763 & 0.1408 & 1.0000 \\
		&						& $p\vert\hat{A}_{12}$ & 0.02$\vert$0.45 & 0.00$\vert$0.09 & 0.00$\vert$0.00 & 0.00$\vert$0.00 & 0.00$\vert$0.00 & 0.00$\vert$0.00 \\
		\addlinespace[0.5em] 
		& \multirow{3}{*}{\textbf{$\Delta$}}	& \textbf{OPAM} & 1.0000 & 0.7899 & \cellcolor{blue!30}\textbf{0.4640} & \cellcolor{blue!30}\textbf{0.4077} & \cellcolor{blue!30}\textbf{0.5083} & 0.9433 \\
		&						& \textbf{RS} & 0.9766 & \cellcolor{gray!20}\textbf{0.5910} & 0.6114 & 0.7052 & 0.7448 & 1.0000 \\
		&						& $p\vert\hat{A}_{12}$ & 0.16$\vert$0.52 & 0.00$\vert$0.84 & 0.00$\vert$0.22 & 0.00$\vert$0.02 & 0.00$\vert$0.11 & 0.08$\vert$0.47 \\
\midrule 
\multirow{9}{*}{\rotatebox{90}{
        \parbox{8.1em}{\centering{$\mathbf{T}^{500}_{r}$ \\ (random, size 500)}}
}}
		& \multirow{3}{*}{\textbf{HV}}	& \textbf{OPAM} & \cellcolor{blue!30}\textbf{1.0000} & \cellcolor{blue!30}\textbf{0.6936} & \cellcolor{blue!30}\textbf{0.9742} & \cellcolor{blue!30}\textbf{0.9481} & \cellcolor{blue!30}\textbf{0.9810} & \cellcolor{blue!30}\textbf{0.9999} \\
		&						& \textbf{RS} & 0.9000 & 0.6401 & 0.8138 & 0.6904 & 0.5183 & 0.0000 \\
		&						& $p\vert\hat{A}_{12}$ & 0.02$\vert$0.55 & 0.00$\vert$0.85 & 0.00$\vert$1.00 & 0.00$\vert$1.00 & 0.00$\vert$1.00 & 0.00$\vert$1.00 \\
		\addlinespace[0.5em] 
		& \multirow{3}{*}{\textbf{GD+}}	& \textbf{OPAM} & \cellcolor{blue!30}\textbf{0.0000} & \cellcolor{blue!30}\textbf{0.0169} & \cellcolor{blue!30}\textbf{0.0031} & \cellcolor{blue!30}\textbf{0.0041} & \cellcolor{blue!30}\textbf{0.0062} & \cellcolor{blue!30}\textbf{0.0134} \\
		&						& \textbf{RS} & 0.0883 & 0.0394 & 0.0914 & 0.0794 & 0.1420 & 1.0000 \\
		&						& $p\vert\hat{A}_{12}$ & 0.02$\vert$0.45 & 0.00$\vert$0.03 & 0.00$\vert$0.00 & 0.00$\vert$0.00 & 0.00$\vert$0.00 & 0.00$\vert$0.00 \\
		\addlinespace[0.5em] 
		& \multirow{3}{*}{\textbf{$\Delta$}}	& \textbf{OPAM} & 1.0000 & 0.7415 & \cellcolor{blue!30}\textbf{0.4637} & \cellcolor{blue!30}\textbf{0.4077} & \cellcolor{blue!30}\textbf{0.4854} & 0.9433 \\
		&						& \textbf{RS} & 0.9766 & \cellcolor{gray!20}\textbf{0.5251} & 0.6358 & 0.7042 & 0.7535 & 1.0000 \\
		&						& $p\vert\hat{A}_{12}$ & 0.16$\vert$0.52 & 0.00$\vert$0.80 & 0.00$\vert$0.20 & 0.00$\vert$0.03 & 0.00$\vert$0.09 & 0.08$\vert$0.47 \\
		\bottomrule
		\addlinespace[0.5em]
		\multicolumn{9}{l}{\parbox[t]{0.95\linewidth}{
		\colorbox{blue!30}{\textbf{n.nnnn}}: OPAM outperforms RS \quad\quad \colorbox{gray!20}{\textbf{n.nnnn}}: RS outperforms OPAM}} \\
\end{tabularx}
\end{center}
\label{tbl:rq1QIs}
\end{table}

Given the Pareto priority assignments obtained by OPAM and RS, we further assessed the quality values of the solutions by evaluating them with different sets of task-arrival sequences. As described in Section~\ref{subsec:design}, we created six test sets of task-arrival sequences for each subject by varying the sequence generation methods and the number of task-arrival sequences in a set (see $\mathbf{T}_{a}^{10}$, $\mathbf{T}_{w}^{10}$, $\mathbf{T}_{r}^{10}$, $\mathbf{T}_{a}^{500}$, $\mathbf{T}_{w}^{500}$, and $\mathbf{T}_{r}^{500}$ described in Section~\ref{subsec:design}). Table~\ref{tbl:rq1QIs} reports the average quality values measured by HV, GD+, and $\Delta$ based on 50 runs of OPAM and RS with the different test sets of task-arrival sequences. The results indicate that OPAM significantly outperforms RS in most comparison cases. Specifically, out of a total of 108 comparisons, OPAM outperforms RS 87 times (see the blue-colored cells related to OPAM in Table~\ref{tbl:rq1QIs}). Regarding $\Delta$, RS outperforms OPAM for the CCS subject (see the gray-colored cells related to RS in Table~\ref{tbl:rq1QIs}). As shown in Table~\ref{tbl:subjects}, CCS has only 3 aperiodic tasks and RS was therefore able to find better solutions with respect to $\Delta$ for such a simple subject.

\begin{mdframed}[style=RQFrame]
\emph{The answer to {\bf RQ1} is that} OPAM significantly outperforms RS with respect to HV and GD+. In particular, OPAM performs considerably better than RS when more aperiodic tasks are involved. 
\end{mdframed}

\begin{figure}[!th]
\begin{center}
    \subfloat[ICS]{
        \includegraphics[width=0.45\columnwidth]{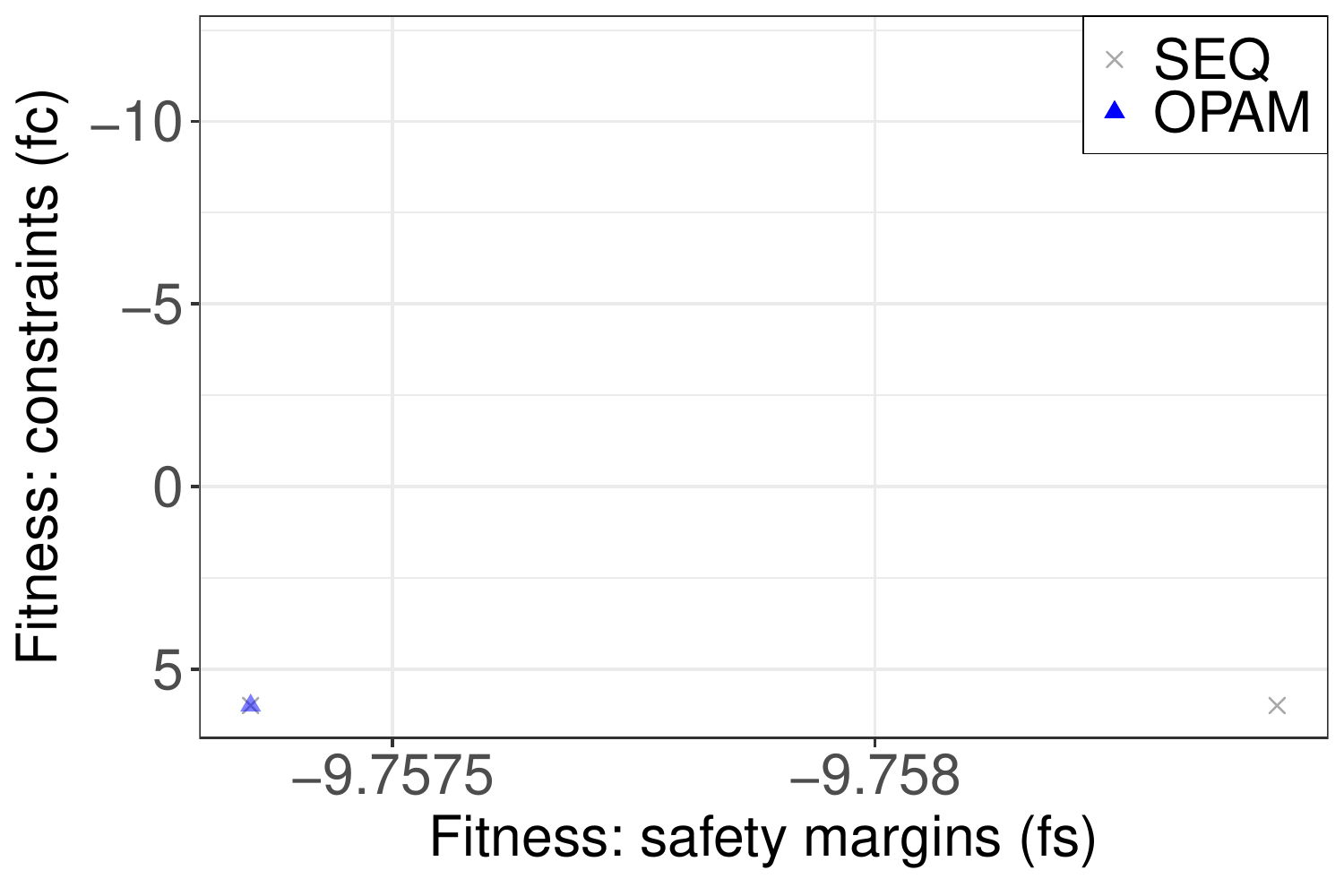}
        \label{fig:rq2scatter ics}
    }
    \hspace*{\fill}
    \subfloat[CCS]{
        \includegraphics[width=.45\columnwidth]{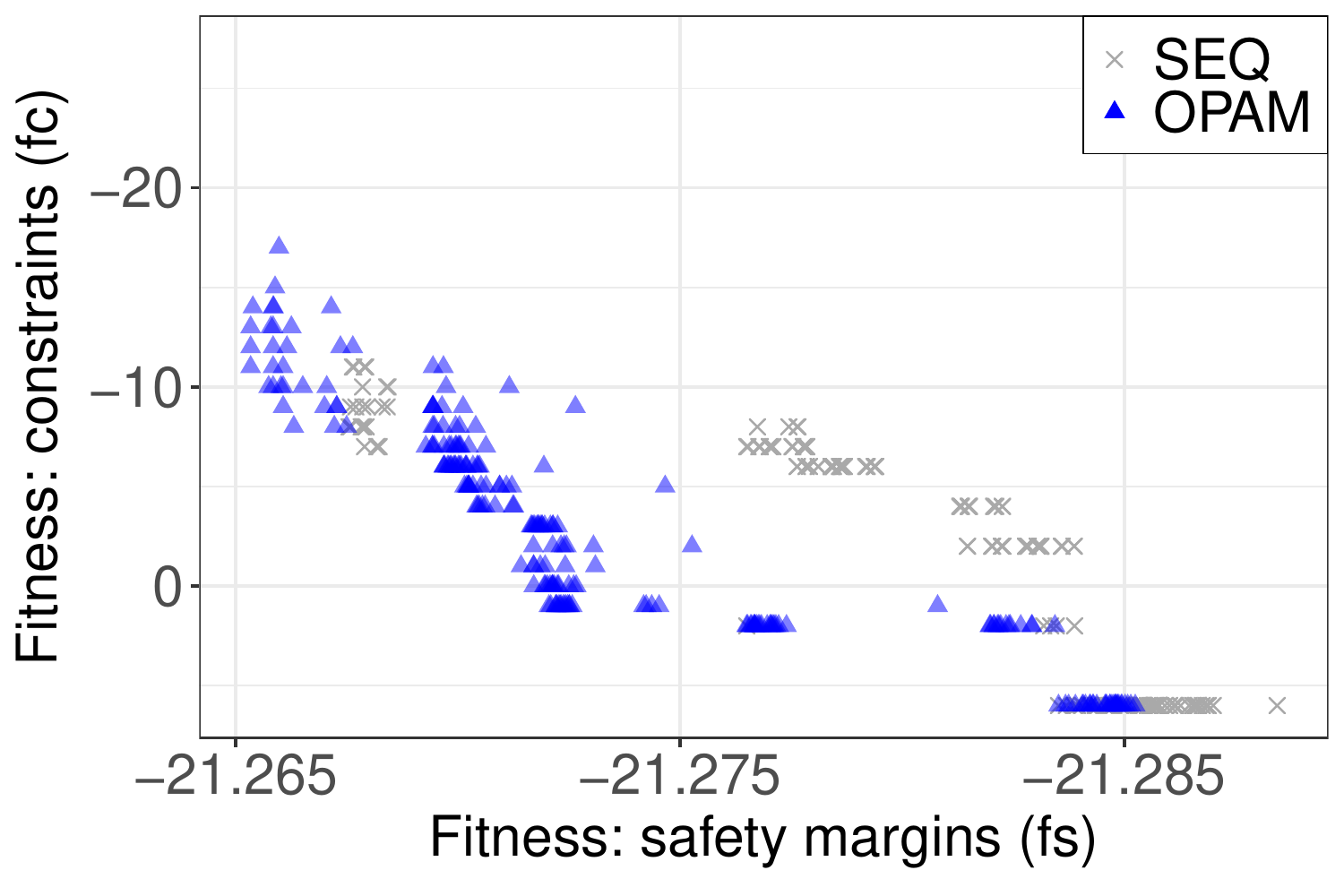}
        \label{fig:rq2scatter ccs}
    }
    \hfill
    \subfloat[UAV]{
    \includegraphics[width=.45\columnwidth]{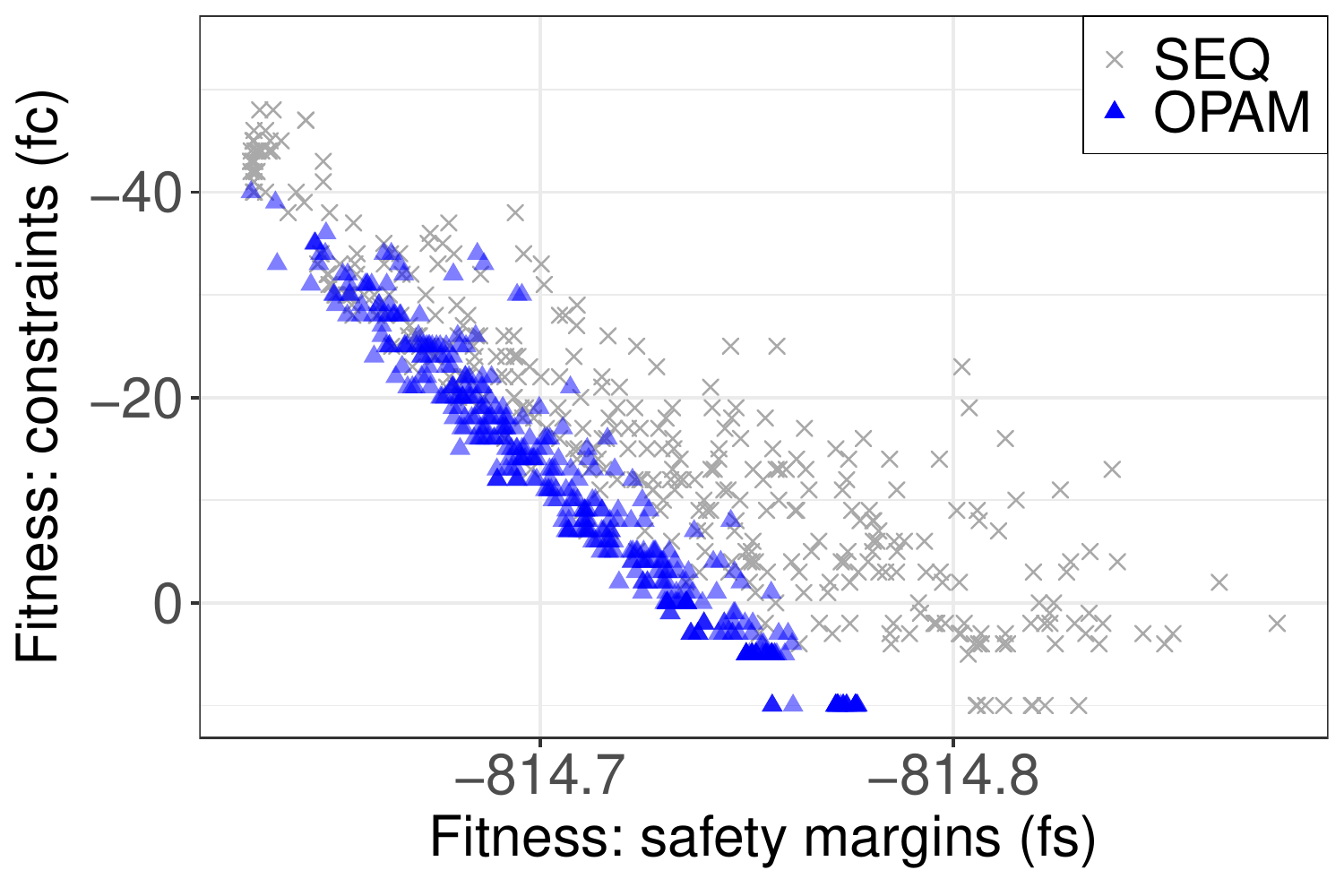}
    \label{fig:rq2scatter uav}
    }
    \hspace*{\fill}
    \subfloat[GAP]{
        \includegraphics[width=.45\columnwidth]{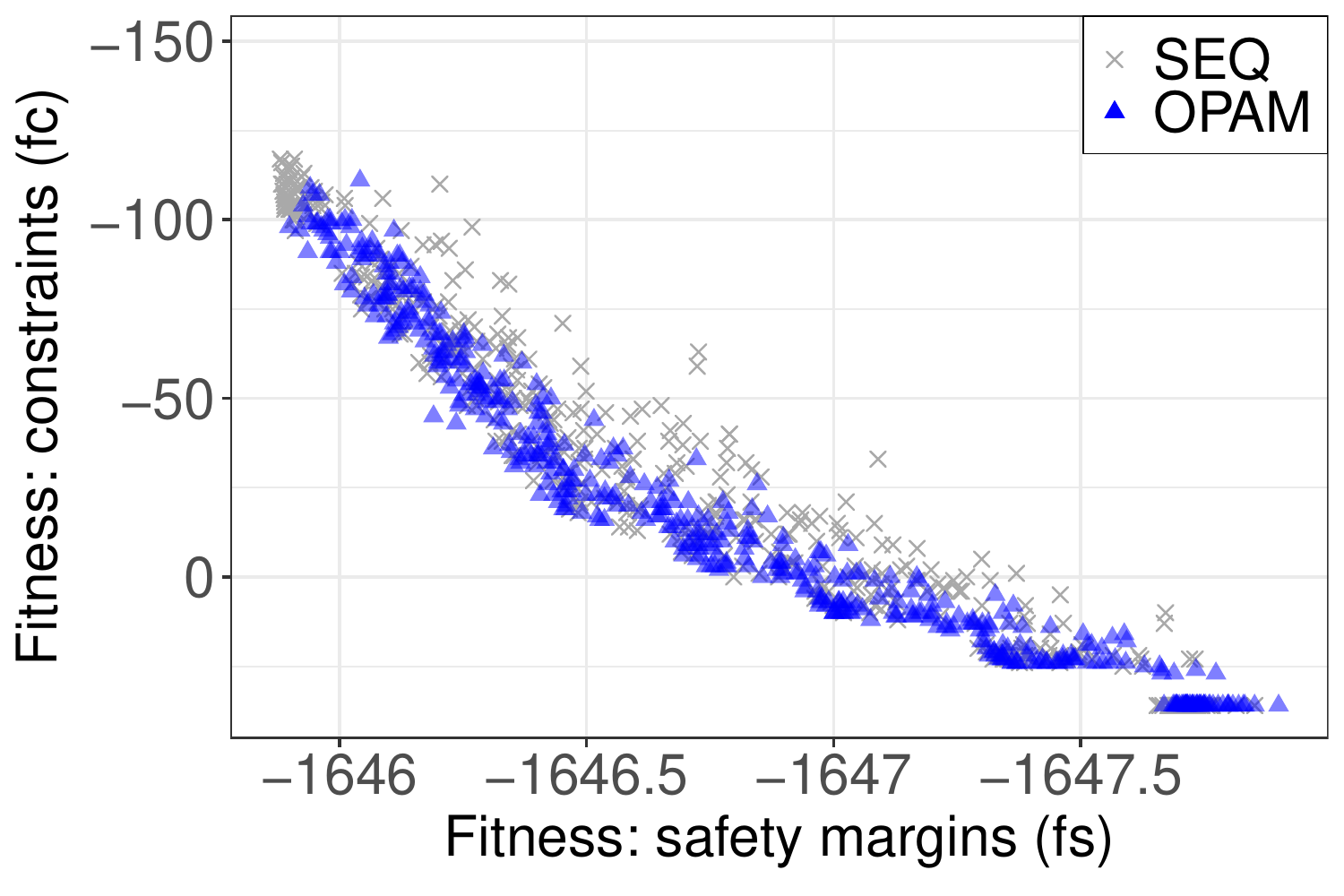}
        \label{fig:rq2scatter gap}
    }
    \hfill
    \subfloat[HPSS]{
        \includegraphics[width=.45\columnwidth]{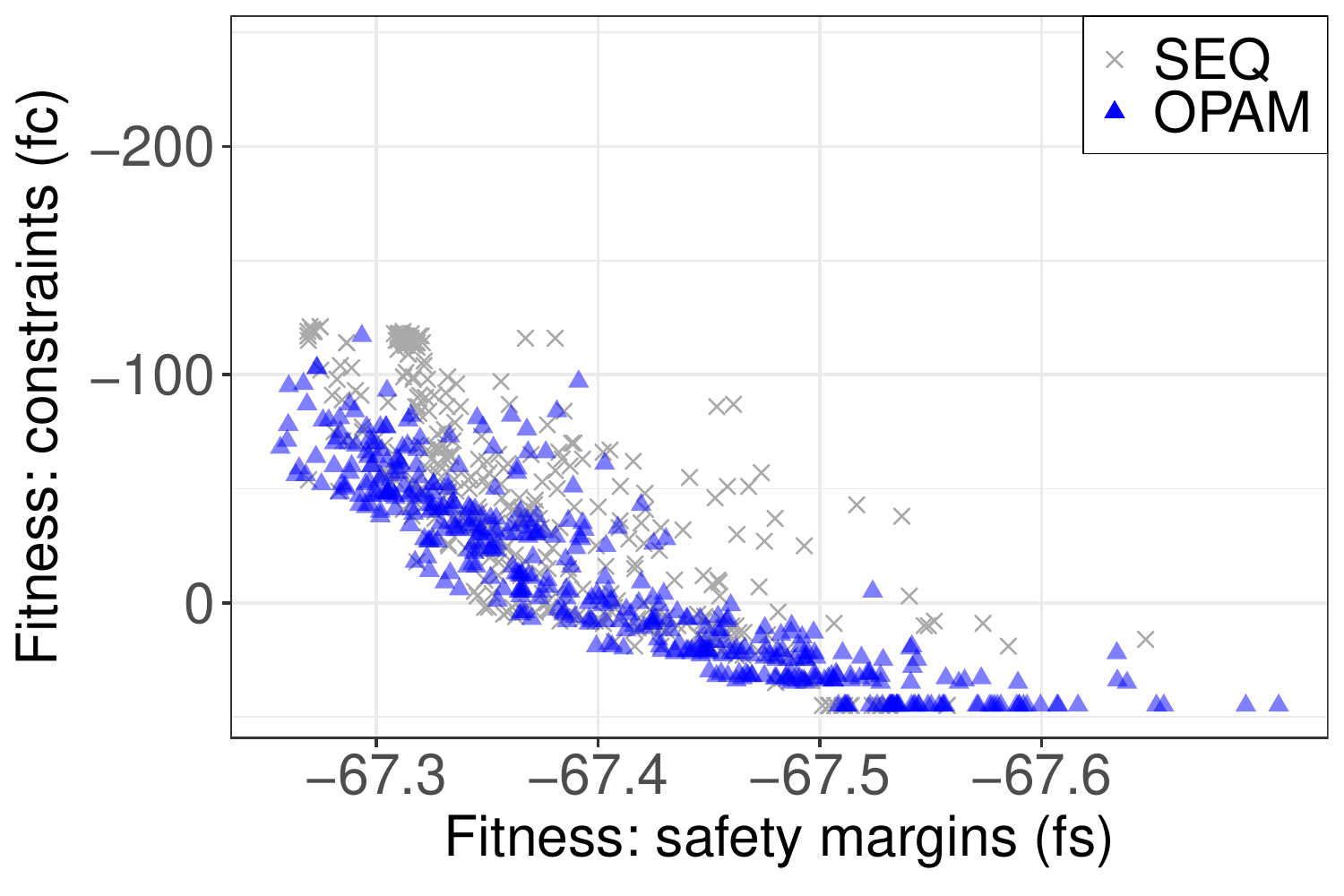}
        \label{fig:rq2scatter hpss}
    }
    \hspace*{\fill}
    \subfloat[ESAIL]{
        \includegraphics[width=0.45\columnwidth]{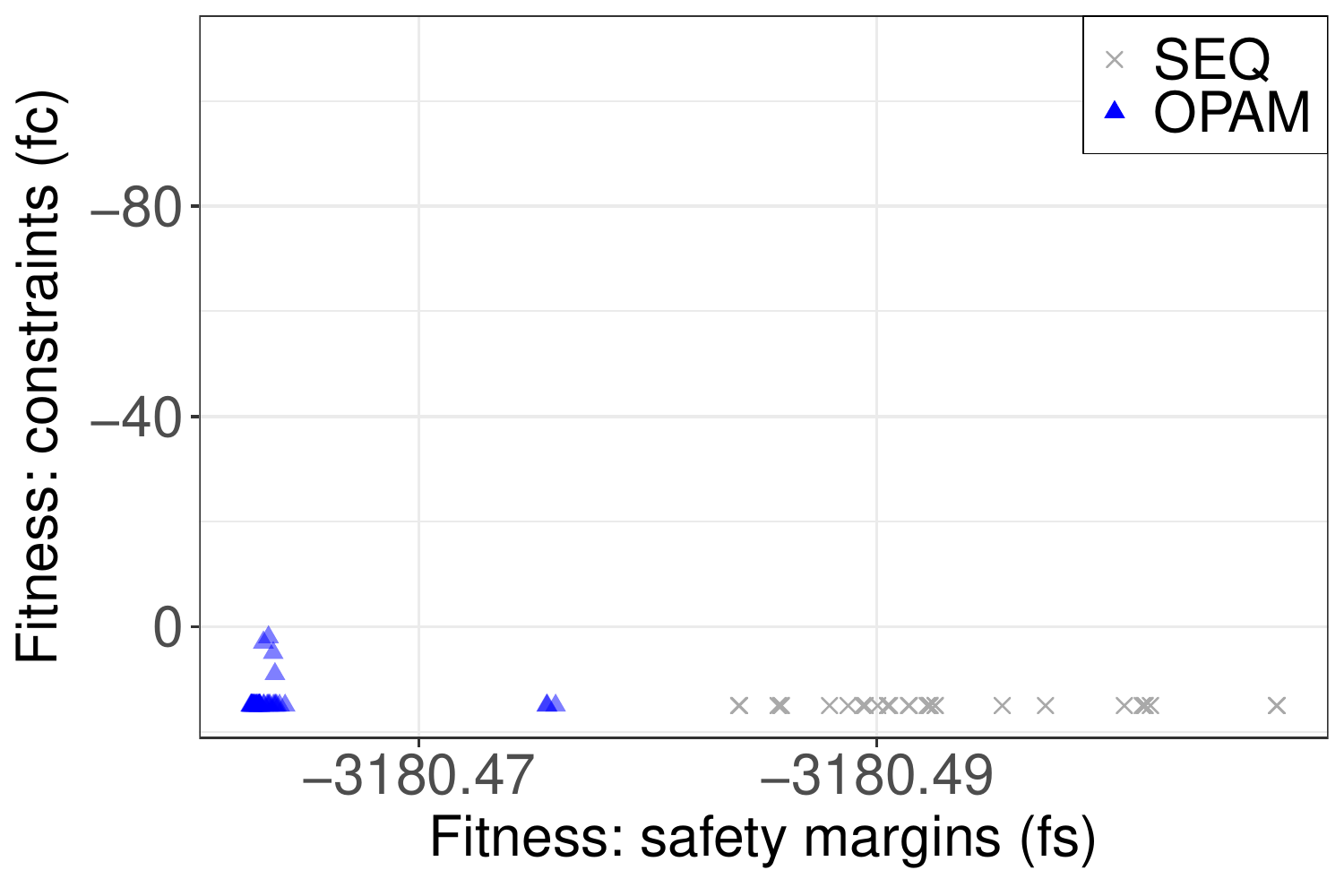}
        \label{fig:rq2scatter esail}
    }
\end{center}
\caption{Pareto fronts obtained by OPAM and SEQ for the six industrial subjects: (a)~ICS, (b)~CCS, (c)~UAV, (d)~GAP, (e)~HPSS, and (f)~ESAIL. The fitness values are computed based on each subject's set $\mathbf{T}_{a}^{500}$ of task-arrival sequences (see Section~\ref{subsec:design}). The points located closer to the bottom left of each plot are considered to be better priority assignments when compared to points closer to the top right.}
\label{fig:rq2scatter}
\end{figure}

\noindent\textbf{RQ2.}
To compare OPAM and SEQ, we first visually inspect the best Pareto fronts obtained from 50 runs of OPAM and SEQ for the six study systems described in Section~\ref{subsec:industrial subjects} by varying the test sets of task-arrival sequences for each subject (see $\mathbf{T}_{a}^{10}$, $\mathbf{T}_{w}^{10}$, $\mathbf{T}_{r}^{10}$, $\mathbf{T}_{a}^{500}$, $\mathbf{T}_{w}^{500}$, and $\mathbf{T}_{r}^{500}$ described in Section~\ref{subsec:design}), which are created independently from OPAM and SEQ. Overall, we observed that OPAM finds significantly better priority assignments in most cases. For example, Figure~\ref{fig:rq2scatter} depicts the best Pareto fronts obtained by OPAM and SEQ when the fitness values are computed based on each subject's test set $\mathbf{T}_{a}^{500}$ of 500 task-arrival sequences, which are generated with adaptive random search. The results clearly show that OPAM outperforms SEQ with respect to producing more optimal Pareto fronts for ICS, CCS, UAV, HPSS, and ESAIL. For GAP, the visual inspection is not sufficient to provide any conclusions. Hence, we further compare OPAM and SEQ based on the quality indicators described in Section~\ref{subsec:metrics}.

\begin{table}[p]
	\caption{Comparing OPAM and SEQ using the three quality indicators: HV, GD+, and $\Delta$. Average quality values computed based on 50 runs of OPAM and SEQ using the different sets of task-arrival sequences (see Section~\ref{subsec:design}).}
	\vspace{-1.2em}
	\fontsize{8}{8}\selectfont
	\def\arraystretch{0.5}
\begin{center}
\begin{tabularx}{\columnwidth}{m{2.5em}@{}c@{\hspace{0.3em}}r@{\hspace{1em}} RRRRRR}
		\toprule
		\addlinespace[0.5em]
		\multicolumn{3}{c}{} & \multicolumn{1}{c}{\textbf{ICS}} & \multicolumn{1}{c}{\textbf{CCS}}
		& \multicolumn{1}{c}{\textbf{UAV}} & \multicolumn{1}{c}{\textbf{GAP}} & \multicolumn{1}{c}{\textbf{HPSS}} & \multicolumn{1}{c}{\textbf{ESAIL}} \\
		\addlinespace[0.2em]
\midrule 
\multirow{15}{*}{\rotatebox{90}{
        \parbox{8em}{\centering{$\mathbf{T}^{10}_{a}$ \\ (adaptive, size 10)}}
}}
		& \multirow{3}{*}{\textbf{HV}}	& \textbf{OPAM} & 0.0000 & \cellcolor{blue!30}\textbf{0.6052} & \cellcolor{blue!30}\textbf{0.6011} & \cellcolor{blue!30}\textbf{0.6088} & \cellcolor{blue!30}\textbf{0.6290} & \cellcolor{blue!30}\textbf{0.9808} \\
		&						& \textbf{SEQ} & 0.0000 & 0.4172 & 0.5354 & 0.5868 & 0.6086 & 0.4470 \\
		&						& $p\vert\hat{A}_{12}$ & 1.00$\vert$0.50 & 0.00$\vert$1.00 & 0.00$\vert$0.95 & 0.00$\vert$0.76 & 0.02$\vert$0.63 & 0.00$\vert$1.00 \\
		\addlinespace[0.5em]
		& \multirow{3}{*}{\textbf{GD+}}	& \textbf{OPAM} & \cellcolor{blue!30}\textbf{0.0000} & \cellcolor{blue!30}\textbf{0.0244} & \cellcolor{blue!30}\textbf{0.0175} & \cellcolor{blue!30}\textbf{0.0148} & \cellcolor{blue!30}\textbf{0.0529} & \cellcolor{blue!30}\textbf{0.0249} \\
		&						& \textbf{SEQ} & 0.2191 & 0.0835 & 0.0350 & 0.0201 & 0.0625 & 0.1887 \\
		&						& $p\vert\hat{A}_{12}$ & 0.00$\vert$0.01 & 0.00$\vert$0.00 & 0.00$\vert$0.01 & 0.00$\vert$0.25 & 0.00$\vert$0.26 & 0.00$\vert$0.03 \\
		\addlinespace[0.5em]
		& \multirow{3}{*}{\textbf{$\Delta$}}	& \textbf{OPAM} & 1.0000 & 0.7653 & 0.4239 & 0.3343 & 0.5297 & 0.9444 \\
		&						& \textbf{SEQ} & \cellcolor{gray!20}\textbf{0.0200} & \cellcolor{gray!20}\textbf{0.5656} & \cellcolor{gray!20}\textbf{0.3628} & \cellcolor{gray!20}\textbf{0.2875} & 0.5706 & \cellcolor{gray!20}\textbf{0.8285} \\
		&						& $p\vert\hat{A}_{12}$ & 0.00$\vert$0.99 & 0.00$\vert$0.81 & 0.01$\vert$0.64 & 0.01$\vert$0.65 & 0.33$\vert$0.44 & 0.00$\vert$0.75 \\
\midrule 
\multirow{15}{*}{\rotatebox{90}{
        \parbox{8em}{\centering{$\mathbf{T}^{10}_{w}$ \\ (worst, size 10)}}
}}
		& \multirow{3}{*}{\textbf{HV}}	& \textbf{OPAM} & 0.0000 & \cellcolor{blue!30}\textbf{0.7345} & \cellcolor{blue!30}\textbf{0.6258} & \cellcolor{blue!30}\textbf{0.6290} & \cellcolor{blue!30}\textbf{0.7460} & \cellcolor{blue!30}\textbf{0.9059} \\
		&						& \textbf{SEQ} & 0.0000 & 0.6794 & 0.5933 & 0.5928 & 0.6856 & 0.5046 \\
		&						& $p\vert\hat{A}_{12}$ & 1.00$\vert$0.50 & 0.00$\vert$0.82 & 0.00$\vert$0.82 & 0.00$\vert$0.88 & 0.00$\vert$0.87 & 0.00$\vert$1.00 \\
		\addlinespace[0.5em]
		& \multirow{3}{*}{\textbf{GD+}}	& \textbf{OPAM} & 0.0000 & 0.0912 & \cellcolor{blue!30}\textbf{0.0191} & \cellcolor{blue!30}\textbf{0.0131} & \cellcolor{blue!30}\textbf{0.0340} & \cellcolor{blue!30}\textbf{0.0724} \\
		&						& \textbf{SEQ} & 0.0000 & \cellcolor{gray!20}\textbf{0.0695} & 0.0272 & 0.0211 & 0.0667 & 0.1720 \\
		&						& $p\vert\hat{A}_{12}$ & 1.00$\vert$0.50 & 0.00$\vert$0.86 & 0.00$\vert$0.12 & 0.00$\vert$0.14 & 0.00$\vert$0.03 & 0.00$\vert$0.07 \\
		\addlinespace[0.5em]
		& \multirow{3}{*}{\textbf{$\Delta$}}	& \textbf{OPAM} & 1.0000 & 0.7009 & 0.4835 & 0.3616 & \cellcolor{blue!30}\textbf{0.4695} & 0.9470 \\
		&						& \textbf{SEQ} & 1.0000 & \cellcolor{gray!20}\textbf{0.5376} & \cellcolor{gray!20}\textbf{0.3111} & \cellcolor{gray!20}\textbf{0.3054} & 0.5453 & \cellcolor{gray!20}\textbf{0.7547} \\
		&						& $p\vert\hat{A}_{12}$ & 1.00$\vert$0.50 & 0.00$\vert$0.74 & 0.00$\vert$0.83 & 0.01$\vert$0.66 & 0.01$\vert$0.35 & 0.00$\vert$0.67 \\
\midrule 
\multirow{15}{*}{\rotatebox{90}{
        \parbox{8em}{\centering{$\mathbf{T}^{10}_{r}$ \\ (random, size 10)}}
}}
		& \multirow{3}{*}{\textbf{HV}}	& \textbf{OPAM} & 0.0000 & \cellcolor{blue!30}\textbf{0.8720} & \cellcolor{blue!30}\textbf{0.8653} & \cellcolor{blue!30}\textbf{0.6340} & 0.7714 & \cellcolor{blue!30}\textbf{0.9055} \\
		&						& \textbf{SEQ} & 0.0000 & 0.5478 & 0.7246 & 0.5879 & 0.7935 & 0.1139 \\
		&						& $p\vert\hat{A}_{12}$ & 1.00$\vert$0.50 & 0.00$\vert$0.99 & 0.00$\vert$1.00 & 0.00$\vert$0.92 & 0.06$\vert$0.39 & 0.00$\vert$1.00 \\
		\addlinespace[0.5em]
		& \multirow{3}{*}{\textbf{GD+}}	& \textbf{OPAM} & 0.0000 & \cellcolor{blue!30}\textbf{0.0911} & \cellcolor{blue!30}\textbf{0.0205} & \cellcolor{blue!30}\textbf{0.0160} & \cellcolor{blue!30}\textbf{0.0472} & \cellcolor{blue!30}\textbf{0.0718} \\
		&						& \textbf{SEQ} & 0.0000 & 0.1358 & 0.0882 & 0.0277 & 0.0646 & 0.2838 \\
		&						& $p\vert\hat{A}_{12}$ & 1.00$\vert$0.50 & 0.00$\vert$0.01 & 0.00$\vert$0.00 & 0.00$\vert$0.10 & 0.00$\vert$0.19 & 0.00$\vert$0.06 \\
		\addlinespace[0.5em]
		& \multirow{3}{*}{\textbf{$\Delta$}}	& \textbf{OPAM} & 1.0000 & 0.8605 & 0.4644 & 0.3825 & 0.4658 & 0.9456 \\
		&						& \textbf{SEQ} & 1.0000 & \cellcolor{gray!20}\textbf{0.5896} & \cellcolor{gray!20}\textbf{0.4072} & \cellcolor{gray!20}\textbf{0.3253} & 0.4620 & \cellcolor{gray!20}\textbf{0.9670} \\
		&						& $p\vert\hat{A}_{12}$ & 1.00$\vert$0.50 & 0.00$\vert$0.82 & 0.02$\vert$0.64 & 0.01$\vert$0.66 & 0.90$\vert$0.49 & 0.00$\vert$0.67 \\
\midrule 
\multirow{15}{*}{\rotatebox{90}{
        \parbox{8.5em}{\centering{$\mathbf{T}^{500}_{a}$ \\ (adaptive, size 500)}}\hspace{-0.3em}
}}
		& \multirow{3}{*}{\textbf{HV}}	& \textbf{OPAM} & 0.0000 & \cellcolor{blue!30}\textbf{0.6781} & \cellcolor{blue!30}\textbf{0.7134} & \cellcolor{blue!30}\textbf{0.6261} & \cellcolor{blue!30}\textbf{0.7332} & \cellcolor{blue!30}\textbf{0.9744} \\
		&						& \textbf{SEQ} & 0.0000 & 0.4854 & 0.6179 & 0.5981 & 0.7056 & 0.3571 \\
		&						& $p\vert\hat{A}_{12}$ & 1.00$\vert$0.50 & 0.00$\vert$1.00 & 0.00$\vert$1.00 & 0.00$\vert$0.83 & 0.00$\vert$0.73 & 0.00$\vert$1.00 \\
		\addlinespace[0.5em]
		& \multirow{3}{*}{\textbf{GD+}}	& \textbf{OPAM} & \cellcolor{blue!30}\textbf{0.0000} & \cellcolor{blue!30}\textbf{0.0174} & \cellcolor{blue!30}\textbf{0.0140} & \cellcolor{blue!30}\textbf{0.0134} & \cellcolor{blue!30}\textbf{0.0320} & \cellcolor{blue!30}\textbf{0.0285} \\
		&						& \textbf{SEQ} & 0.2191 & 0.0727 & 0.0549 & 0.0197 & 0.0565 & 0.2153 \\
		&						& $p\vert\hat{A}_{12}$ & 0.00$\vert$0.01 & 0.00$\vert$0.00 & 0.00$\vert$0.00 & 0.00$\vert$0.20 & 0.00$\vert$0.08 & 0.00$\vert$0.04 \\
		\addlinespace[0.5em]
		& \multirow{3}{*}{\textbf{$\Delta$}}	& \textbf{OPAM} & 1.0000 & 0.7833 & 0.4964 & 0.3588 & \cellcolor{blue!30}\textbf{0.4564} & 0.9442 \\
		&						& \textbf{SEQ} & \cellcolor{gray!20}\textbf{0.0200} & 0.7319 & \cellcolor{gray!20}\textbf{0.4002} & 0.3315 & 0.5312 & \cellcolor{gray!20}\textbf{0.8554} \\
		&						& $p\vert\hat{A}_{12}$ & 0.00$\vert$0.99 & 0.23$\vert$0.57 & 0.00$\vert$0.72 & 0.07$\vert$0.60 & 0.02$\vert$0.36 & 0.00$\vert$0.75 \\
\midrule 
\multirow{8}{*}{\rotatebox{90}{
        \parbox{8em}{\centering{$\mathbf{T}^{500}_{w}$ \\ (worst, size 500)}}
}}
		& \multirow{3}{*}{\textbf{HV}}	& \textbf{OPAM} & 0.0000 & 0.4732 & \cellcolor{blue!30}\textbf{0.6330} & \cellcolor{blue!30}\textbf{0.6181} & \cellcolor{blue!30}\textbf{0.6990} & \cellcolor{blue!30}\textbf{0.8755} \\
		&						& \textbf{SEQ} & 0.0000 & \cellcolor{gray!20}\textbf{0.5564} & 0.5958 & 0.5792 & 0.6800 & 0.1183 \\
		&						& $p\vert\hat{A}_{12}$ & 1.00$\vert$0.50 & 0.00$\vert$0.04 & 0.00$\vert$0.85 & 0.00$\vert$0.90 & 0.00$\vert$0.70 & 0.00$\vert$1.00 \\
		\addlinespace[0.5em]
		& \multirow{3}{*}{\textbf{GD+}}	& \textbf{OPAM} & \cellcolor{blue!30}\textbf{0.0000} & 0.0511 & \cellcolor{blue!30}\textbf{0.0141} & \cellcolor{blue!30}\textbf{0.0135} & \cellcolor{blue!30}\textbf{0.0258} & \cellcolor{blue!30}\textbf{0.0911} \\
		&						& \textbf{SEQ} & 0.2191 & \cellcolor{gray!20}\textbf{0.0343} & 0.0267 & 0.0226 & 0.0336 & 0.2849 \\
		&						& $p\vert\hat{A}_{12}$ & 0.00$\vert$0.01 & 0.00$\vert$0.96 & 0.00$\vert$0.05 & 0.00$\vert$0.11 & 0.00$\vert$0.24 & 0.00$\vert$0.06 \\
		\addlinespace[0.5em]
		& \multirow{3}{*}{\textbf{$\Delta$}}	& \textbf{OPAM} & 1.0000 & 0.7569 & 0.4950 & 0.3751 & 0.5379 & 0.9469 \\
		&						& \textbf{SEQ} & \cellcolor{gray!20}\textbf{0.0200} & 0.7259 & \cellcolor{gray!20}\textbf{0.3315} & \cellcolor{gray!20}\textbf{0.3139} & 0.5102 & \cellcolor{gray!20}\textbf{0.8957} \\
		&						& $p\vert\hat{A}_{12}$ & 0.00$\vert$0.99 & 0.43$\vert$0.55 & 0.00$\vert$0.82 & 0.01$\vert$0.66 & 0.20$\vert$0.57 & 0.00$\vert$0.67 \\
\midrule 
\multirow{8}{*}{\rotatebox{90}{
        \parbox{8.1em}{\centering{$\mathbf{T}^{500}_{r}$ \\ (random, size 500)}}
}}
		& \multirow{3}{*}{\textbf{HV}}	& \textbf{OPAM} & 0.0000 & \cellcolor{blue!30}\textbf{0.6646} & \cellcolor{blue!30}\textbf{0.8446} & \cellcolor{blue!30}\textbf{0.6321} & \cellcolor{blue!30}\textbf{0.7087} & \cellcolor{blue!30}\textbf{0.8782} \\
		&						& \textbf{SEQ} & 0.0000 & 0.4876 & 0.7242 & 0.5839 & 0.6786 & 0.1965 \\
		&						& $p\vert\hat{A}_{12}$ & 1.00$\vert$0.50 & 0.00$\vert$1.00 & 0.00$\vert$1.00 & 0.00$\vert$0.93 & 0.00$\vert$0.72 & 0.00$\vert$1.00 \\
		\addlinespace[0.5em]
		& \multirow{3}{*}{\textbf{GD+}}	& \textbf{OPAM} & \cellcolor{blue!30}\textbf{0.0000} & \cellcolor{blue!30}\textbf{0.0184} & \cellcolor{blue!30}\textbf{0.0172} & \cellcolor{blue!30}\textbf{0.0165} & \cellcolor{blue!30}\textbf{0.0327} & \cellcolor{blue!30}\textbf{0.0900} \\
		&						& \textbf{SEQ} & 0.2191 & 0.0684 & 0.0791 & 0.0285 & 0.0580 & 0.2620 \\
		&						& $p\vert\hat{A}_{12}$ & 0.00$\vert$0.01 & 0.00$\vert$0.00 & 0.00$\vert$0.00 & 0.00$\vert$0.09 & 0.00$\vert$0.06 & 0.00$\vert$0.06 \\
		\addlinespace[0.5em]
		& \multirow{3}{*}{\textbf{$\Delta$}}	& \textbf{OPAM} & 1.0000 & 0.7449 & 0.5059 & 0.3960 & \cellcolor{blue!30}\textbf{0.4502} & 0.9472 \\
		&						& \textbf{SEQ} & \cellcolor{gray!20}\textbf{0.0200} & 0.6798 & \cellcolor{gray!20}\textbf{0.4156} & \cellcolor{gray!20}\textbf{0.3341} & 0.5148 & \cellcolor{gray!20}\textbf{0.8546} \\
		&						& $p\vert\hat{A}_{12}$ & 0.00$\vert$0.99 & 0.19$\vert$0.58 & 0.00$\vert$0.71 & 0.01$\vert$0.66 & 0.03$\vert$0.38 & 0.00$\vert$0.67 \\
		\bottomrule
		\addlinespace[0.5em]
		\multicolumn{9}{l}{\parbox[t]{0.95\linewidth}{
		\colorbox{blue!30}{\textbf{n.nnnn}}: OPAM outperforms SEQ \quad\quad \colorbox{gray!20}{\textbf{n.nnnn}}: SEQ outperforms OPAM}} \\
\end{tabularx}
\end{center}
\label{tbl:rq2QI-SEQ}
\end{table}

Table~\ref{tbl:rq2QI-SEQ} compares the quality values measured by HV, GD+, and $\Delta$ for the six study subjects. To fairly compare the priority assignments obtained by OPAM and SEQ, we assess them with the test sets of task-arrival sequences for each subject (see $\mathbf{T}_{a}^{10}$, $\mathbf{T}_{w}^{10}$, $\mathbf{T}_{r}^{10}$, $\mathbf{T}_{a}^{500}$, $\mathbf{T}_{w}^{500}$, and $\mathbf{T}_{r}^{500}$ described in Section~\ref{subsec:design}). Table~\ref{tbl:rq2QI-SEQ} reports the average quality values computed based on 50 runs of OPAM and SEQ. In Table~\ref{tbl:rq2QI-SEQ}, the statistical comparison of the two corresponding distributions are reported using p-values and $\hat{A}_{12}$ values.

As shown in Table~\ref{tbl:rq2QI-SEQ}, we compared OPAM and SEQ 108 times by varying the study subjects, the quality indicators, the number of task-arrival sequences, and the task-arrival sequence generation methods. Out of 108 comparisons, OPAM significantly outperforms SEQ 63 times. Specifically, out of 36 HV comparisons, OPAM obtains better HV values than SEQ 28 times. For ICS (6 HV comparisons), the differences in HV values between OPAM and SEQ are not statistically significant. In only one HV comparison for CCS, SEQ outperforms OPAM (see the gray-colored cell related to HV and CCS in Table~\ref{tbl:rq2QI-SEQ}). To interpret these results, one must recall from Table~\ref{tbl:subjects} that ICS and CCS have only three aperiodic tasks that impact the degree of uncertainty in task arrivals and  therefore represent simple cases. Out of 36 GD+ comparisons, OPAM outperforms SEQ 32 times. SEQ outperforms OPAM only two times for CCS. Hence, overall, the results indicate that OPAM outperforms SEQ, in terms of generating more optimal Pareto fronts, when the subjects feature a considerable degree of uncertainty in task arrivals and therefore make our search problem more complex. Otherwise differences are not statistically or practically significant.  Regarding $\Delta$, which focuses on the diversity of solutions on the Pareto front, SEQ outperforms OPAM 24 times out of 36 comparisons (see the gray-colored cells related to $\Delta$ in Table~\ref{tbl:rq2QI-SEQ}). However, since OPAM produces enough alternative priority assignments spreading across Pareto fronts (as visible from the solutions obtained by OPAM in Figure~\ref{fig:rq2scatter}), these differences in $\Delta$ have limited implications in practice.

\begin{mdframed}[style=RQFrame]
\emph{The answer to {\bf RQ2} is that} OPAM significantly outperforms SEQ with respect to HV and GD+ when in the presence of more than a few aperiodic tasks and therefore higher uncertainty in terms of task arrivals. OPAM therefore generate solutions on a Pareto front that is closer to the unknown, optimal one. In other words,  coevolution is a suitable and successful strategy for finding better priority assignments in complex systems.
\end{mdframed}

\begin{table}[t]
\caption{Execution times and memory usage required to run OPAM for the six industrial subjects. Average values computed based on 50 runs of OPAM are reported.}
\begin{center}
\begin{tabular}{lrrrrr}
    \toprule
    \multicolumn{1}{c}{Subject} & \multicolumn{1}{c}{Execution time (s)} & \multicolumn{1}{c}{Memory usage (MB)} \\
    \midrule
    ICS & 104.34 & 89.97 \\
    CCS & 165.50 & 111.85 \\
    UAV & 1455.35 & 312.85 \\
    GAP & 2819.03 & 730.29 \\
    HPSS & 226.98 & 127.77 \\
    ESAIL & 55844.23 & 2879.79 \\
    \bottomrule
\end{tabular}
\end{center}
\label{tab:rq3industrial}
\end{table}

\noindent\textbf{RQ3.}
Table~\ref{tab:rq3industrial} reports the average execution times and memory usage required to run OPAM for the six industrial subjects, over 50 runs. As shown in Table~\ref{tab:rq3industrial}, finding optimal priority assignments for ESAIL requires the largest execution time ($\approx$15.5h) and memory usage ($\approx$2.9GB), compared to the other subjects. We note that such execution time and memory usage are acceptable as OPAM can be executed offline in practice. 

\begin{figure}[t]
\begin{center}
    \subfloat[Number of tasks ($n$)]{
        \includegraphics[width=0.45\columnwidth]{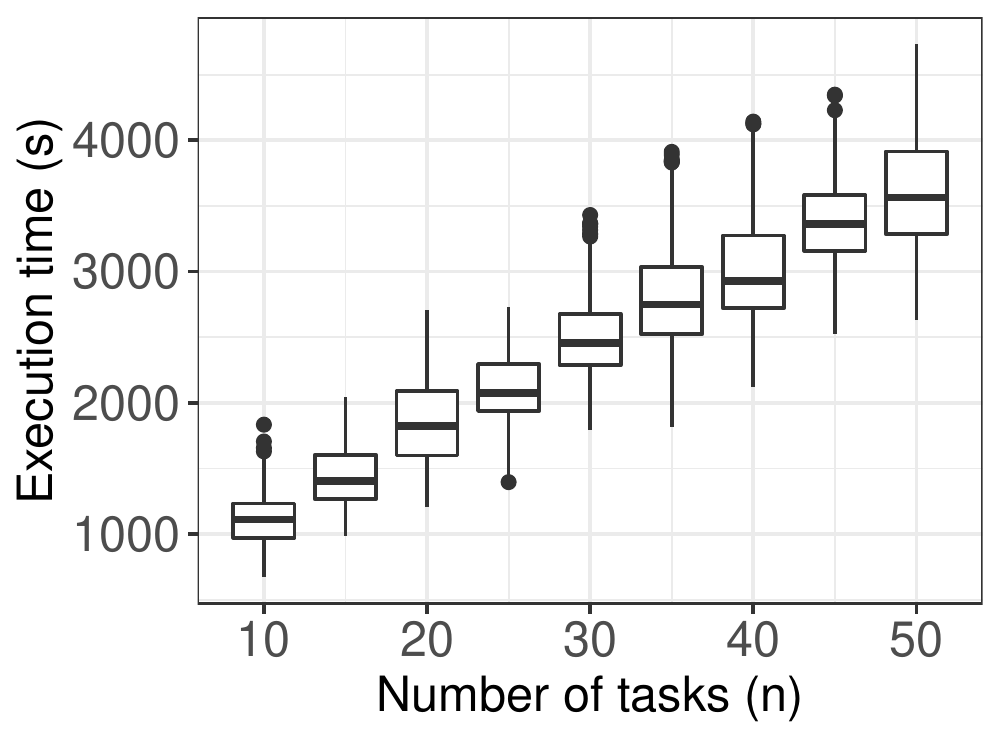}
        \label{fig:rq3time exp1}
    }
    \hspace*{\fill}
    \subfloat[Ratio of aperiodic tasks($\gamma$)]{
        \includegraphics[width=.45\columnwidth]{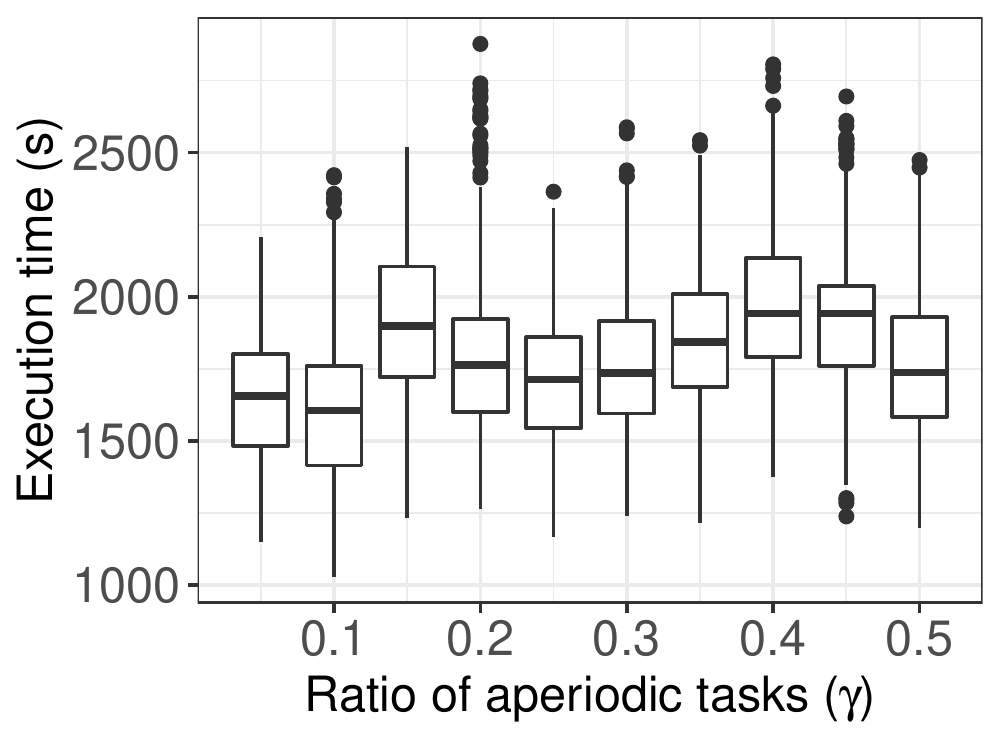}
        \label{fig:rq3time exp2}
    }
    \hfill
    \subfloat[Range factor ($\mu$)]{
        \includegraphics[width=.45\columnwidth]{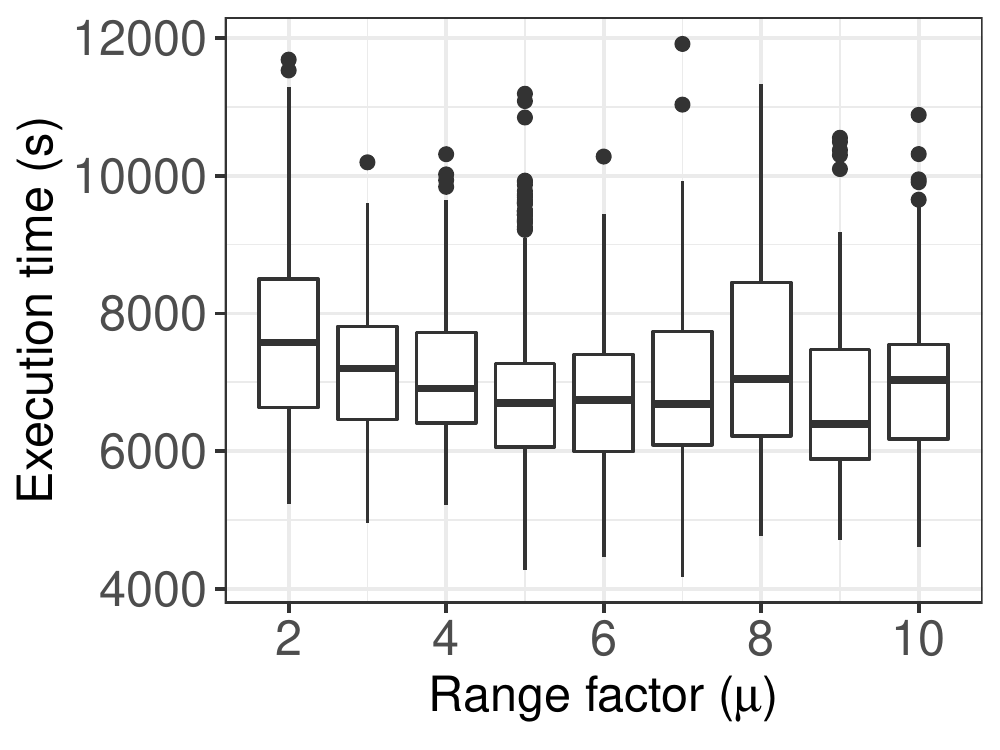}
        \label{fig:rq3time exp3}
    }
    \hspace*{\fill}
    \subfloat[Simulation time ($T$)]{
        \includegraphics[width=.45\columnwidth]{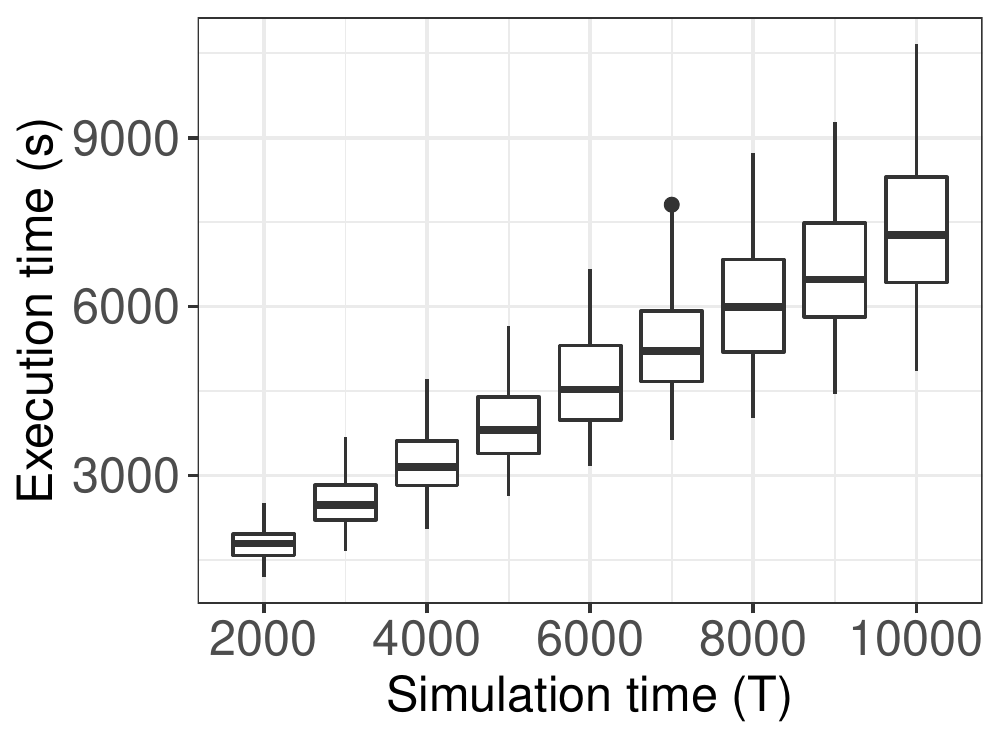}
        \label{fig:rq3time exp4}
    }
\end{center}
\caption{Execution times of OPAM when varying the values of the following parameters: (a)~number of tasks $n$, (b)~ratio of aperiodic tasks $\gamma$, (c)~range factor $\mu$, and (d)~simulation time $T$. The boxplots (25\%-50\%-75\%) show the execution times obtained from 500 runs of OPAM, i.e., 50 runs for each of the 10 synthetic subjects with the same configuration.}
\label{fig:rq3time}
\end{figure}

\begin{figure}[htp]
\begin{center}
    \subfloat[Number of tasks ($n$)]{
        \includegraphics[width=0.45\columnwidth]{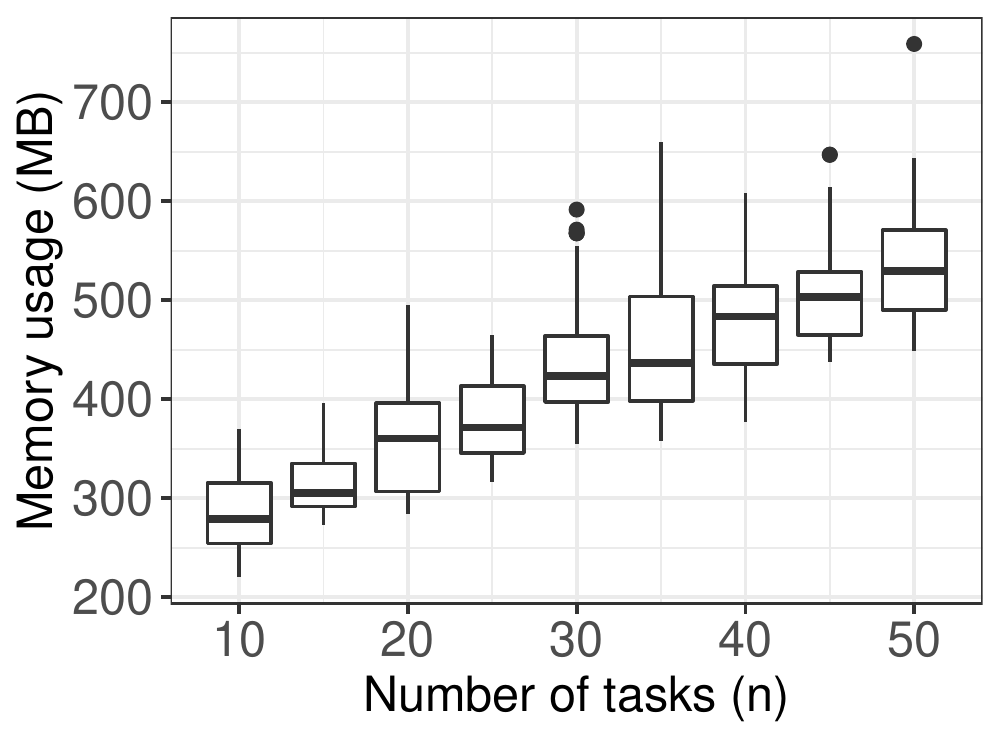}
        \label{fig:rq3memory exp1}
    }
    \hspace*{\fill}
    \subfloat[Ratio of aperiodic tasks($\gamma$)]{
        \includegraphics[width=.45\columnwidth]{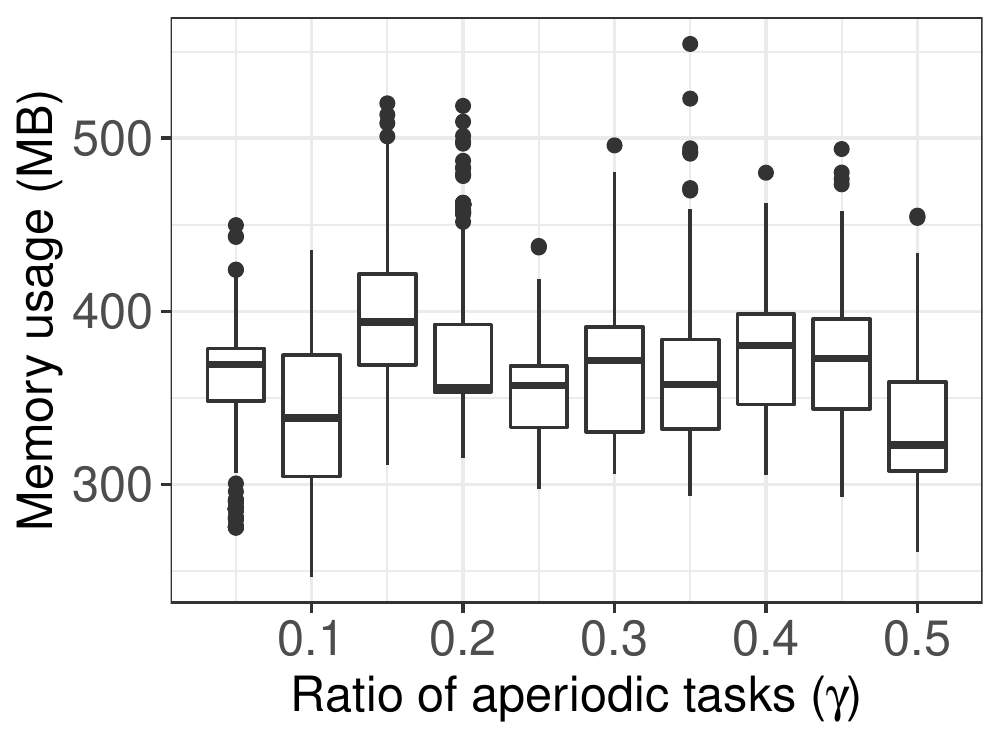}
        \label{fig:rq3memory exp2}
    }
    \hfill
    \subfloat[Range factor ($\mu$)]{
        \includegraphics[width=.45\columnwidth]{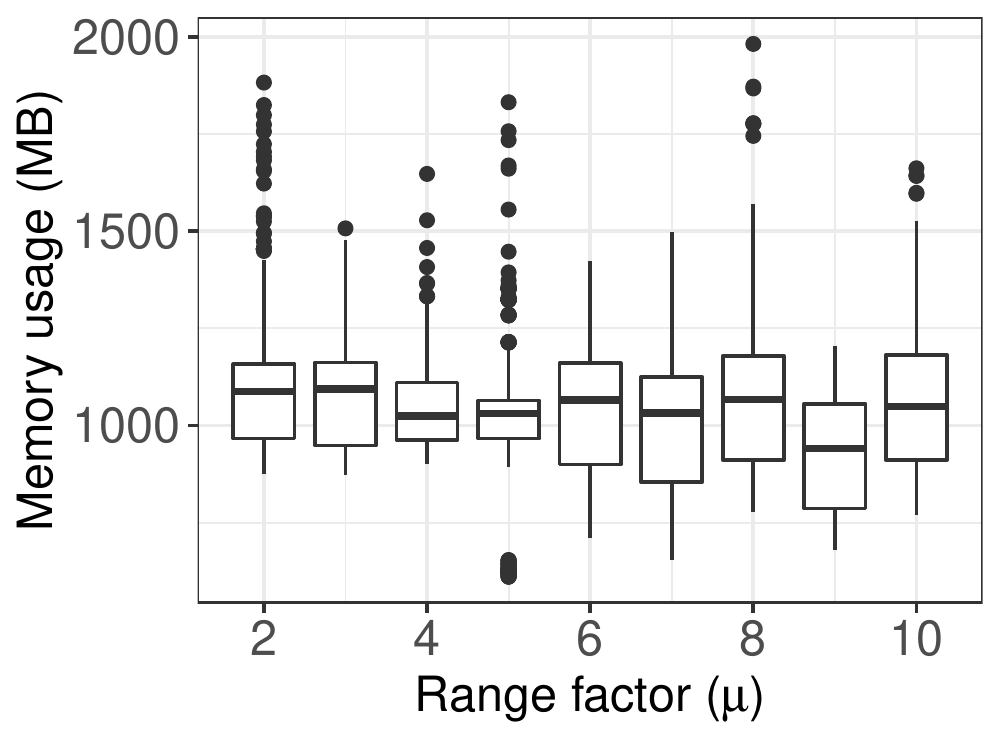}
        \label{fig:rq3memory exp3}
    }
    \hspace*{\fill}
    \subfloat[Simulation time ($T$)]{
        \includegraphics[width=.45\columnwidth]{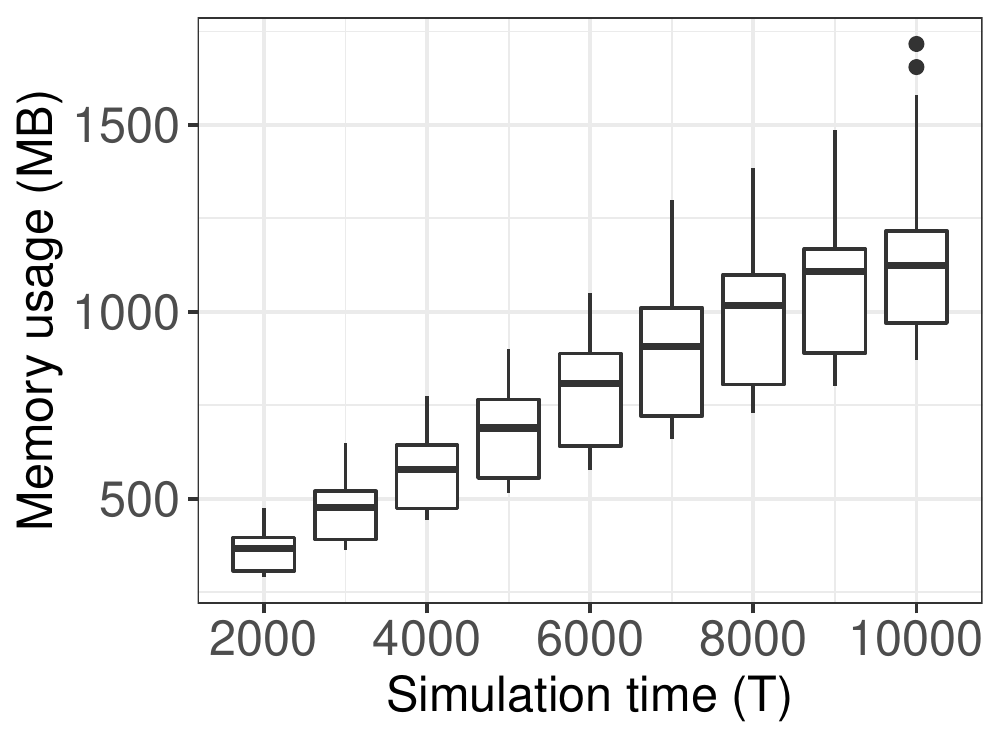}
        \label{fig:rq3memory exp4}
    }
\end{center}
\caption{Memory usage of OPAM when varying the values of the following parameters: (a)~number of tasks $n$, (b)~ratio of aperiodic tasks $\gamma$, (c)~range factor $\mu$, and (d)~simulation time $T$. The boxplots (25\%-50\%-75\%) show the memory usage obtained from 500 runs of OPAM, i.e., 50 runs for each of the synthetic subjects with the same configuration.}
\label{fig:rq3memory}
\end{figure}

Figures~\ref{fig:rq3time} and \ref{fig:rq3memory} show, respectively, the execution times and  memory usage from EXP3.1 (a), EXP3.2 (b), EXP3.3 (c), and EXP3.4 (d), described in Section~\ref{subsec:design}. The boxplots in the figures show distributions (25\%-50\%-75\%) obtained from 50 $\times$ 10 runs of OPAM for a set of 10 synthetic subjects, which are created with the same experimental setting. Regarding the execution time of OPAM, Figures~\ref{fig:rq3time exp1} and \ref{fig:rq3time exp4} show that the execution time of OPAM is linear both in the number of tasks and simulation time. As for the memory usage of OPAM, results in Figures~\ref{fig:rq3memory exp1} and \ref{fig:rq3memory exp4} indicate that memory usage is linear both in the number of tasks and in the simulation time. However, the results depicted in Figures~\ref{fig:rq3time exp2}, \ref{fig:rq3time exp3}, \ref{fig:rq3memory exp2}, and \ref{fig:rq3memory exp3} indicate that there are no correlations between OPAM execution time and memory usage and the following two parameters: ratio of aperiodic tasks and range factor. Therefore, we expect OPAM to scale well as the numbers of tasks and simulation time increase.
 
\noindent\parbox{\textwidth}{
\begin{mdframed}[style=RQFrame]
\emph{The answer to {\bf RQ3} is that} the execution time and memory usage of OPAM are linear in the number of tasks and simulation time, thus scaling to industrial systems. Further, across our experiments, OPAM takes at most 15.5h using 2.9GB of memory to optimize priority assignments, an acceptable result since this is done offline.
\end{mdframed}
}

\begin{figure*}[!th]
\begin{center}
    \subfloat[ICS]{
        \includegraphics[width=0.45\columnwidth]{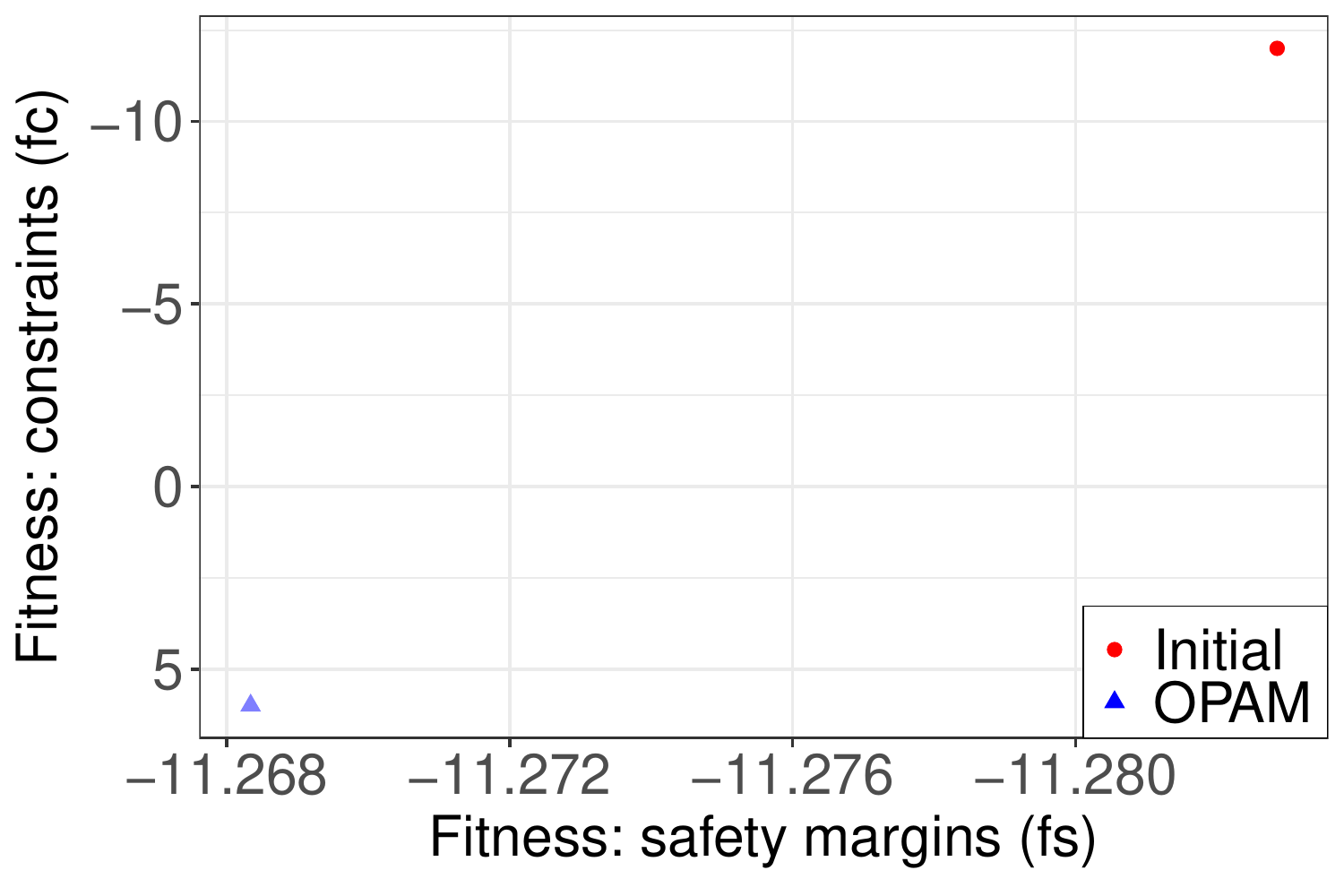}
        \label{fig:rq4scatter ics}
    }
    \hspace*{\fill}
    \subfloat[CCS]{
        \includegraphics[width=.45\columnwidth]{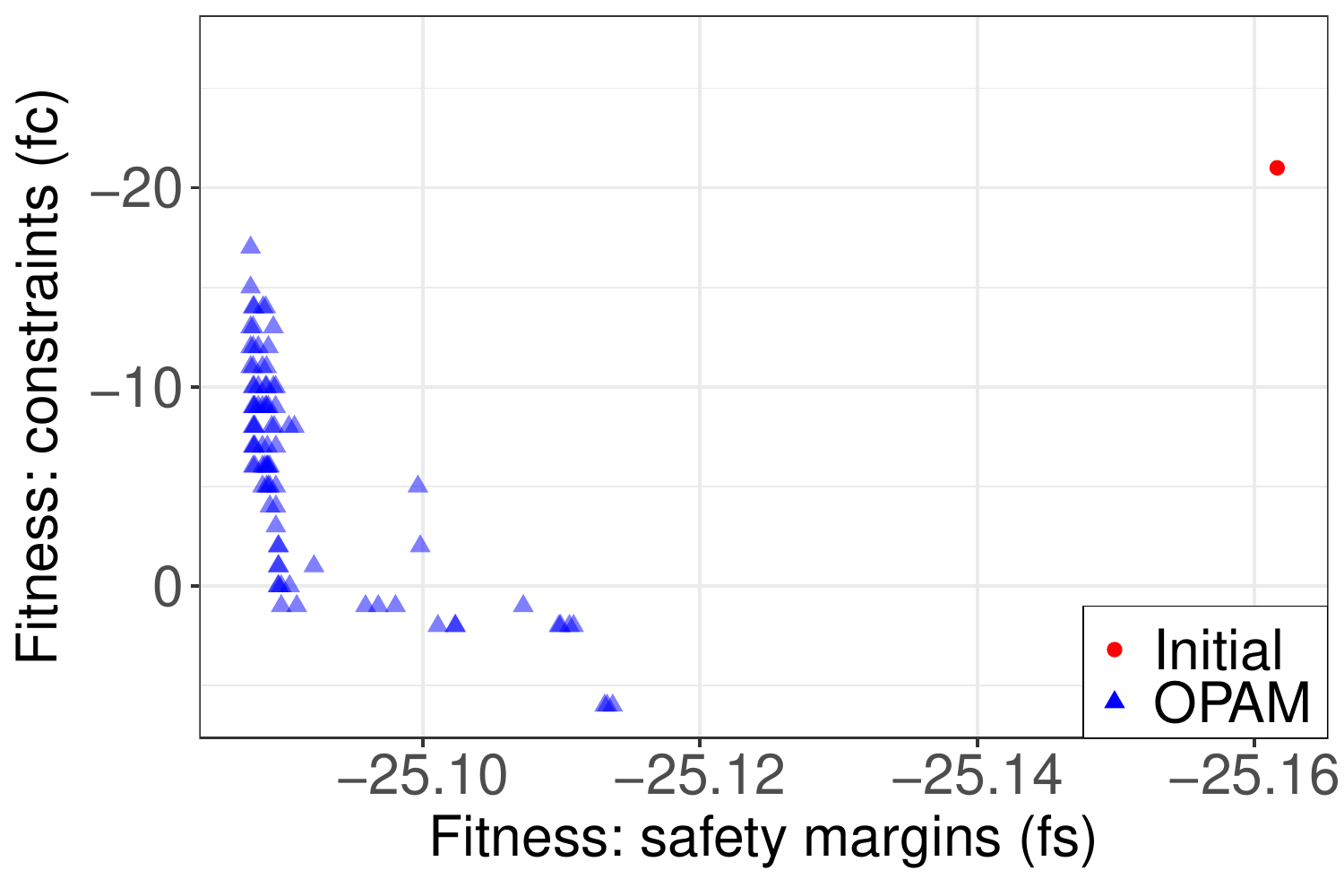}
        \label{fig:rq4scatter ccs}
    }
    \hfill
    \subfloat[UAV]{
        \includegraphics[width=.45\columnwidth]{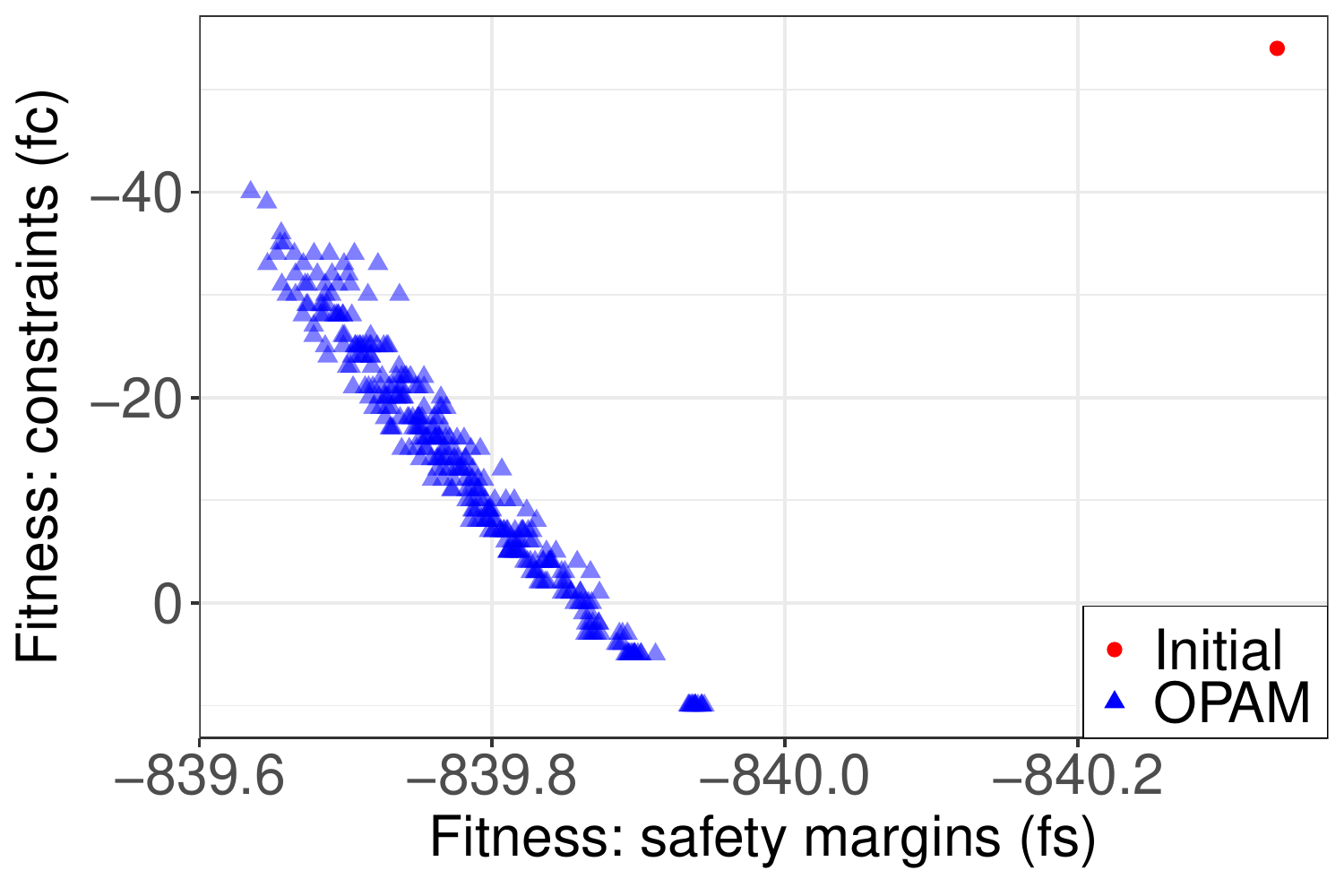}
        \label{fig:rq4scatter uav}
    }
    \hspace*{\fill}
    \subfloat[GAP]{
        \includegraphics[width=.45\columnwidth]{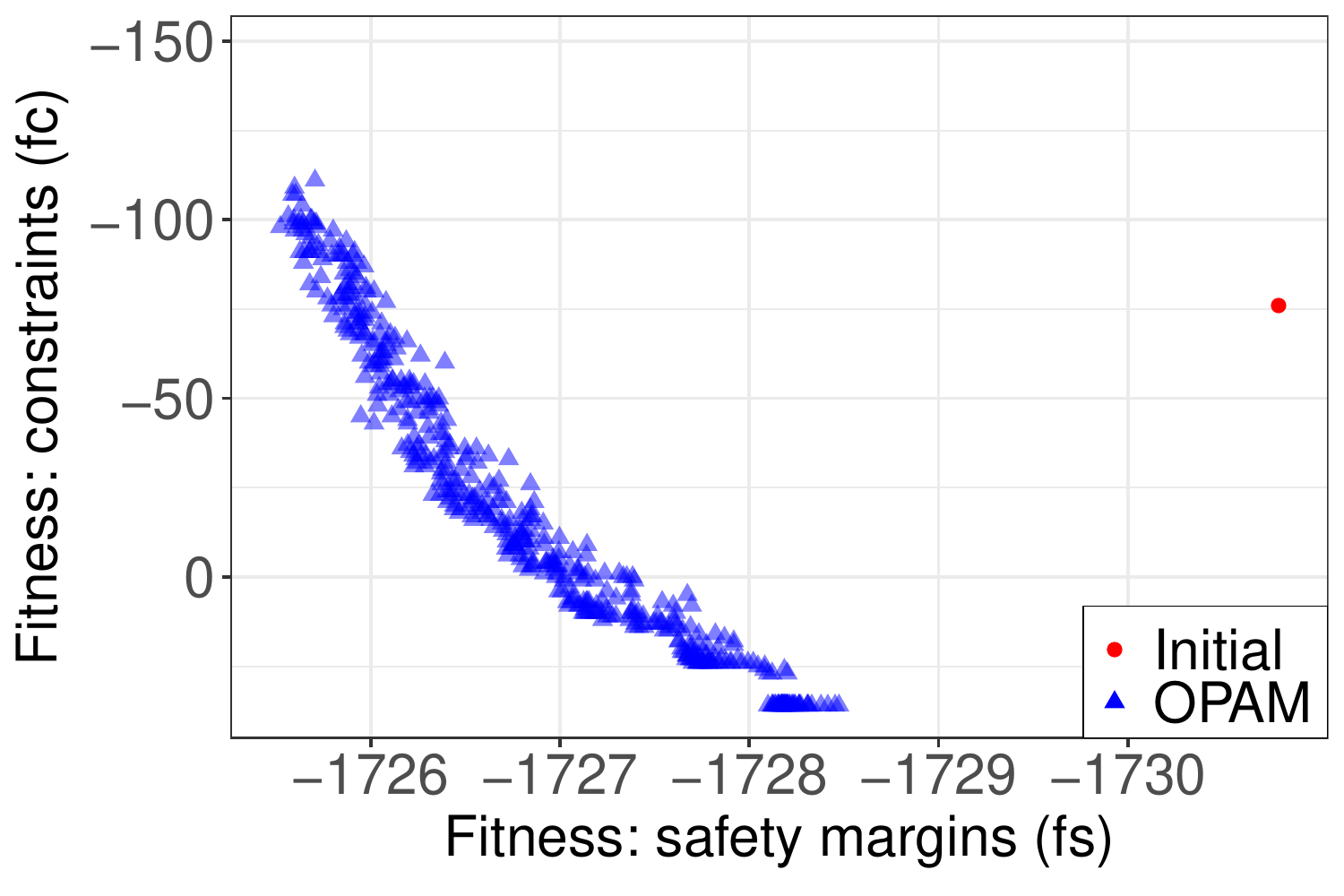}
        \label{fig:rq4scatter gap}
    }
    \hfill
    \subfloat[HPSS]{
        \includegraphics[width=.45\columnwidth]{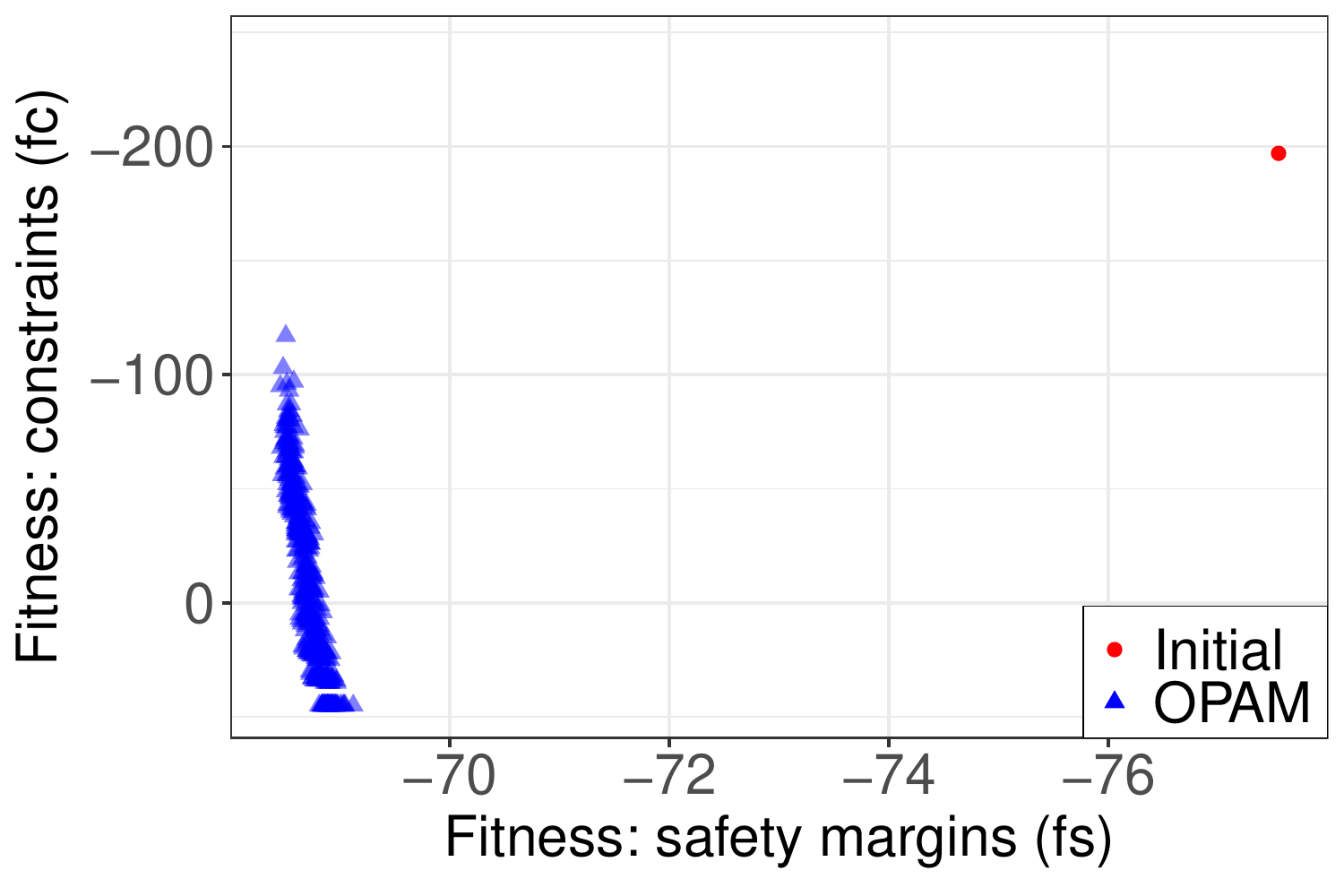}
        \label{fig:rq4scatter hpss}
    }
    \hspace*{\fill}
    \subfloat[ESAIL]{
        \includegraphics[width=0.45\columnwidth]{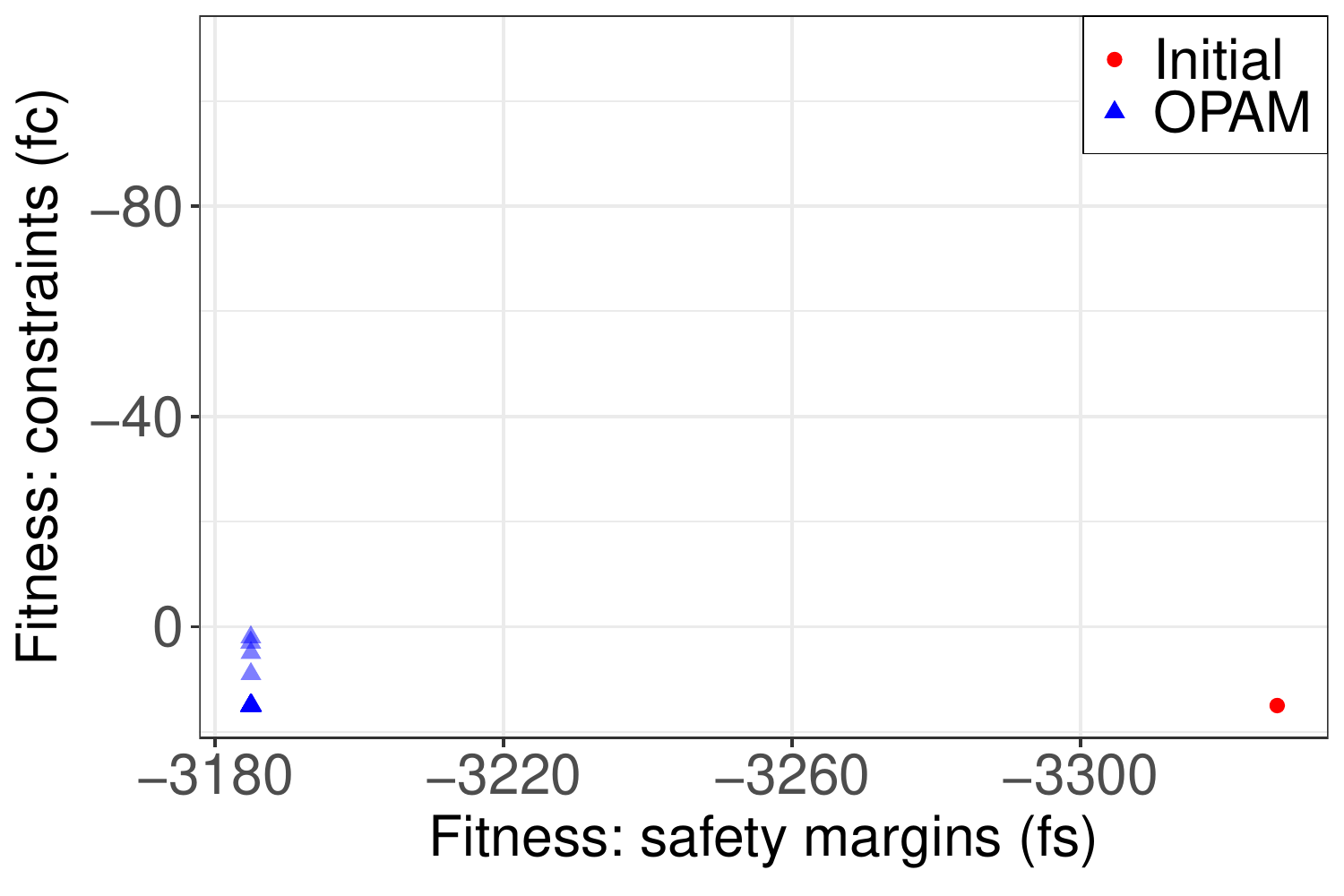}
        \label{fig:rq4scatter esail}
    }
\end{center}
\caption{Comparing Pareto solutions obtained by OPAM and priority assignments defined by engineers for the six industrial subjects: (a)~ICS, (b)~CCS, (c)~UAV, (d)~GAP, (e)~HPSS, and (f)~ESAIL. The points located closer to the bottom left of each plot are considered to be better priority assignments when compared to points closer to the top right.}
\label{fig:rq4scatter}
\end{figure*}

\noindent\textbf{RQ4.}
Figure~\ref{fig:rq4scatter} compares, with respect to external fitness (see the $\fun{fs}()$ and $\fun{fc}()$ fitness functions and the set $\mathbf{E}$ of sequences of task arrivals described in Section~\ref{subsec:external}), the Pareto solutions obtained by OPAM against the priority assignments defined by engineers for the six industrial subjects: ICS (Figure~\ref{fig:rq4scatter ics}), CCS (Figure~\ref{fig:rq4scatter ccs}), UAV (Figure~\ref{fig:rq4scatter uav}), GAP (Figure~\ref{fig:rq4scatter gap}), HPSS (Figure~\ref{fig:rq4scatter hpss}), and ESAIL (Figure~\ref{fig:rq4scatter esail}). 

As shown in the figure, the solutions obtained by OPAM clearly outperform the priority assignments defined by engineers regarding the two external objectives: the magnitude of safety margins and the extent to which constraints are satisfied.

\begin{table}[t]
\caption{Comparing safety margins from the task executions of ESAIL when using our optimized priority assignment and the one defined by engineers.}
	\begin{center}
	\begin{tabular}{ccrrr}
        \toprule
        \multicolumn{1}{l}{} & \multicolumn{1}{l}{} & \multicolumn{1}{c}{Periodic tasks} & \multicolumn{1}{c}{Aperiodic tasks} & \multicolumn{1}{c}{All tasks} \\
        \midrule
        \multirow{4}{*}{Engineer} & Min& -44.5& 9.4& -44.5 \\
            & Max& 1879.7& 59710.3& 59710.3 \\
            & Avg.& 126.6& 52.6& 78.1 \\
            & Median& 82.1& 9.4& 48.1 \\
        \midrule
        \multirow{4}{*}{OPAM} & Min& 48.1& 9.4& 9.4 \\
            & Max& 1879.7& 59707.2& 59707.2 \\
            & Avg.& 129.8& 57.2& 82.3 \\
            & Median& 85.7& 9.4& 48.1 \\
        \midrule
        \multirow{4}{*}{\% Difference} & Min& 208.09\%& 0.00\%& 121.12\% \\
            & Max& 0.00\%& -0.01\%& -0.01\% \\
            & Avg.& 2.53\%& 8.89\%& 5.33\% \\
            & Median& 4.38\%& 0.00\%& 0.00\% \\
        \bottomrule
        \multicolumn{5}{l}{\footnotesize $\ast$ Unit of time: ms}
	\end{tabular}
	\end{center}
\label{tbl:rq4margins}
\end{table}

Table~\ref{tbl:rq4margins} summarizes safety margins from the task executions of ESAIL when using one of our priority assignments optimized by OPAM and the one defined by engineers at LuxSpace. Note that we focus on ESAIL as it is not possible to access the engineers who developed the other five industrial subjects reported in the literature~\citep{Locke1990,Traore2006,Frati2008,Marius2010,Anssi2011}. For comparison, we chose the bottom-left solution in Figure~\ref{fig:rq4scatter esail} since it is optimal for the constraint fitness, which is the same as the fitness value of the priority assignment defined by engineers, and the differences in safety margin fitness among our solutions are negligible.

As shown in Table~\ref{tbl:rq4margins}, our optimized priority assignment significantly outperforms the one of engineers. Our solution increases safety margins, on average, by 5.33\% compared to the engineers' solution. For aperiodic tasks, our solution decreases safety margins by 0.01\% (4.2ms difference) when the safety margins being compared are the maximum margins observed in both solutions (see the maximum safety margins, 59710.3ms obtained by engineers' solution and 59707.2ms obtained by OPAM, in Table~\ref{tbl:rq4margins}). Such a small decrease is however negligible in the context of ESAIL as the maximum safety margin obtained by our solution is still large, i.e., $\approx$1m. For periodic tasks, we note that our solution increases safety margins by 208.09\% when the safety margins being compared are the minimum margins observed in both solutions (see the minimum safety margins, -44.5ms obtained by engineers' solution and 48.1ms obtained by OPAM, in Table~\ref{tbl:rq4margins}). Note that the minimum safety margin of -44.5ms obtained with the engineers' solution indicates that a task violates its deadline. In the context of ESAIL, which is a mission-critical system, such gain in safety margins in the executions of periodic tasks is important because the hard deadlines of periodic tasks are more critical than the soft deadlines of aperiodic tasks.

Investigating practitioners' perceptions of the benefits of OPAM is necessary to adopt OPAM in practice. To do so, we draw on the qualitative reflections of three software engineers at LuxSpace, with whom we have been collaborating on this research. They have had four to seven years of experience developing satellite systems at LuxSpace, with more than 50 years of collective experience in companies. All the reflections are based on observations made throughout our interactions. The engineers at LuxSpace deemed OPAM to be an improvement over their current practice as it allows them to perform domain-specific trade-off analysis among Pareto solutions and is useful in practice to support decision making with respect to their task design. Encouraged by the promising results, we are now applying OPAM to new systems in collaboration with LuxSpace.

\noindent\parbox{\textwidth}{
\begin{mdframed}[style=RQFrame]
\emph{The answer to {\bf RQ4} is that} OPAM helps optimize priority assignments such that they outperform those manually defined by engineers based on domain expertise. Our results show that OPAM, compared to current practice, increases safety margins, on average, by 5.33\%.
\end{mdframed}
}
\subsection{Threats to Validity}
\label{subsec:threats}

To mitigate the main threats that arise from not accounting for random variation, we compared OPAM against RS under identical parameter settings. We present all the underlying parameters and provide the full package of our experiments to facilitate replication. Also, we ran OPAM 50 times for each study subject and compared results using statistical analysis, i.e., Mann-Whitney U-test and Vargha and Delaney's $\hat{A}_{12}$.

We note that there are prior studies that aim at optimizing priority assignments such as OPA~\citep{Audsley1991} and RPA~\citep{Davis2007}. However, to our knowledge, none of the existing works offer ways to analyze trade-offs among equally viable priority assignments with respect to safety margins and the satisfaction of constraints. Nevertheless, we attempted to compare OPAM with an extension of an existing method, e.g., RPA~\citep{Davis2007}. To do so, we first applied an exhaustive schedulability analysis technique to the ESAIL subject -- our motivating case study -- in order to verify whether the ESAIL tasks are schedulable for a given priority assignment. Note that existing priority assignment techniques are built on such schedulability analysis methods, which are therefore a prerequisite. We chose UPPAAL~\citep{Behrmann2004}, a model checker, for schedulability analysis as it has been used in real-time system studies~\citep{Marius2010,Yu2010,Yalcinkaya2019}. However, our experiment results using UPPAAL for ESAIL showed that it was not able to complete the analysis task, even after 5 days of execution, for a single priority assignment. We were therefore not able to perform experimental comparisons with existing priority assignment methods. Since this evaluation is not the main focus of this article, we point the reader to the UPPAAL specification of ESAIL available online~\citep{Artifacts}.

Recall from Section~\ref{subsec:simulation} that OPAM assigns tasks' WCETs to their execution times when it simulates the worst-case executions of tasks while varying task arrival times. In many real-time systems studies~\citep{Briand2005, Guan2009, Lin2009, Anssi2011, Zeng2014, Alesio2015, Durr2019}, static WCETs are often used instead of varying task execution times for the purpose of real-time analysis. For example, practitioners typically use WCETs to estimate the lowest bound of CPU utilization required to properly apply the rate monotonic scheduling policy~\citep{Fineberg1967} to their systems. \textcolor{rev3}{Similarly, OPAM assumes that near-worst-case schedule scenarios can be simulated by assigning tasks' WCETs to their execution times and varying tasks' arrival times using search. A near-worst-case schedule scenario entails that the magnitude of deadline misses is maximized when tasks execute as per this scenario. Under this working assumption, we were able to empirically evaluate the sanity, coevolution, scalability, and usefulness aspects of OPAM (see Section~\ref{sec:eval}). The results indicate that OPAM is a promising and useful tool.} However, the formal proof of whether or not the WCET assumption holds in the system model described in Section~\ref{sec:problem} requires complex analysis, accounting for varying task arrival times, triggering relationships, resource dependencies, and multiple cores. When task execution times need to be varied during simulation, engineers can adapt OPAM by utilizing  Monte-Carlo simulation~\citep{Kroese2014} to account for such variations.

The main threat to external validity is that our results may not generalize to other systems. We mitigate potential biases and errors in our experiments by drawing on real industrial subjects from different domains and several synthetic subjects. Specifically, we selected two subjects from the aerospace domain, two from the automotive domain, and two from the avionics domain. The positive feedback obtained from LuxSpace and the encouraging results from our industrial case studies indicate that OPAM is a scalable and practical solution. \textcolor{rev3}{Furthermore, we believe OPAM introduces a promising avenue for addressing the problem of priority assignment by applying coevolutionary algorithms, even for systems that use other scheduling policies, e.g., priority inheritance. In order for OPAM to support different scheduling policies, the main requirement is to replace the existing simulator (described in Section~\ref{sec:approach}) with a new simulator supporting the desired scheduling policy. In our approach, the coevolution part of OPAM is separated from the scheduling policy, which is contained in the simulator. Hence, we deem the expected changes for the coevolution part of OPAM to be minimal. Future studies are nevertheless necessary to investigate how OPAM can be adapted to find near-optimal priority assignments for other real-time systems in different contexts.
}

\section{Conclusion}
\label{sec:conclusion}

We developed OPAM, a priority assignment method for real-time systems, that aims to find equally viable priority assignments that maximize the magnitude of safety margins and the extent to which engineering constraints are satisfied. OPAM uses a novel approach, based on multi-objective, competitive coevolutionary search, that simultaneously evolves different species, i.e., populations of priority assignments and stress test scenarios, that compete with one another with opposite objectives, the former trying to minimize chances of deadline misses while the latter attempts to maximize them. We evaluated OPAM on a number of synthetic systems as well as six industrial systems from different domains. The results indicate that OPAM is able to find significantly better solutions than both those manually defined by engineers based on expert knowledge and those obtained by our baselines: random search and sequential search. Further, OPAM scales linearly with the number of tasks in a system and the time required to simulate task executions. Execution times on our industrial systems are practically acceptable. 

\textcolor{rev3}{In the future, we will continue to study the problem of optimal priority assignment by accounting for (1)~priority assignments that change dynamically, (2)~WCET value ranges that account for non-deterministic computation times, (3)~interrupt handling routines that execute differently compared to real-time tasks, and (4)~hybrid scheduling policies that combine multiple standard policies.} We also plan to develop a real-time task modeling language to specify task characteristics such as resource dependencies, triggering relationships, engineering constraints, and behaviors of real-time tasks and to facilitate real-time system analysis, e.g., optimal priority assignment and schedulability analysis. 
In addition, we would like to incorporate additional analysis capabilities into OPAM in order to verify whether or not a system satisfies the required properties, e.g., schedulability of tasks and absence of deadlocks, for a given priority assignment. For example, statistical model checking~\citep{Legay2010} may allow us to verify whether tasks meet their deadlines for a given priority assignment with a probabilistic guarantee.
In the long term, we plan to more conclusively validate the usefulness of OPAM by applying it to additional case studies in different application domains. 

\begin{acknowledgements}
We thank Yago Isasi Parache, LuxSpace, for his support in conducting our industrial case study. This project has received funding from the European Research Council (ERC) under the European Union's Horizon 2020 research and innovation programme (grant agreement No 694277), and NSERC of Canada under the Discovery and CRC programs.
The experiments presented in this paper were carried out using the HPC facilities of the University of Luxembourg~\citep{Varrette2014}
{\small -- see \href{http://hpc.uni.lu}{hpc.uni.lu}}.
\end{acknowledgements}

%
%
\balance
\bibliography{ref}

\end{document}